\documentclass[twocolumn, reprint, aps,superscriptaddress,floatfix,letterpaper,nofootinbib]{revtex4-1}
\usepackage{nicefrac}
\usepackage{braket}
\usepackage{amsmath, amssymb, graphicx}
\usepackage{changes}
\usepackage{xcolor}
\usepackage{mathtools}
\usepackage[caption=false]{subfig}
\usepackage{comment}
\usepackage{subfig}
\usepackage{qcircuit}

\newcommand{\IBM}{IBM Quantum, IBM T.J. Watson Research Center, Yorktown Heights, NY 10598, USA}
\newcommand{\PNNL}{Pacific Northwest National Laboratory, Richland, WA 99354, USA}
\newcommand{\BNL}{Brookhaven National Laboratory, Upton, NY 11973, USA}
\newcommand{\MIT}{Department of Physics, Massachusetts Institute of
Technology, Cambridge, MA 02139, USA}
\newcommand{\VT}{Department of Physics, Virginia Tech, Blacksburg, VA 24061, USA}
\newcommand{\PrincetonCS}{Department of Computer Science, Princeton University, Princeton, NJ 08544, USA}
\newcommand{\PrincetonEE}{Department of Electrical Engineering, Princeton University, Princeton, USA} 
\newcommand{\YaleAP}{Departments of Applied Physics and Physics, Yale University, New Haven, CT 06520, USA}
\newcommand{\YaleEE}{Departments of Electrical Engineering, Yale University, New Haven, CT 06511, USA}
\newcommand{\YaleQI}{Yale Quantum Institute, Yale University, New Haven, CT 06511, USA}
\newcommand{\Pitt}{Department of Physics and Astronomy, University of Pittsburgh, Pittsburgh, PA 15260, USA}
\newcommand{\AmesIowa}{Ames National Laboratory and Iowa State University, Ames, Iowa 50011, USA}
\newcommand{\NASA}{Quantum Artificial Intelligence Laboratory (QuAIL), Exploration Technology Directorate, NASA Ames Research Center, Moffett Field, California 94035, USA}
\newcommand{\KBR}{KBR, 601 Jefferson St., Houston, Texas 77002, USA}
\newcommand{\UT}{Department of Computer Science, University of Toronto, Toronto, ON M5S1J7, Canada}
\newcommand{\Caltech}{Thomas J. Watson, Sr., Laboratory of Applied Physics, California Institute of Technology, Pasadena, CA, USA}
\newcommand{\UW}{Department of Physics, University of Washington, Seattle WA 98195, USA}
\newcommand{\UMassA}{Department of Physics, University of Massachusetts-Amherst, Amherst, MA, 01003, USA}

\newcommand{\mywidthL}{3.8cm}
\newcommand{\mywidthR}{4.2cm}

\begin{document}
\begin{titlepage}

\title{
Architectures for Multinode Superconducting Quantum Computers
}

\author{James Ang}
\affiliation{\PNNL}
\author{Gabriella Carini}
\affiliation{\BNL}
\author{Yanzhu Chen}
\affiliation{\VT}
\author{Isaac Chuang}
\affiliation{\MIT}
\author{Michael Austin DeMarco}
\email{mdemarco@bnl.gov}
\affiliation{\BNL}
\affiliation{\MIT}
\author{Sophia E. Economou}
\affiliation{\VT}
\author{Alec Eickbusch}
\affiliation{\YaleQI}
\affiliation{\YaleAP}
\author{Andrei Faraon}
\affiliation{\Caltech}
\author{Kai-Mei Fu}
\affiliation{\UW}
\author{Steven M. Girvin}
\affiliation{\YaleQI}
\author{Michael Hatridge}
\affiliation{\Pitt}
\author{Andrew Houck}
\affiliation{\PrincetonEE}
\author{Paul Hilaire}
\affiliation{\VT}
\author{Kevin Krsulich}
\affiliation{\IBM}
\author{Ang Li}
\affiliation{\PNNL}
\author{Chenxu Liu}
\affiliation{\VT}
\author{Yuan Liu}
\affiliation{\MIT}
\author{Margaret Martonosi}
\affiliation{\PrincetonCS}
\author{David C. McKay}
\affiliation{\IBM}
\author{James Misewich}
\affiliation{\BNL}
\author{Mark Ritter}
\affiliation{\IBM}
\author{Robert J. Schoelkopf}
\affiliation{\YaleAP}
\author{Samuel A. Stein}
\affiliation{\PNNL}
\author{Sara Sussman}
\affiliation{\PrincetonEE}
\author{Hong X. Tang}
\affiliation{\YaleEE}
\affiliation{\YaleQI}
\author{Wei Tang}
\affiliation{\PrincetonCS}
\author{Teague Tomesh} 
\affiliation{\PrincetonCS}
\author{Norm M. Tubman}
\affiliation{\NASA}
\author{Chen Wang}
\affiliation{\UMassA}
\author{Nathan Wiebe}
\affiliation{\PNNL}
\affiliation{\UT}
\author{Yong-Xin Yao}
\affiliation{\AmesIowa}
\author{Dillon C. Yost}
\affiliation{\NASA}
\affiliation{\KBR}
\author{Yiyu Zhou}
\affiliation{\YaleQI}
\affiliation{\YaleEE}

\begin{abstract}

Many proposals to scale quantum technology rely on modular or distributed designs  where individual quantum processors, called nodes, are linked together to form one large multinode quantum computer (MNQC). One scalable method to construct an MNQC is using superconducting quantum systems with optical interconnects. However, a key limiting factor of these machines will be internode gates, which may be two to three orders of magnitude noisier and slower than local operations. Surmounting the limitations of  internode gates will require a range of techniques, including improvements in entanglement generation, the use of entanglement distillation, and optimized software and compilers, and it remains unclear how improvements to these components interact to affect overall system performance, what performance from each is required, or even how to quantify the performance of a component. 
In this paper, we employ a `co-design' inspired approach to quantify overall MNQC performance in terms of hardware models of internode links, entanglement distillation, and local architecture.
In the particular case of superconducting MNQCs with microwave-to-optical interconnects, we uncover a tradeoff between entanglement generation and distillation that threatens to degrade MNQC performance. We show how to navigate this tradeoff in the context of algorithm performance, layout how compilers and software should optimize the balance between local gates and internode gates, and discuss when noisy quantum internode links have an advantage over purely classical links. Using these results, we introduce a research roadmap for the realization of early MNQCs, which illustrates potential improvements to the hardware and software of MNQCs and outlines criteria for evaluating the improvement landscape, from progress in entanglement generation to the use of quantum memory in entanglement distillation and dedicated algorithms such as distributed quantum phase estimation. While we focus on superconducting devices with optical interconnects, our approach is general across MNQC implementations. 
\end{abstract}

\maketitle

\end{titlepage}

\section{Introduction} 

Modular, distributed, or multinode quantum computers (MNQCs) \cite{2022arXiv220906841B, IBMRoadmap, 2012arXiv1208.0391M, 2019NatCo..10.4692Z, LaRacuente:2022xqq, 2014PhRvA..89b2317M, Zhou:2021cri, doi:10.1126/science.abg1919}, wherein smaller devices or ``nodes'' are networked together~\cite{PRXQuantum.2.017002} to make a unified multinode quantum computer, are considered a leading approach to building large scale systems~\cite{Gyongyosi:2021yol} without the associated difficulties of producing large monolithic devices \cite{LaRacuente:2022xqq}. Leading platforms include ion-trap computers with multiple traps~\cite{2019ApPRv...6b1314B, 2021Natur.592..209P, 2020AVSQS...2a4101K, 2021Natur.592..209P}, solid-state systems ~\cite{2013Natur.497...86B, 2016NatSR...626284N, 2021Sci...372..259P}, atomic systems~\cite{2007Natur.449...68M, 2012Natur.484..195R, 2022ApPhB.128..151Y}, and superconducting devices~\cite{IBMRoadmap, 2021npjQI...7..142G, 2018QS&T....3a4005F, 2021PRXQ....2c0321B, Magnard2020, 2016PhRvX...6c1036N, 2013Sci...339.1169D, 2019arXiv190513641K, 2004cond.mat.11174D, Zhou:2021cri}. 

In superconducting devices, which we focus on in this paper, a motivation for MNQCs is not only the complexities associated with building larger devices, but the limitations set by the individual capacity of the cryogenic dilution refrigerator required to cool the device \cite{2018arXiv180607862K}. Building links between devices in different refrigerators is thus a key capability \cite{PhysRevX.4.041041}. Early-stage MNQCs with cryogenic links between refrigerators have been demonstrated~\cite{Magnard2020}, and when cryogenic links can be feasibly constructed they are a leading candidate for small MNQCs~\cite{Yan2022}. On the other hand, future large quantum systems may involve many nodes distributed over tens or even hundreds of meters, at which scale both serviceability requirements \cite{2022Cryo..12103390Z} and cable loss \cite{Axline2018, Campagne2018, Kurpiers2018,Leung2019,Magnard2020, Burkhart2021, Yan2022} become an issue. Rather than using cryogenic links, a system composed of transmon array devices \cite{Place2021,Wang2022} housed in separate refrigerators with room-temperature microwave-to-optical (M2O) quantum internode links \cite{han2021microwave, kurizki2015quantum, lambert2020coherent, lauk2020perspectives, chu2020perspective, clerk2020hybrid, 2020Natur.588..599M} between them is a more scalable proposal for building future MNQCs. 

Despite many promising experimental platforms, internode links in these systems are likely to be much noisier and slower than local gates and thus threaten the viability of MNQCs \cite{2019arXiv191206642A, Qiao:2022rvy}. These weak internode links hamper performance both by directly causing errors and by creating a computational bottleneck which allows decoherence \cite{Place2021,Wang2022} to degrade quantum information. While true across platforms, this problem is particularly pronounced in superconducting devices with M2O links, where the conversion faces serious limitations due to the weakness of the nonlinear conversion process, fiber-to-chip coupling, thermal added noise, and other hardware difficulties~\cite{han2021microwave, kurizki2015quantum, lambert2020coherent, lauk2020perspectives, chu2020perspective, clerk2020hybrid}. 
In order to be viable, systems with quantum internode links must outperform not only monolithic quantum systems but also systems with only classical `circuit cutting' links between nodes \cite{tang2021cutqc, tang2022scaleqc}. 

To guide the development of M2O MNQCs, we must quantify the available performances of internode links using M2O hardware, determine how these performances affect algorithm execution performance, and determine how hardware and software should jointly navigate design space tradeoffs. 

However, evaluating the performance of MNQCs becomes complicated because MNQCs must balance three expensive resources: local two-qubit gates, internode gates, and classical circuit cutting links. In particular, any internode gate may always be cut and replaced with a circuit cutting link, thereby exchanging the noise and time of an internode link for the multiple executions required by circuit cutting~\cite{bravyi2016,peng2020simulating,mitarai2020,sutter2022}. On the other hand, compiler and algorithm design can often trade an internode gate for increased local computation, thus exchanging an internode link for a longer and deeper circuit~\cite{ferrari2020compiler, wu2022autocomm}. Thus, in order to understand the attainable performance of an MNQC, we must not only understand the computational cost of internode gates, but also the relative costs of local computation and circuit cutting vis-a-vis internode gates.

\begin{figure}
    \centering
    \includegraphics[width = \columnwidth]{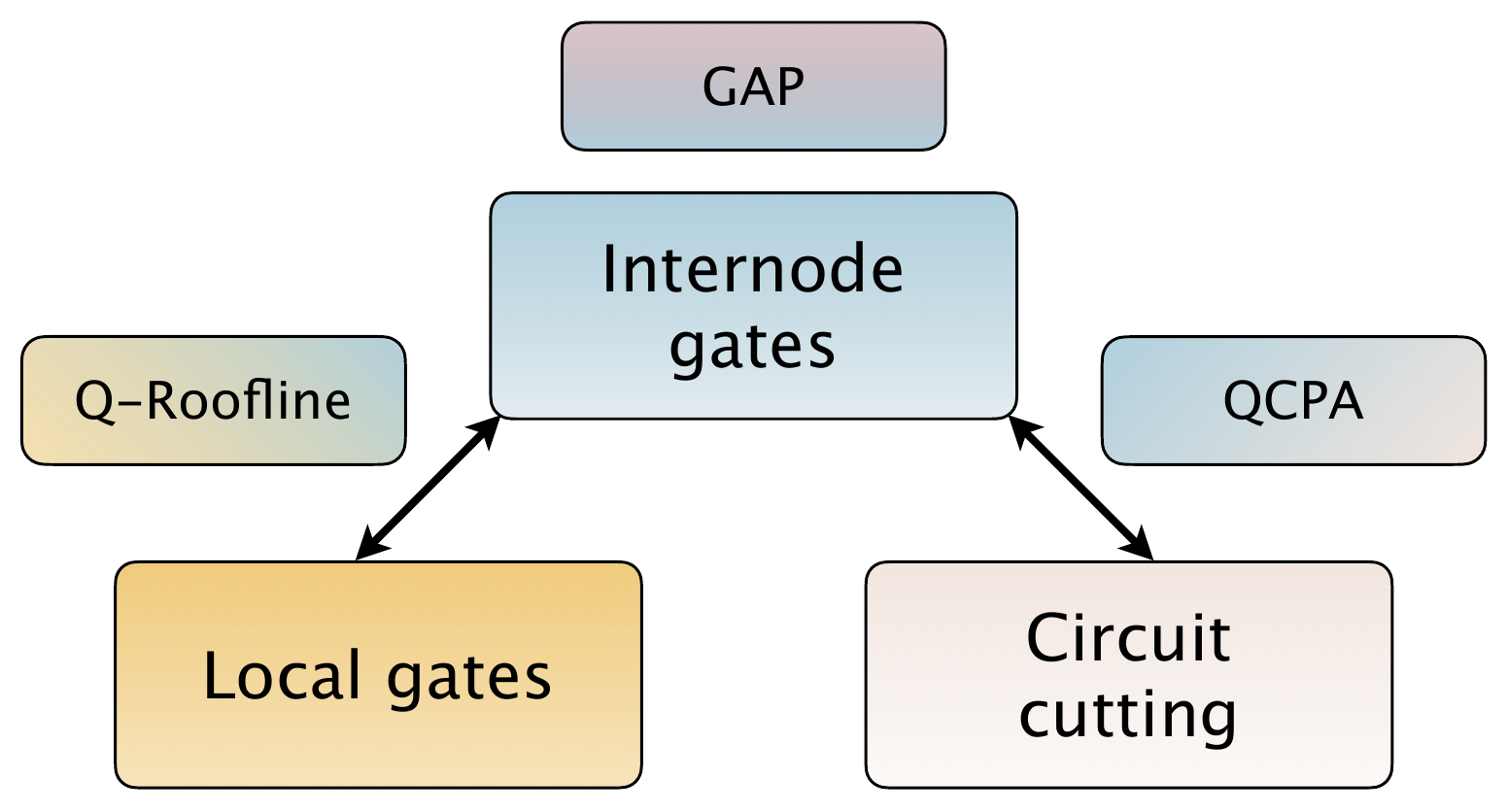}
    \caption{Multinode Quantum Computers (MNQCs) must balance three key resources: internode gates, local computation, and circuit cutting gates. To evaluate the costs of these resources, we introduce three models. The Gate-Algorithm Performance (GAP) model determines the overall performance of internode gates in terms of detailed models of the internode hardware and local distillation and compares this to the demands of algorithms. Building on this analysis, the Quantum Roofline (Q-Roofline) model compares the relative costs of internode gates and local computation and guides compiler balancing between the two. Similarly, the Quantum vs. Classical Performance Analysis (QCPA) evaluates the relative cost of internode gates and circuit cutting gates to optimize the usage of both.}
    \label{fig:models_overview}
\end{figure}

Evaluation of internode gate performance itself presents a challenge of complexity and scale. 
Rather than directly transmitting quantum information between nodes, MNQCs can use heralded protocols to distribute entangled pairs (EPs)~\cite{2019arXiv191206642A, rueda2016efficient, mckenna2020cryogenic} between nodes, distill them~\cite{Dur2003, Yan2022, Bennett1996}, and execute teleported gates \cite{krastanov2021optically}. However, modeling this as a unified system quickly becomes untenable: modeling the internode link including the entanglement generation \cite{krastanov2021optically} and up to just six rounds of distillation requiring $2^6$ EPs quickly scales to require a minimum of 18 qubits just in the internode link, rendering the simulation of a just a 10-qubit MNQC equivalent to the simulation of a $2^{28}\times 2^{28}$ density matrix, well beyond what can be simulated today~\cite{li2020density, o2017density}. Furthermore, understanding the relative costs of internode gates, ciruit cutting, and local computation will require integrating the already large internode link performance models with an understanding of compiler frameworks \cite{wu2022collcomm, ferrari2020compiler, wu2022autocomm, dadkhah2022reordering, beals2013efficient} and network architectures that are able to optimize around the weak links and scale with system sizes. Thus to address the scaling behavior of large MNQCs, we must determine simple and scalable ways to evaluate circuit cutting and local computation against internode gate performance.  Without a clear understanding of how all the components of an MNQC, from entanglement generation and distillation to compiler and algorithm design, interact to affect system performance, future improvements in hardware and software may be incompatible and lead to reduced or even no improvements in MNQC performance \cite{10.1145/3477206.3477464}. 

Classical computing has navigated similarly complex design constraints to build high-performance multi-node systems \cite{tanenbaum2007distributed, asanovic2006landscape, van2016brief}. Early multicomputers including the ALEWIFE \cite{524544, 747864} and BEOWULF \cite{sterling2003beowulf} systems utilized existing state-of-the art hardware to lay the foundations of classical multinode networks with distributed memory and communication systems that evolved into the interconnection architectures such as the Infiniband and Slingshot networks \cite{2020arXiv200808886D, Infiniband, gara2005overview} used by contemporary high-performance computing systems. Physical hardware and software constraints were key to building modern interconnection architectures, from `fat-tree' networks \cite{6312192} that balance network bandwidth against the size of an architecture to adaptive routing \cite{Scott96thecray} and complex network architectures \cite{4556717} that maximize system performance while minimizing wiring overhead. From an architectural perspective, these tradeoffs are captured in `Roofline' models \cite{williams2009roofline} which quantifies the relative burden of local computation and memory communication. Taken together, this approach of navigating tradeoffs by designing hardware and software jointly came to be called `co-design' \cite{10.1145/2818950.2818959} and has played a significant role in the design of modern high-performance and exascale computing \cite{co-design1, foster2011high}.

On the other hand, considerable research has been directed towards the design of networked quantum systems, of which MNQCs would be a subset. Building on early proposals for quantum networks \cite{1997PhRvL..78.3221C} and quantum internet \cite{2008Natur.453.1023K}, recent works have elaborated a vision for the development of a truly distributed quantum ecosystem \cite{2020arXiv200211808C, 2018Sci...362.9288W,2022arXiv220906841B}, although hardware which is capable of delivering the requisite performance largely remains to be developed \cite{RevModPhys.83.33, Ruihong_2019}. Layered link protocols \cite{Loke:2022bcq, 2019arXiv190309778D, 2019NJPh...21c3003P} focused on the preparation of nonlocal entanglement \cite{PhysRevApplied.17.054021, Loke:2022bcq} modeled from the classical internet have also been elaborated. However, how these will interact with highly constrained platforms has only begun to be understood, with progress on routing optimization \cite{2020arXiv201111644V} and dedicated compilers and frameworks \cite{Cuomo:2021hdi, Wu:2022snb, 2021AVSQS...3c0501B, Wu:2022sfh, 2020arXiv200512259B}. With substantial progress envisioned in the realization of high-performance quantum interlinks~\cite{PRXQuantum.2.017002}, joint co-design of hardware and software will be key to enabling quantum networks~\cite{tomesh2021codesign}. 

In this paper, we use a co-design layer architecture to simplify the problem of internode links and thus quantify the full range of available internode performance with present technology; we then integrate these results into models that determine the relative costs of local, internode, and circuit cutting links to model algorithm performance and map out tradeoffs; and finally we lay out a research roadmap that proposes improvements and quantifies their possible effects in terms of these models. Our analysis reveals a key tradeoff between the time of execution and the fidelity of internode gates and we show how to navigate this using M2O pump power settings, entanglement distillation, and error mitigation techniques. We present simulation results on sample hardware demonstrating this tradeoff and quantify the available performance of internode links. Using these results, we evaluate algorithm execution performance using internode links using a `Gate-Algorithm Performance' (GAP) model, introduce a `Quantum Roofline' (Q-Roofline) model which determines the relative costs of internode and local computation and guides compiler resource balancing, and perform a `Quantum-Classical Performance Analysis' (QCPA) to demonstrate the relative costs of error-mitigated internode links against circuit cutting links. Finally, we discuss proposed improvements to each layer of the MNQC, from entanglement generation and distallation to algorithm and compiler design, and use the GAP, Q-Roofline, and QCPA models to discuss their effects and relative tradeoffs. While we focus on superconducting systems with M2O interlinks in this paper, our approach is generic to any physical MNQC implementation which uses entanglement generation to execute remote operations or with links that may be characterized by the time and fidelity of operations, including quantum networks \cite{2020arXiv200211808C, 2018Sci...362.9288W} and cryogenic microwave links~\cite{bravyi2022,Zhou:2021cri, Yan2022,2021npjQI...7..142G}. 

The next section presents an overview of the specific platform we consider, namely superconducting transmon devices with M2O interlinks. In Section \ref{sec:Architecture}, we discuss the co-design architecture that allows the problem of internode links to be drastically simplified by splitting the internode link into distinct layers. Section \ref{sec:LayerModels} then presents and analyzes models of each of the layers. Section \ref{sec:FullModels} unifies these models into a full stack model, and presents the GAP, Q-Roofline, and QCPA analyses. In Section \ref{sec:Roadmap}, we present a research roadmap for the development of highly performant MNQCs and discuss advances in light of the MNQC architecture and analyses. Finally, Section \ref{sec:Outlook} discusses potential applications of our methodology to other quantum platforms.

\section{Superconducting Devices with M2O Interlinks}\label{sec:SCM2O}

Over the past two decades, the superconducting circuit has become an established platform for large-scale quantum information processing. While systems with several hundred superconducting qubits have been built, and systems with thousands are planned \cite{IBMRoadmap,Chamberland2022}, scaling remains a serious challenge. Available cryogenic capacity and qubit control infrastructure are two major limitations for achieving devices at the scale required for cutting-edge applications. Furthermore, today's large superconducting devices are monolithic so smaller subsystems cannot be tested in isolation or individually replaced to improve system performance. Hence we consider a multinode superconducting quantum computer with each `node' comprised of a refrigerator holding a superconducting circuit which handles intermediate-scale computation tasks via local state preparation, gate operation, and measurement. For simplicity, we assume the classical communication between nodes to be fast, reliable and well-synchronized to a single clock. The key to creating an effective multinode architecture is then establishing quantum links between nodes. While short cryogenic microwave interlinks have been built \cite{Magnard2020}, the challenges of scaling such links between distant refrigerators lead us to consider room temperature links based on M2O transduction. In this section, we review the progression of remote entanglement distribution experiments done with superconducting qubits, and discuss the use of M2O protocols to link them. In particular, we discuss how the transmon decoherence rate sets a lower bound on the M2O transduction rate, which will be an engineering challenge for MNQCs.

Although various superconducting processor designs are being prototyped~\cite{Bao2022,Chamberland2022}, the most widely used architecture in both academic and industry settings is a two-dimensional lattice of nearest-neighbor coupled transmons~\cite{Koch2007,Acharya2022} cooled to milli-Kelvin temperatures in dilution refrigerators. Transmons are robust quantum computing building blocks, fortified by recent engineering breakthroughs in high-fidelity two-qubit gates~\cite{Foxen2020,Sung2021,Wei2022} and individual qubit coherence, which can approach 0.5 ms ~\cite{Place2021,Wang2022}. Transmon processors with as many as 433 qubits are available~\cite{collins_nay} and steady progress is being made towards the goal of fault-tolerant quantum computing using stabilizer codes with 49 qubits~\cite{Acharya2022}, 23 qubits~\cite{Sundaresan2022}, and 17 qubits~\cite{Zhao2022,Krinner2022}. 

Remote entanglement distribution experiments with transmon-qubit-based processors connected by cold microwave links have evolved after a period of intensive engineering. Almost a decade ago, a series of heralding-based probabilistic entanglement distribution experiments were performed between qubits on separate chips within a fridge \cite{Roch2014, 2016PhRvX...6c1036N, Dickel2018}. Then, deterministic entanglement distribution experiments were done between chips separated by 1-5 m of cable, one of which housed the chips in separate fridges \cite{Magnard2020}, and the resulting fidelities were largely limited by cable loss ~\cite{Axline2018, Campagne2018, Kurpiers2018,Leung2019,Magnard2020}. The next generation of deterministic entanglement distribution experiments took care to minimize cable loss so transfer and process infidelities were instead limited by qubit loss \cite{Zhong2019, Chang2020}. Heralded deterministic entanglement experiments have also been done \cite{Kurpiers2019} where sophisticated techniques were developed to mitigate cable loss \cite{Burkhart2021}.
Nonetheless, today's state-of-the-art deterministic entanglement distribution experiments are again limited by cable loss \cite{Zhong2021,Yan2022}. In all of these experiments, the quantum links were cryogenic and their length was on the order of a few meters, which poses a challenge for scaling to a many node system distributing entanglement across tens or hundreds of meters.

While microwave photon loss poses a significant obstacle to scaling MNQCs, optical photons at the telecommunication wavelengths are promising candidates for mediating information exchange due to the extremely low loss and negligible thermal photon noise of optical fibers at room temperature. For medium or long-distance quantum communication between superconducting chips, it is more promising to transduce quantum information from the microwave regime to optical wavelengths and generate entanglement through heralded schemes. Recently, electro-optomechanical transducers were integrated into transmon qubit systems and used for qubit readout \cite{mirhosseini2020superconducting, delaney2022superconducting, lecocq2021control}, but these converters are not yet efficient and broadband enough for use in a remote entanglement distribution experiment. A high-fidelity M2O converter will be an essential component for realizing large-scale distributed superconducting quantum computing.

\begin{figure}[t]
    \centering
    \includegraphics[width=\linewidth]{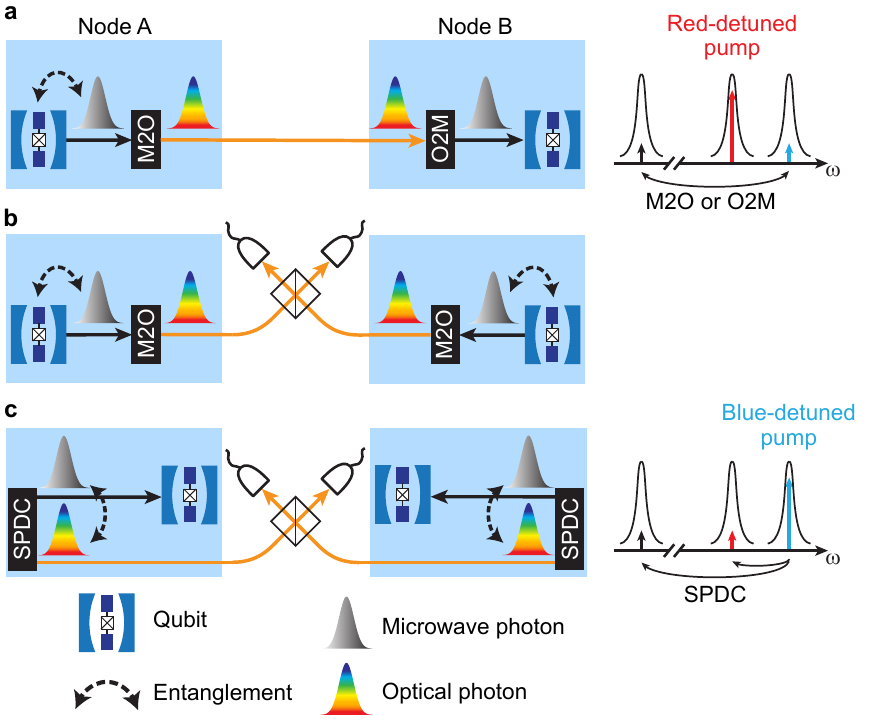}
    \caption{Schemes for entanglement generation between remote nodes. (a), entanglement between a qubit and a microwave photon is prepared at node A. By applying an M2O and a subsequent O2M converter (see the right panel), the state of the microwave photon can be transferred to the qubit at node B.
    (b), the direct conversion heralded scheme. Entangled qubit and microwave photon states $\ket{g0} + \ket{e1}$ are prepared at two nodes. The microwave photons are up-converted to optical photons and interfere at a beamsplitter. The single-photon detector click heralds the generation of entangled qubits $\ket{ge}\pm \ket{eg}$ at two nodes.
    (c), the SPDC heralded scheme. Entangled microwave-optical photon pairs are generated at two nodes when an M2O converter is used as a SPDC source by pumping at the blue-detuned resonance frequency (see the right panel). By erasing the which-path information with a beamsplitter, a click from the detectors heralds the generation of microwave entangled Bell state $\ket{10}\pm \ket{01}$ at two nodes. The microwave photon states are transferred to qubits, which leads to entangled qubit state $\ket{ge}\pm \ket{eg}$.}
    \label{fig:M2O_scheme}
\end{figure}

While an ideal M2O converter should have unity quantum conversion efficiency, in practice the intrinsic weakness of optical nonlinearity poses an extreme challenge for high-efficiency M2O conversion. Various schemes have been proposed and experimentally demonstrated, including cavity electro-optics~\cite{mckenna2020cryogenic, holzgrafe2020cavity, xu2021bidirectional, fan2018superconducting, fu2021cavity, youssefi2021cryogenic, soltani2017efficient, hease2020bidirectional, sahu2022quantum}, opto-magnonics~\cite{zhang2016optomagnonic, hisatomi2016bidirectional, zhu2020waveguide, zhang2014strongly}, electro-optomechanics~\cite{andrews2014bidirectional, higginbotham2018harnessing, arnold2020converting, delaney2022superconducting,brubaker2022,kumar2022quantum,vainsencher2016bi, jiang2020efficient, forsch2020microwave, han2020cavity, mirhosseini2020superconducting}, cold atoms~\cite{tu2022high, vogt2019efficient, covey2019microwave} and rare-earth ions~\cite{o2014interfacing, fernandez2015coherent, fernandez2019cavity, bartholomew2020chip, everts2019microwave}. Reviews of recent experimental advances in M2O conversion can be found in Refs.~\cite{han2021microwave, kurizki2015quantum, lambert2020coherent, lauk2020perspectives, chu2020perspective, clerk2020hybrid}. 
The conversion efficiency achieved in state-of-the-art experiments, however, remains far less than unity. Despite the relatively high on-chip conversion efficiency, the total efficiency can be significantly lower due to the inevitable fiber-to-chip coupling loss and the optical-pump-rejection filtering loss. In addition, because a high-power optical pump is needed to boost the conversion efficiency, thermal microwave photons generated by the optical-pump-induced heat can be combined with transduced signal in the optical output channel as `added noise'. The performances of state-of-the-art M2O converters are summarized in Table~\ref{tab:M2O_efficiency} of Appendix \ref{app:M2O}.

In order to generate entanglement between separated refrigerators, one straightforward method is to locally generate entangled qubit-microwave photon pair at one node and subsequently apply an M2O and an O2M converter to deliver the microwave photon to another node as shown in Fig.~\ref{fig:M2O_scheme}(a). However, this scheme is sensitive to the low M2O conversion efficiency and thus suffers from a low entanglement fidelity. Alternatively, direct M2O conversion could be used in a heralded scheme. Analogous to the optical photon heralded schemes \cite{Cabrillo1999, minar2008phase, humphreys2018deterministic}, the superconducting qubit is first entangled with a microwave photon at each node as $\ket{g0}+\ket{e1}$. The microwave photons at both nodes then undergo direct M2O conversion and the optical photons are then routed and detected as shown in Fig.~\ref{fig:M2O_scheme}(b) (referred to as the direct conversion heralded scheme). The optical photons from both nodes interfere at a beamsplitter, and a click from the optical detector heralds the generation of entangled qubit state $\ket{ge} \pm \ket{eg}$. In addition, a remote entanglement generation scheme previously developed for atomic ensembles \cite{Cabrillo1999, Bose1999, Duan2001, Protsenko2002, Barrett2005, Martin2015} is another option for superconducting platforms \cite{krastanov2021optically} to obtain high-fidelity entanglement generation in the presence of low M2O conversion efficiency.
As shown in Fig.~\ref{fig:M2O_scheme}(c), an M2O converter can be pumped at the blue-detuned resonance frequency and thus be used as a spontaneous parametric down conversion (SPDC) source to generate entangled microwave-optical photon pairs (referred to as the SPDC heralded scheme). The optical photons generated at two fridges interfere in a beamsplitter to erase the which-path information, and the click from the optical single-photon detectors heralds the generation of an entangled microwave Bell state $\ket{01}\pm \ket{10}$ between the two fridges. This scheme has the benefit that the entanglement fidelity is less sensitive to the M2O conversion efficiency, while the entanglement generation rate depends on the conversion efficiency as well as the bandwidth. 


Ultimately, the performance of MNQCs made of superconducting qubits and M2O converters will depend strongly on the conversion efficiency and bandwidth of the M2O converters, which is very slow and noisy compared to the local operations. We estimate that the on-chip entangled pair generation rate of the state-of-the-art M2O converters is of the order 1 MHz with infidelity of 0.2 (see Sec.~\ref{sec:LayerModels}). In comparison, transmon two-qubit gate infidelity has already been engineered down to less than 0.002~\cite{Wei2022}, with two-qubit gate times on the order of 100 ns. 
Hence we see that the internode operations are the major limitation on MNQC performance, and future MNQCs will need to marshal a range of techniques to surmount the weak internode links. 


\section{Multinode Quantum Computing Architecture}\label{sec:Architecture}

A central task in evaluating MNQC performance is to understand the available performance of internode gates. However, the evaluation of internode gates in multinode systems is itself the evaluation of a multipart quantum system: MNQCs using superconducting devices with M2O links will need to compensate for weak internode links using a combination of entanglement generation \cite{2019arXiv191206642A, rueda2016efficient, mckenna2020cryogenic} settings, entanglement distillation \cite{Dur2003, Yan2022, Bennett1996}, and compiler optimization \cite{wu2022collcomm, ferrari2020compiler, wu2022autocomm, dadkhah2022reordering, beals2013efficient}. One direct approach might be to simply conduct a simulation of the full system, treating M2O conversion, entanglement distillation, remote gate execution, local operations, and measurement in one large analysis. However, this calculation quickly grows too large even for relatively simple MNQCs. With current density matrix simulations limited to $\mathcal{O}(20)$ qubits \cite{li2020density, o2017density}, allotting just a few qubits for entanglement distillation, measurement ancillae, and treating M2O conversion with a quantum framework costs approximately 18 qubits per internode link. This quickly limits the system to algorithms performed on a single-digit-number of qubits even with the best simulation algorithms. What is needed is a framework for organizing these components and their interactions into a structure that can be treated quantitatively. This framework must simplify the complex interactions between components into a few quantities that describe the relevant interaction while identifying key tradeoffs in the operation of components.

Classical computing network architecture has long faced similarly complex systems and developed approaches to tackle them. A foundational example is the Open Systems Interconnection (OSI) model which tackles the complex problem of networks and distributed systems by splitting the system into `layers' in a `stack'. Each layer has its role in the system, often referred to as the `service' it provides to layers above it in the stack. While the OSI model is foundational, more contemporary classical analogs of MNQCs, including the architectures underlying the ARES and InfiniBand network systems which directly provide network services for multinode computers \cite{2020arXiv200808886D, Infiniband, gara2005overview}, often use a 5 layer network stack comprising a Physical layer which transmits signals, a Link layer which manages packet transmission, a Network layer which provides routing and network management, a Transport layer which is responsible for the reliable transition of data, and Upper (or Application) layers where the users operate (for a detailed discussion of these layers, see~\cite{tanenbaum2007distributed}). A key concept behind the operation of multinode systems is \emph{transparency} \cite{van2016brief}: modern parallel classical platforms seek to offer users a seamless transition between single-node and multinode operations, with the multinode system appearing to the user as a single unified system. The chief service of the network stack is then to manage the execution of internode operations for the compiler in a transparent way.

\begin{figure}
    \centering
    \includegraphics[width=0.7\columnwidth]{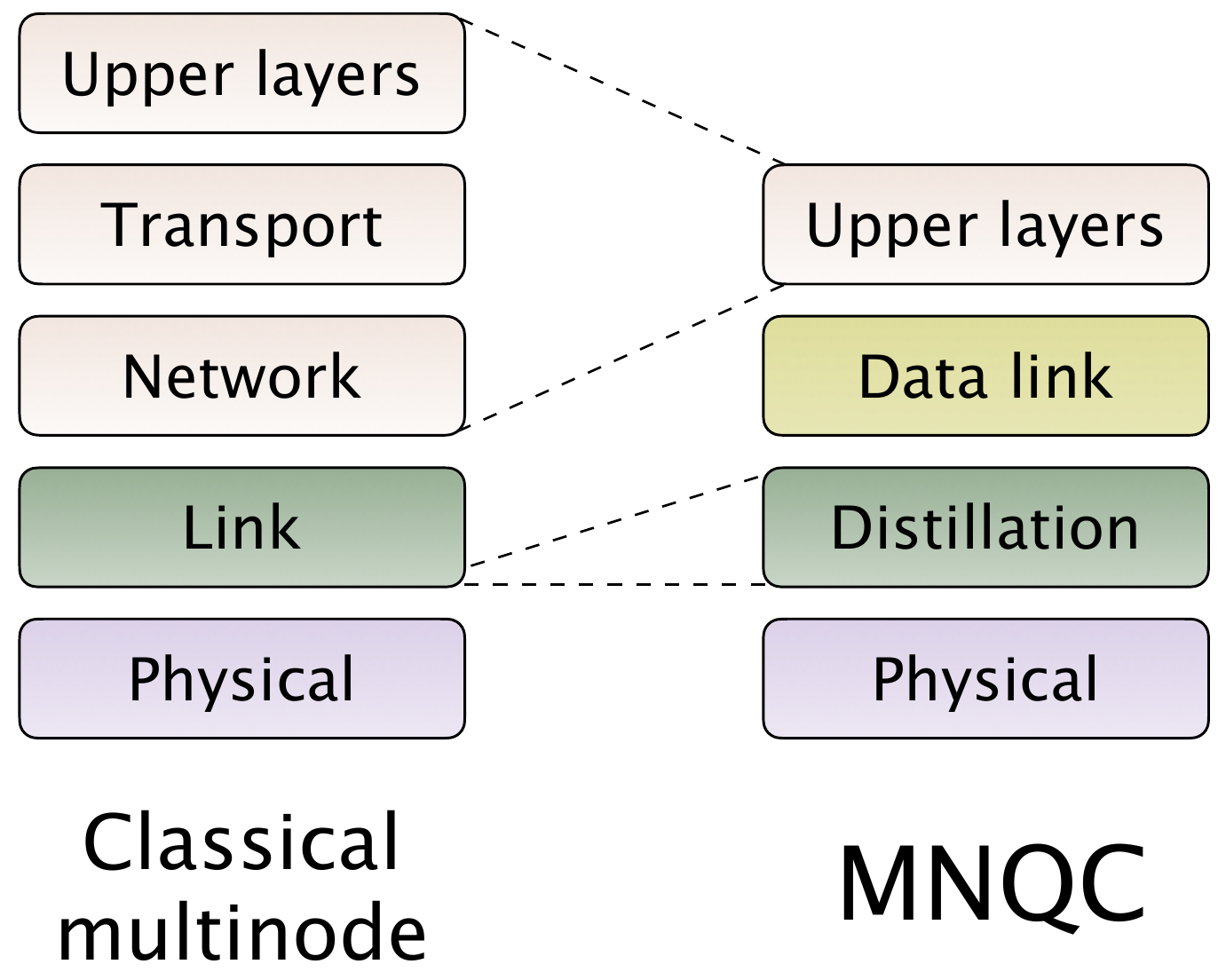}
    \caption{Comparison of the layer stack for classical multinode architectures \cite{Infiniband, tanenbaum2007distributed} and our MNQC architecture. Weak, noisy internode link merits a dedicated Distillation layer, while the low-level access of quantum compilers and high cost of network resources breaks the abstraction between the Upper layers and the Transport and Network layers.}
    \label{fig:classical_quantum_stack}
\end{figure}

\begin{figure*}
\centering
\subfloat[\centering]{{\includegraphics[width=0.35\textwidth]{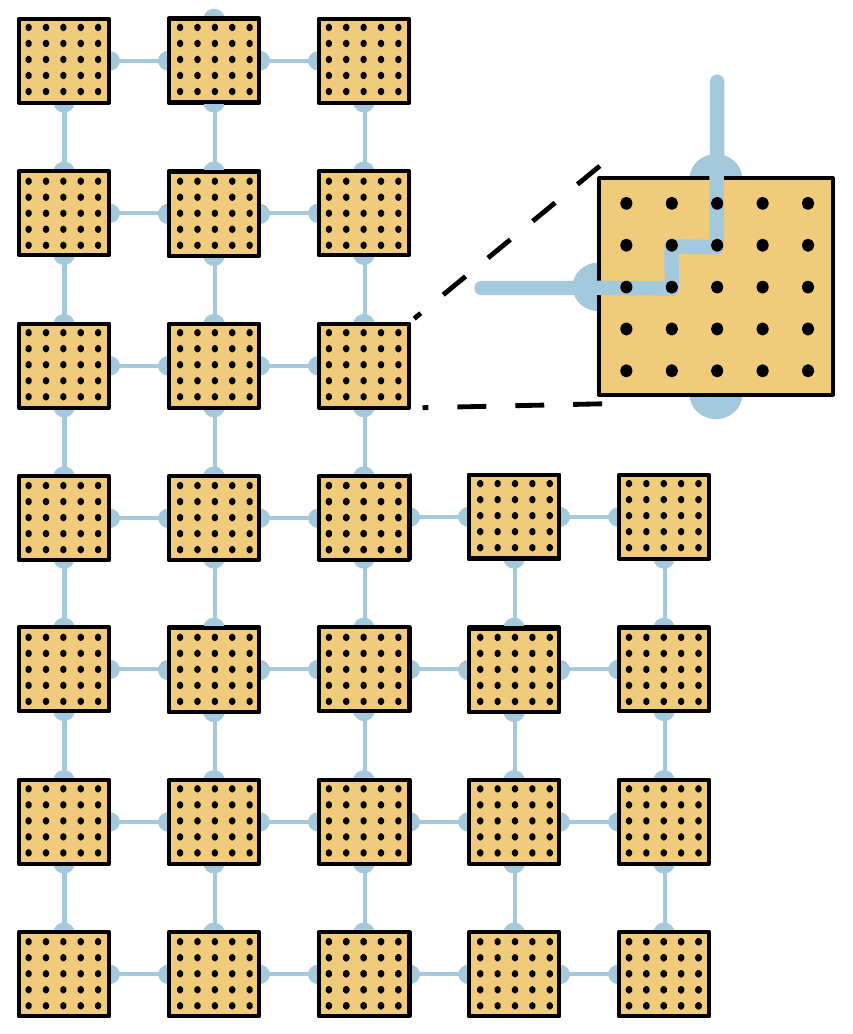}} }%
\qquad
\subfloat[\centering]{{\includegraphics[width=0.55\textwidth]{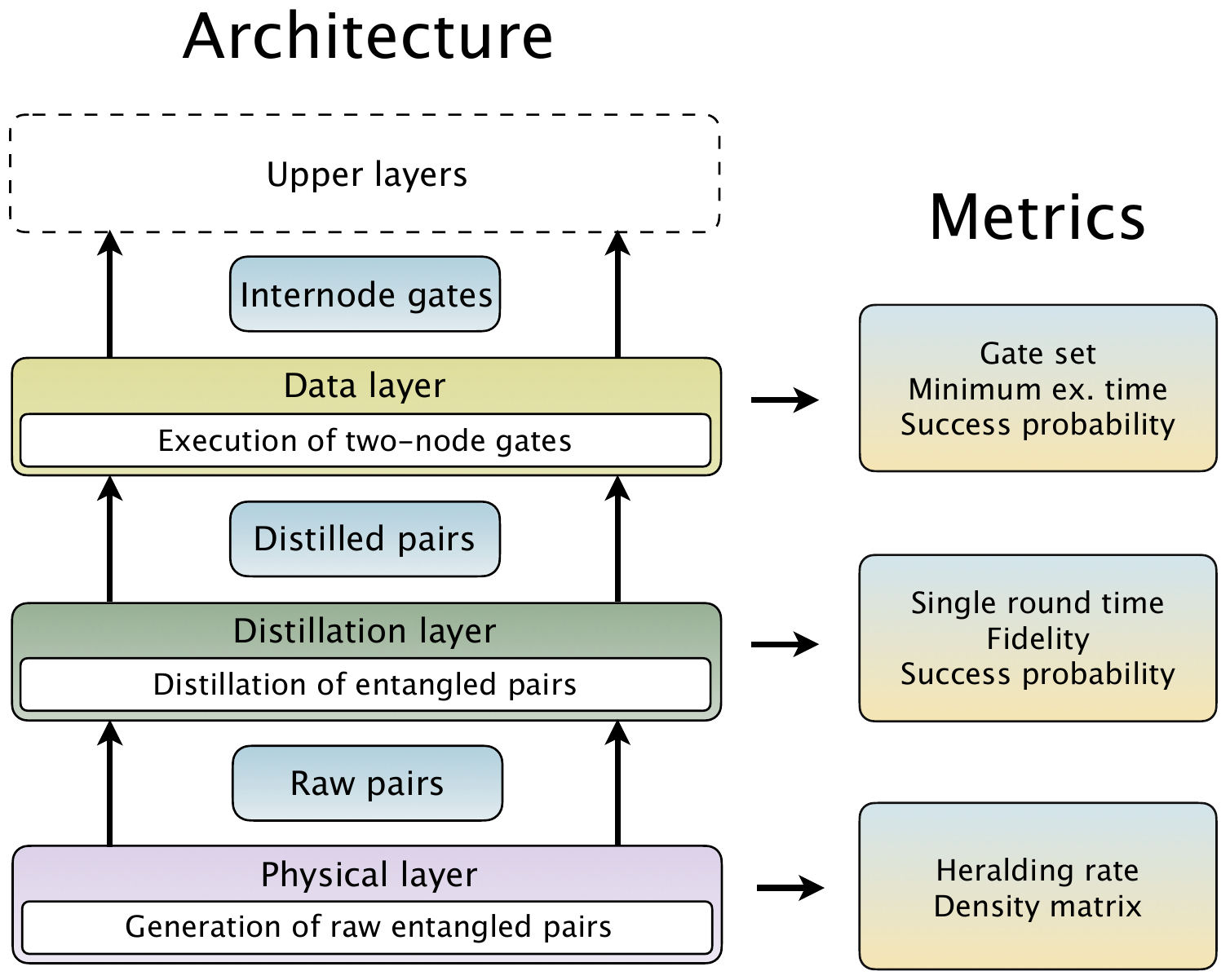} }}%
\caption{\emph{(color online). (a):} Schematic depiction of a multinode quantum system. Nodes (orange boxes) are connected with quantum links (blue lines) that allow internode gates between remote qubits (black dots). Internode gates in an MNQC will be slow and noisy, which leads us to focus on two-node, nearest-neighbor internode gates. This breaks the usual abstraction between the network and local hardware components. \emph{(b):}Schematic description of the layers, functions, interfaces, and key metrics of the MNQC model. Our focus is on the MNQC network stack (the Physical, Distillation, and Data) layers and how their performance affects overall MNQC performance.}
\label{fig:sQOSI_Schematic}
\end{figure*}

However, there are several key differences between quantum MNQCs and their classical counterparts (see Figure \ref{fig:classical_quantum_stack}). First, quantum internode communication suffers from far higher error rates than those in comparable classical architectures, and MNQCs will need dedicated resources to compensate for an extreme noise environment. 
Connected to the problem of internode noise is the presence of within-node noise that accumulates with time. Executing operations more slowly is not sufficient to improve performance, as the time of execution is itself a source of noise that will need to be accounted for. Furthermore, the prevalence of noise has led modern quantum compilers to operate at the level of gates or even pulse sequences, a much lower level than their classical counterparts. We expect links to be largely optimized to generate fast and high fidelity entanglement between just two nodes, independent of other links (Figure~\ref{fig:sQOSI_Schematic}a). Any switching will be costly and is to occur within the local compute nodes, and the compiler is directly issuing commands for two-qubit, two node gates, and any non-nearest-neighbor gates will need to be transpiled directly into nearest-neighbor gates. Finally, the principle of transparency of quantum systems means that the function of the quantum network stack should be to offer internode links to the compiler in the same way as local links, simply with longer gate times, lower fidelities, and probabilistic success.

An efficient method for the execution of remote gates is to use EPs produced from the M2O SPDC process, local operations, and classical internode communication to execute remote gates \cite{1999Natur.402..390G, PhysRevA.76.062323}. However, the low rate and high infidelity of EPs may lead to low fidelity of internode gates. This performance may be improved by using entanglement distillation \cite{Deutsch1996}, which consumes raw (not distilled) EPs to produce distilled EPs, which may then be used for remote gates. The function of the network stack is then to produce raw EPs, distill them, and manage the execution of internode gates, offering internode gates as a resource to the upper layers while abstracting away the details of their execution. 

Taken together, raw EPs, distilled EPs, and internode gates form a chain of key resources for remote gates, each produced from the previous. 
We may use these key resources to construct an MNQC network stack in analogy with the classical approaches by devoting a `layer' of the system to the production of each key resource. At the `bottom' of the stack, M2O SPDC hardware produces raw EPs. In analogy with the classical approach, we call this the `Physical layer'. Next, a `Distillation layer' converts raw EPs into distilled EPs at a lower generation rate. Finally, a `Data layer' manages the execution of internode gates, and exposes them as a resource to the compiler. A comparison of this MNQC architecture with that from modern classical interconnection architectures is given in Fig. \ref{fig:classical_quantum_stack}.
Similar models for network architectures have been proposed in several pioneering works \cite{2019NJPh...21c3003P, 2019arXiv190309778D, Loke:2022bcq, 2017arXiv170807142P}. These papers lay out criteria for quantum networks, and also find that a stack based on classical interconnection architectures, with the added function of distillation, is an effective way to structure the network. While these are general studies, a hardware-focused model has been proposed for networks using NV centers \cite{2019arXiv190309778D}. An excellent general overview of planned progress in quantum interlink technologies may be found in~\cite{PRXQuantum.2.017002}; here we focus on a particular technology (M2O interlinks) and provide detailed studies of MNQC algorithm performance. 
To our knowledge, this paper is the first to present a detailed hardware-based model of multinode (or networked) architectures using superconducting devices with M2O interconnects.  

An important property of the network stack is that each layer interfaces only with the layers above and below it in the stack. For example, all raw M2O EPs are passed to the Distillation layer, and there is no need for the Data layer to interact with Physical M2O generation. Similarly, entanglement distillation is hidden from the compiler. Instead, layers only interact by passing instructions and success flags as indicated in Figure \ref{fig:sQOSI_Schematic}. Hence this layer architecture can simplify the conception and operation of MNQCs by reducing the potential complexities of component interactions to a linear layer and stack structure.

For our present purposes, the key property of this MNQC architecture is that it also simplifies the simulation and benchmarking of an MNQC. Because the layers interact only with their neighbors in the stack, we can quantify the operation of each layer in terms of the production quality of key resources. Beginning from the bottom of the stack, the Physical layer works to produce raw M2O EPs, which are quantified in terms of the heralded rate of production and the density matrix of produced EPs. The Distillation layer then is responsible for taking these raw EPs and producing distilled EPs, which are quantified by the minimum time to produce a distilled pair, the density matrix of the produced pair, and the success probability of the operation, each as a function of the number of rounds applied. At the top of the network stack, the Data layer uses distilled EPs to execute internode gates, which are quantified by the set of available gates as well as the minimum time and success probability of each gate. These metrics are denoted in Figure \ref{fig:sQOSI_Schematic}b. 

Each layer then may be treated quantitatively by simulating the metrics of the resource it produces in terms of the incoming resource. In the next section, we delve into models of each layer in order to examine the available performance profiles and lay out a key tradeoff in joint hardware and software operation. 

One significant simplification which we make in this paper is suppressing the probabilistic nature of the entangled pair generation and distillation processes, characterizing each by the average time of the process. This greatly simplifies our models, but the probabilistic processes should be treated fully in future, more detailed studies in order to reveal the complex interplay of real-time application execution. 

\section{Models of the MNQC Network Layers}\label{sec:LayerModels}

The MNQC architecture organizes entangled pair generation, entanglement distillation, and remote gate execution into layers. Furthermore, because the key resources between layers are specified, the operation and performance of each layer can be treated individually and later unified into a whole model. In this section, we examine the available performance of each layer of the MNQC stack using models of expected hardware and software performance. The next section will then unify these layer models into a model of overall MNQC performance. 

Beginning from the bottom of the stack, our first task is to estimate the fidelity and generation rate of EPs created using M2O converters in the Physical layer. In the following, we focus on the direct conversion heralded scheme shown in Fig.~\ref{fig:M2O_scheme}(b), where the qubit-microwave photon pair is initialized in $\sqrt{0.5}\ket{g0}+\sqrt{0.5}\ket{e1}$ at both nodes.
\begin{figure}[t]
    \centering
    \includegraphics[width=\linewidth]{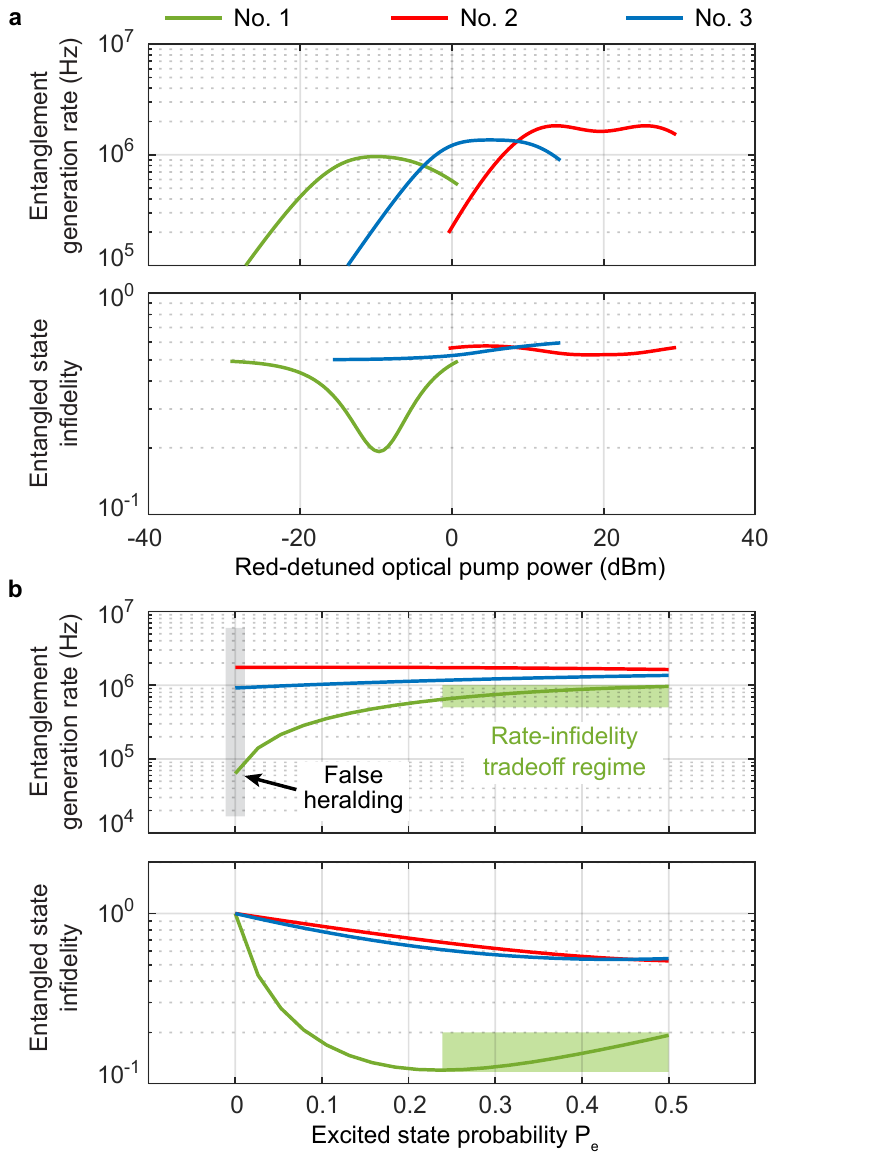}
    \caption{The estimated Bell state generation rate and infidelity (log scales) for the scheme shown in Fig.~\ref{fig:M2O_scheme}(b). (a) the performance as a function of pump power with $P_e$ fixed at 0.5. The simulation is performed using the parameter sets of the No. 1, 2, and 3 converters listed in Appendix~\ref{app:M2O} Table~\ref{tab:M2O_sets}. We assume the intrinsic decay rate of both microwave and optical resonators to be five times lower than the current actual experiment values, which we expect to be available relatively soon. (b) the performance as a function of $P_e$ with fixed pump power such that the cooperativity $C=1$. The rate at $P_e=0$ is the false heralding rate triggered by thermal noise, and a rate-infidelity tradeoff regime (the green shaded area) can be identified for the No.~1 parameter set.}
    \label{fig:M2O_state}
\end{figure}
The simulated entanglement infidelity and generation rate for current experimental platforms are shown in Fig.~\ref{fig:M2O_state}(a) (see details in Appendix~\ref{app:M2O}). Here, we use the parameter sets of three resonator-based M2O converters (see Appendix~\ref{app:M2O} Table~\ref{tab:M2O_sets}) to perform the simulation. We assume the intrinsic decay rate of both microwave and optical resonators to be five times lower than the actual experiment values, which we expect to be achievable relatively soon. The best M2O conversion efficiency is obtained at the pump power that reaches a unity cooperativity $C=1$ \cite{tsang2011cavity}, which consequently leads to the highest generation rate and lowest infidelity. For the No.~1 converter (the green curve), the entangled qubit generation rate can approach 1~MHz with an infidelity near 0.2. However, for the No.~2 (the red curve) and No.~3 (the blue curve) converters, the infidelity remains $>$0.5 because a high pump power is needed to achieve $C=1$, and the microwave thermal added noise induced by the high pump power strongly limits the fidelity. We next set the initial qubit-microwave photon as $\sqrt{1-P_e}\ket{g0} + \sqrt{P_e}\ket{e1}$ where $P_e$ is the probability of the excited qubit state and is experimentally tunable \cite{2016PhRvX...6c1036N}. The pump power is fixed such that $C=1$, and the results with $P_e $ tuned from 0 to 0.5 are shown in Fig.~\ref{fig:M2O_state}(b). The entanglement generation rate at $P_e=0$ is thus the false heralding rate, which dominates for No.~2 and No.~3 parameter sets. The tuning of $P_e$ also reveals a rate-infidelity tradeoff regime, which is highlighted as the green shaded area, where the rate increases but the infidelity also increases with an increasing $P_e$. In this regime, a larger $P_e$ allows more optical photons to be generated, but it also increases the error of having two nodes in the excited states simultaneously. In Section \ref{sec:FullModels}, we will see how the demands of the MNQC must guide the entanglement generation in conjunction with entanglement distillation, to which we now turn.

\begin{figure}
    \centering
    \includegraphics[width = \columnwidth]{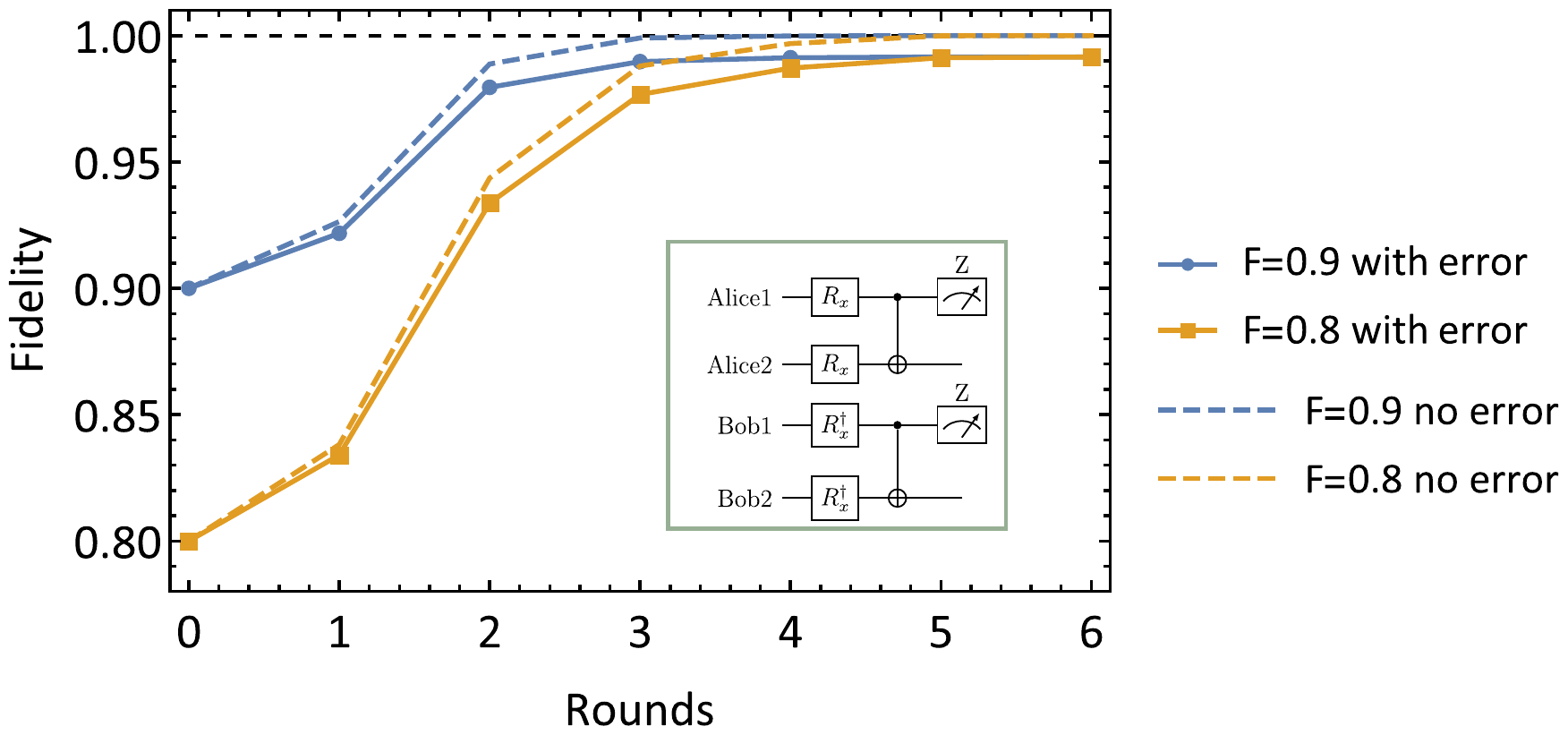}
    \caption{The performance of entanglement distillation under the DEJMPS protocol~\cite{Deutsch1996} with and without imperfections from qubit decay and decoherence and quantum gates. We consider two cases, where the imperfect Bell states with fidelity $F = 0.9$ (blue) and $F = 0.8$ (orange) can be generated across superconducting chips. We perform entanglement purification using DEJMPS protocol. We consider all the imperfect Bell states have been generated at the beginning. The local gates and qubit measurements take $1~\mu$s for each purification step. We assume the local gates between superconducting qubits have a depolarizing error with probability $0.0001$. The solid lines show the fidelity of the Bell states after $n$ rounds of purification, while the qubits are suffered from the imperfections. The qubits are assumed to have $T_1 = T_2 = 1$~ms. The dashed lines shows the corresponding state fidelity without decay and decoherence error. The inset is the quantum circuit for a single round of purification~\cite{Deutsch1996}.}
    \label{fig:purification}
\end{figure}

The distillation layer faces a tradeoff between generation rate and fidelity of EPs as it explicitly consumes raw EPs to produce a smaller number of distilled EPs, thereby exchanging a higher fidelity for a lower generation rate. As detailed in Appendix \ref{app:Distillation}, we simulate the output of entanglement distillation using EPs produced from M2O conversion. We set $T_1=T_2=1$ms and assume all local gates to take $100$ns with a probability of depolarizing errors of $.0001$.  In Fig.~\ref{fig:purification}, we show the fidelity of the Bell state shared between remote superconducting chips after $n$ rounds of recurrent entanglement purification performed using the DEJMPS protocol~\cite{Deutsch1996, 2007RPPh...70.1381D}. Four or five rounds of entanglement distillation can significantly improve the fidelity of the generated EPs, likely leading to improved internode gate performance. However, the improvement quickly suffers diminishing returns, with further rounds yielding only modest increases. This is a significant problem, as every round decreases the rate of distilled entangled pair generation by a factor of 2, thereby slowing internode gates. This problem is further exacerbated by the presence of decoherence, which degrades partially distilled EPs as they wait for more raw EPs, and Figure \ref{fig:purification} shows the performance of entanglement distillation with (solid line) decoherence or (dotted line) an ideal memory that prevents decoherence. 
However, it remains to see what this fidelity improvement can do at the level of internode gates.

In order to translate the output of the Physical and Distillation layers into internode gates, we develop a model of the Data layer, which uses distilled EPs to execute remote internode gates. Since a CX gate provides computationally
complete communication between nodes, we focus on the case of only internode CX gates. Gate teleportation of the CX gate can be accomplished via the consumption of one raw EP, two measurements, and two local CX gates \cite{PhysRevA.76.062323}. Using our simulations of M2O conversion, we numerically calculate the production time and density matrices of the raw EPs from the M2O process. These outputs are fed into the next distillation layer to generate high-fidelity, purified EPs. The time for each round of distillation, the success probability, and the consequent density matrices of the purified EPs are used in the following data layer to simulate the performance of a single internode gate.

\section{Full MNQC Analysis}\label{sec:FullModels}

We now have models of each layer in the MNQC network stack, from the Physical layer with M2O generation models to the Data layer which manages the execution of internode gates. While understanding the available performance and tradeoffs of each of these layers is key to understanding MNQC performance, models of individual layers cannot tell us how the performances of each layer affects overall MNQC performance, how the layers together offer internode gates, or how to navigate tradeoffs that affect multiple layers. Most importantly, they cannot tell us the performance of internode gates, what algorithms require, or how to exchange internode gates with local computation or circuit cutting gates.

We need an overall model of how the layers in an MNQC interact to produce total system performance. In this section, we unite the models of the previous section into a simulation pipeline that models the full MNQC stack, which allows us to perform three quantitative studies of the system. 
First, we introduce a `Gate-Algorithm Performance' (GAP) model which uses the output of the unified model to map out the available internode gate performance in terms of hardware models and compare this to the demands of algorithms. In doing so, we will see the effect of the tradeoffs in average internode gate execution time and fidelity and shows how to navigate them for a small MNQC. Next, we use the unified model output to develop a Quantum Roofline model (Q-Roofline) to show how the compiler can navigate the balance of internode and local computation at scale and identify the effects of hardware and software tradeoffs on internode communication bandwidth. Finally, we compare quantum links with error mitigation to classical circuit cutting links using a Quantum-Classical Performance Analysis (QCPA) to determine at what cost can internode links be exchanged for circuity cutting links.

\subsection{Unification of Layers into an Overall MNQC Model}

The unified model should allow us to quantify overall MNQC performance as a function of the performance of each layer. More specifically, it should accept as inputs hardware and software details of each layer, including the M2O drive strength and Hamiltonian, the entanglement distillation protocol, local operation fidelities and times, qubit $T_1$ and $T_2$, a compiler, and the quantum application to be executed. To characterize the quantum network stack, it should return metrics of the key resource offered by the Data layer to the upper layers, namely the fidelity and average execution time of internode gates. Furthermore, to allow us to determine the needs of algorithms, the unified model should then allow us to study the behavior of the Application and Compilation layers, including performing a full density matrix simulation of algorithms running on the small systems, and providing data on compilation results and estimates of performance on large systems. 

The MNQC layer architecture that we have used in the previous sections to organize our models plays a key role in enabling the unified model. As noted in Section \ref{sec:Architecture}, an MNQC is a complex system, including M2O generation, entanglement distillation, and internode gate distillation in addition to the normal functioning of a quantum computer. Treating the total system at once would quickly outgrow available simulation capabilities. Because we have independent models of each layer, we can link the output of one layer to the input of the next. Roughly speaking, if each layer involves a Hilbert space of size $N_i$, then we must treat a series of $N_i\times N_i$ density matrices. This should be compared to modeling the whole system with a density matrix of size $\sum_{i} N_i \times \sum_i N_i$. In practice, the layer architecture offers even further simplification by allowing us to use heuristic simplifications at the layer interfaces, for example by reducing error channels to the depolarizing channel or abstracting away probabilistically successful processes into an average execution time. 

Let us unify the network stack first, which we can then connect with the Application and Compiler layers. At the top of the network stack, the Data layer supplies internode gates as a key resource to the Compiler and Application layers; our task is thus to quantify the fidelity and execution time of available gates as a function of the outputs of Distillation and Physical layers lower in the stack. Using our simulations of M2O conversion, we numerically calculate the production time and density matrices of the raw EPs from the M2O process. These outputs are fed into the next distillation layer to generate high-fidelity, purified EPs. The time for each round of distillation, the success probability, and the consequent density matrices of the purified EPs are then used in the Data layer simulation to evaluate the performance of a single internode gate. 

Now we connect the MNQC network stack simulation with the Application and Compiler layers to create an overall simulation of the MNQC. Na\"{i}vely, joining the Compiler and Application layers would involve a complex control issue. The compiler could be responsible for managing entanglement distillation and internode gate execution, each of which require multiple measurements, operations, and classical communication, in addition to its function managing local operations. However, the abstractions provided by the MNQC network stack, in particular the principle of transparency articulated in Section \ref{sec:Architecture}, greatly simplify this task. To the compiler, an internode gate is presented in the same way as a local gate, albeit with a longer average execution time and lower fidelity. Transparency thus greatly simplifies the construction, as compilers designed for monolithic systems may be used at the top of the MNQC stack, though they may not be optimal. 

\begin{figure*}
    \centering
    \includegraphics[width=\textwidth]{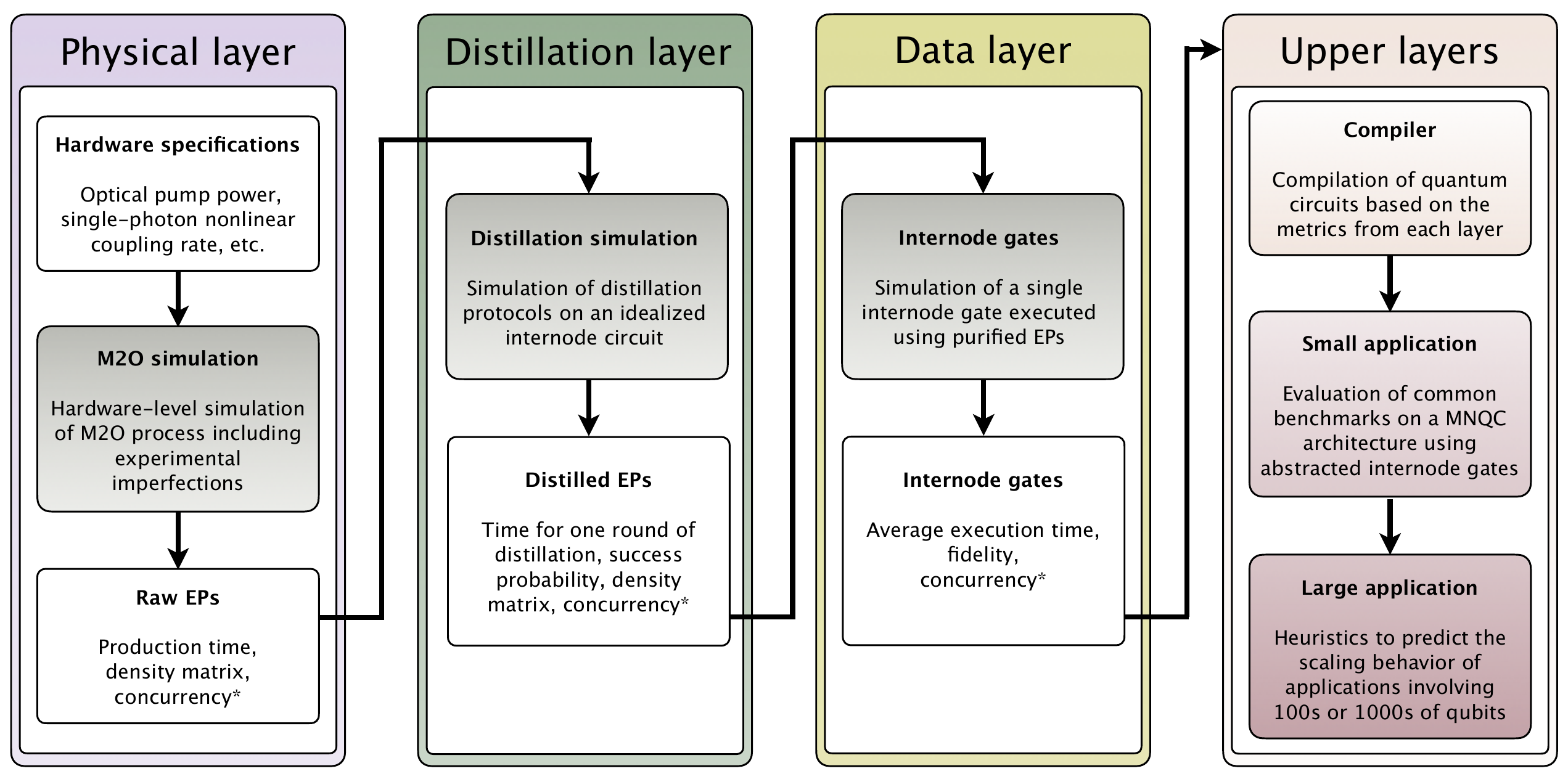}
    \caption{Models of each layer of the MNQC model and the metrics passed between them. Each layer is simulated as detailed in Section \ref{sec:LayerModels}, and the result is united to create an overal simulation of the MNQC. The concurrency metrics marked with an asterisk are set to unity here but can be used in a future generalized model that includes multiplexing of internode links for boosting the remote gate execution rate.}
    \label{fig:Model_Overview}
\end{figure*}

We can construct the full simulation pipeline beginning from the M2O Physical layer simulation, Distillation, and Data layer simulations that we discussed in the previous section and unified in the gate model above to upper layers. Figure \ref{fig:Model_Overview} shows an overview of this simulation pipeline with metrics for each interface. Once the average execution time and fidelity of the internode gate are simulated as in Section \ref{sec:LayerModels}, they are used to evaluate the performance of the upper layers, which includes the compiler layer and the application layer, leading to the full pipeline simulation shown in Figure \ref{fig:Model_Overview}.

\subsection{Gate-Algorithm Performance Models}\label{sec:GAP}

Our first task is to determine how to navigate the tradeoffs in the Physical and Distillation layers identified in Section \ref{sec:LayerModels}. Both of these tradeoffs involve an exchange between the time to create EPs and the infidelity of those EPs. However, they are not independent, as they operate on distinct layers using dependent key resources: the Physical layer can lower M2O pump power setting to decrease the infidelity of generated (raw) EPs at the cost of higher average generation time, while the Distillation layer then uses those raw EPs to create purified EPs, and again may decrease the infidelity of EPs at the cost of slower purified pair production by increasing the number of distillation rounds. To compare these two, we must find a common resource at which to evaluate the performance profile due to their combined effect, and then we must compare this to the demands of algorithms to guide a choice of performance. The unified model achieves this by determining the available internode gate performance produced by the network stack as a function of the operation of the Physical layers and then allowing us to evaluate benchmark algorithms executed on the MNQC using those internode gates. 

\begin{figure}
    \centering
    \includegraphics[width = 1\columnwidth]{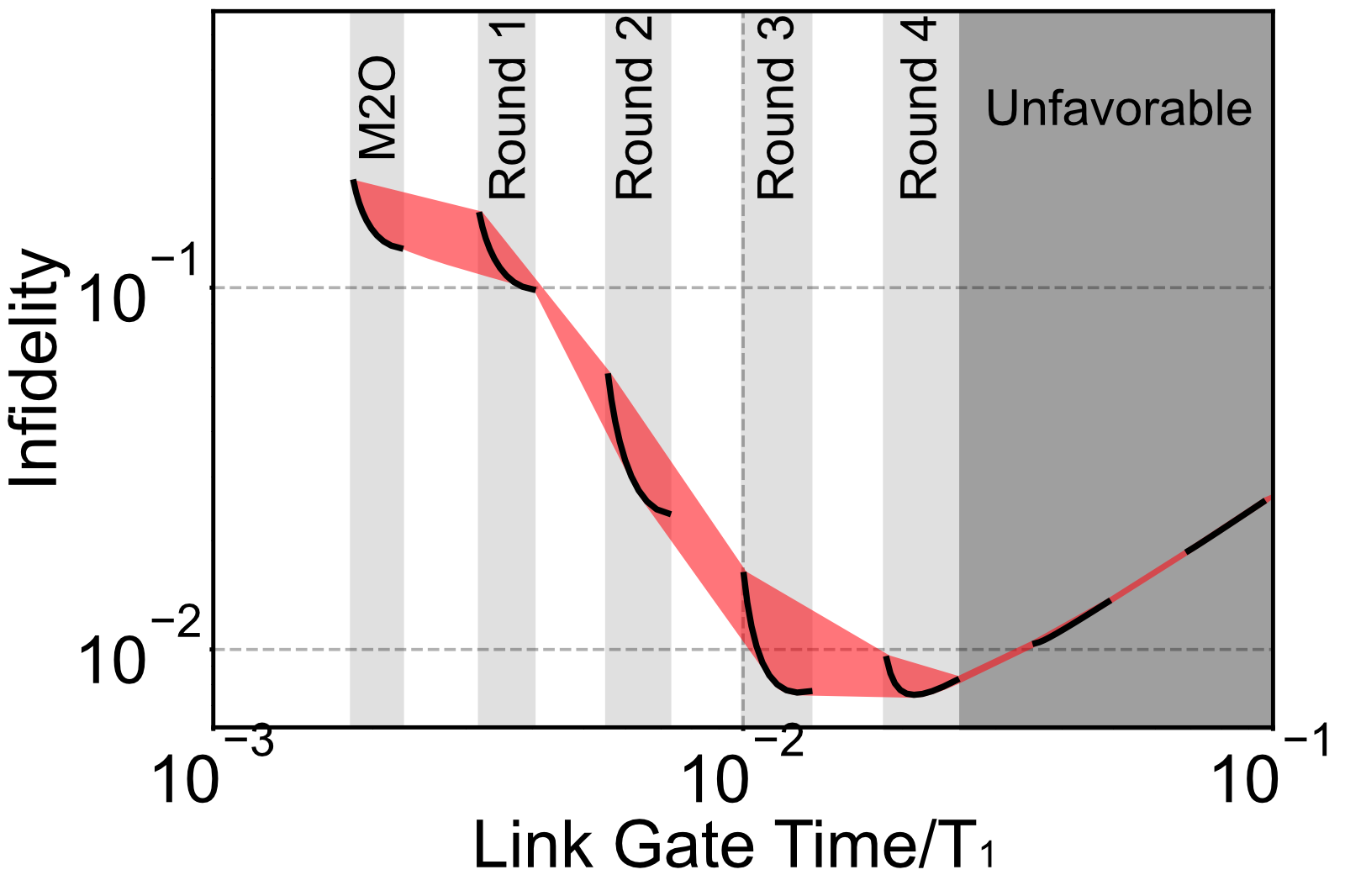}
    \caption{Performance profiles of internode gates using (black line) only raw M2O entangled pairs or (red line) distilled pairs. Raw M2O pairs on the black line may be tuned for higher rate by increasing the excited state probability $P_e$. Successive rounds of distillation are indicated by the markers on the shaded regions. Using only M2O generation leads to the fastest internode gates at the cost of a high infidelity, while distillation reduces the infidelity at the cost of increasing the internode gate time. A limited number of rounds of distillation can be performed before internode errors degrade the qubits.}
    \label{fig:Gates}
\end{figure}

Let us begin by determining the performance of the internode gates offered by the MNQC network stack. Using the unified model pipeline in Figure \ref{fig:Model_Overview}, we can link the models of the previous section in order to evaluate internode gate performance as a function of the Physical, Distillation, and Data layers in the MNQC stack. Figure \ref{fig:Gates} shows the available infidelity and average generation time of internode gates using raw EPs and distilled EPs. The black curves indicate the average execution time and infidelity of internode gates executed using EPs from the M2O conversion process or with successive rounds of entanglement distillation. At the upper left corner of each curve, operating M2O conversion with a high excitation probability $P_e$ creates a higher EP generation rate and thus low execution time, at the cost of higher infidelity. As $P_e$ is decreased, the infidelity decreases but the average execution time increases. This is precisely the tradeoff at the Physical layer identified in Section \ref{sec:LayerModels} and is reflected in the negative slope of the M2O curve. 

Given a particular M2O excitation probability setting, which corresponds to a point along the black curve, the Distillation layer then navigates a similar tradeoff: relative to raw M2O EPs, distilled EPs will have a longer average production time but higher fidelity. At the Data layer, this translates to a longer average execution time but higher fidelity of internode gates. Each round decreases the infidelity, at the cost of increasing the average internode gate time. The interaction of the colored noise affecting the raw entangled pairs as a function of drive power with entanglement distillation leads to complex behavior of the resulting performance. For the fastest gates, no entanglement distillation should be used. For gates with lower infidelity, distillation should be used. The needs of algorithms will then dictate how the internode gate should be executed: given a targeted performance, the number of rounds as well as the excitation probability $P_e$ shape the infidelity and link gate time achievable. If a compiler is able to select from a range of available internode gate times, then the MNQC stack must adjust the excitation probability of M2O generation dynamically to generate the highest fidelity gates for each internode gate time. For example, to achieve gates with lowest infidelity and an internode gate time below $.01 T_1$, one should use the minimal $P_e \approx .25$ and two rounds of distillation. On the other hand, to achieve the lowest infidelity possible at all, one should use $P_e \approx .35$ and four rounds of distillation. Furthermore, depending on the desired execution time, the compiler may wish to select fewer rounds of distillation with a higher $P_e$, or vice versa. This is particularly important as the infidelity is as much as 2x worse when using the incorrect configuration.

\begin{figure}
    \centering
    \includegraphics[width = 0.9\columnwidth]{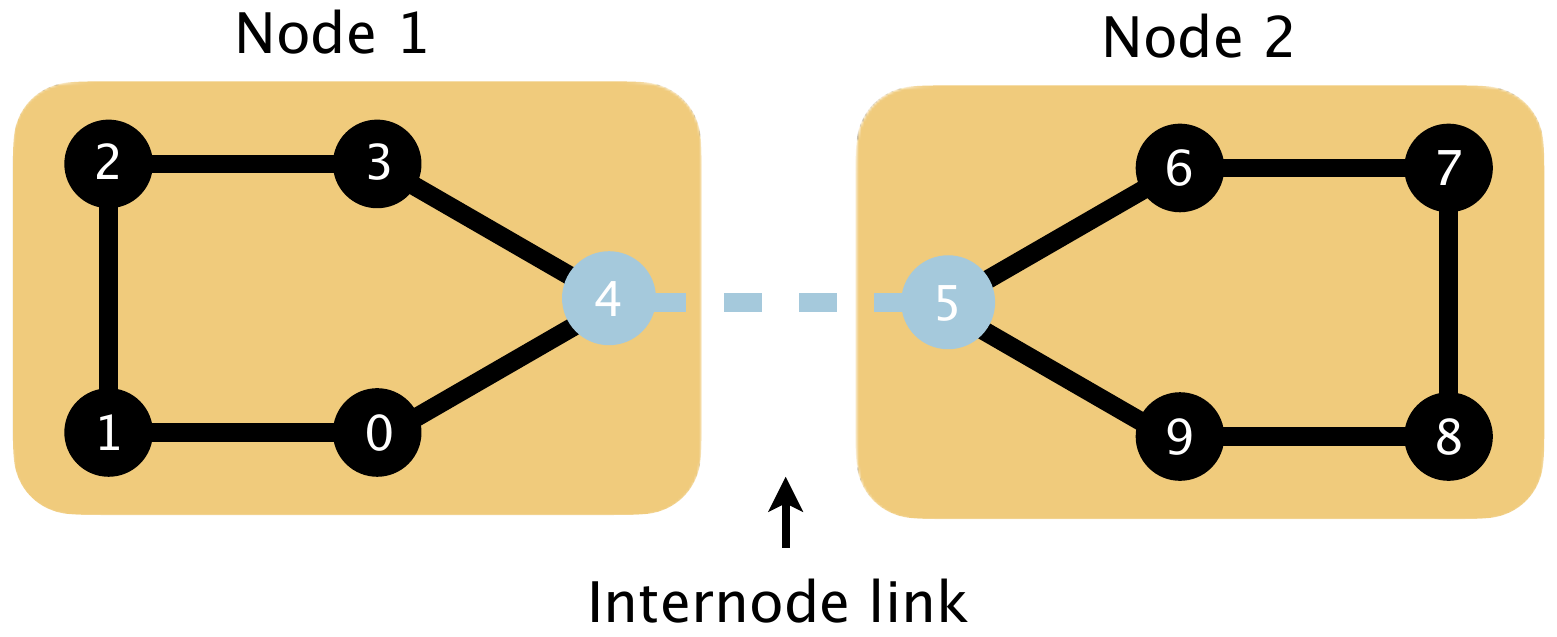}
    \caption{Topology of a small MNQC that we simulate explicitly. The system consists of two five-qubit nodes with a single internode link, with entanglement generation, distillation, and remote gate execution abstracted into the link.}
    \label{fig:dist_qpu}
\end{figure}


Next we turn to see how this internode gate performance affects the performance of algorithms on an MNQC. Beginning with gate performance curves like that of Figure \ref{fig:Gates}, simplified by including only the red bounding curve and removing the unfavorable region, we overlay on them the conditions for successful execution of a successful benchmark to create a `Gate-Algorithm Performance' (GAP) plot. As before, we set $T_1=T_2=1$ms and assume all local gates to take $100$ns with a probability of depolarizing errors of $.0001$. The basis gates for these systems comprise the same basis gates as IBM-Quantum, and each algorithm is transpiled accordingly.

As a first example, we evaluate the effective Quantum Volume (QV) \cite{2019PhRvA.100c2328C}. The QV is a measure of the size of the effective Hilbert space traversed by a quantum system before decoherence occurs. With a perfect internode link, the QV would be $2^{10}$ (Fig \ref{fig:dist_qpu}); with no internode link it would be $2^5$. Hence this benchmark allows us to quantify the degree to which the multinode system outperforms any one of its nodes. To gauge the performance implications of performing distributed quantum computing, we perform a noisy simulation for each algorithm over this architecture, with the inter-node link having the respective gate time and fidelity attained from inter-node gate simulation. 

\begin{figure}
    \centering
    \includegraphics[width = \columnwidth]{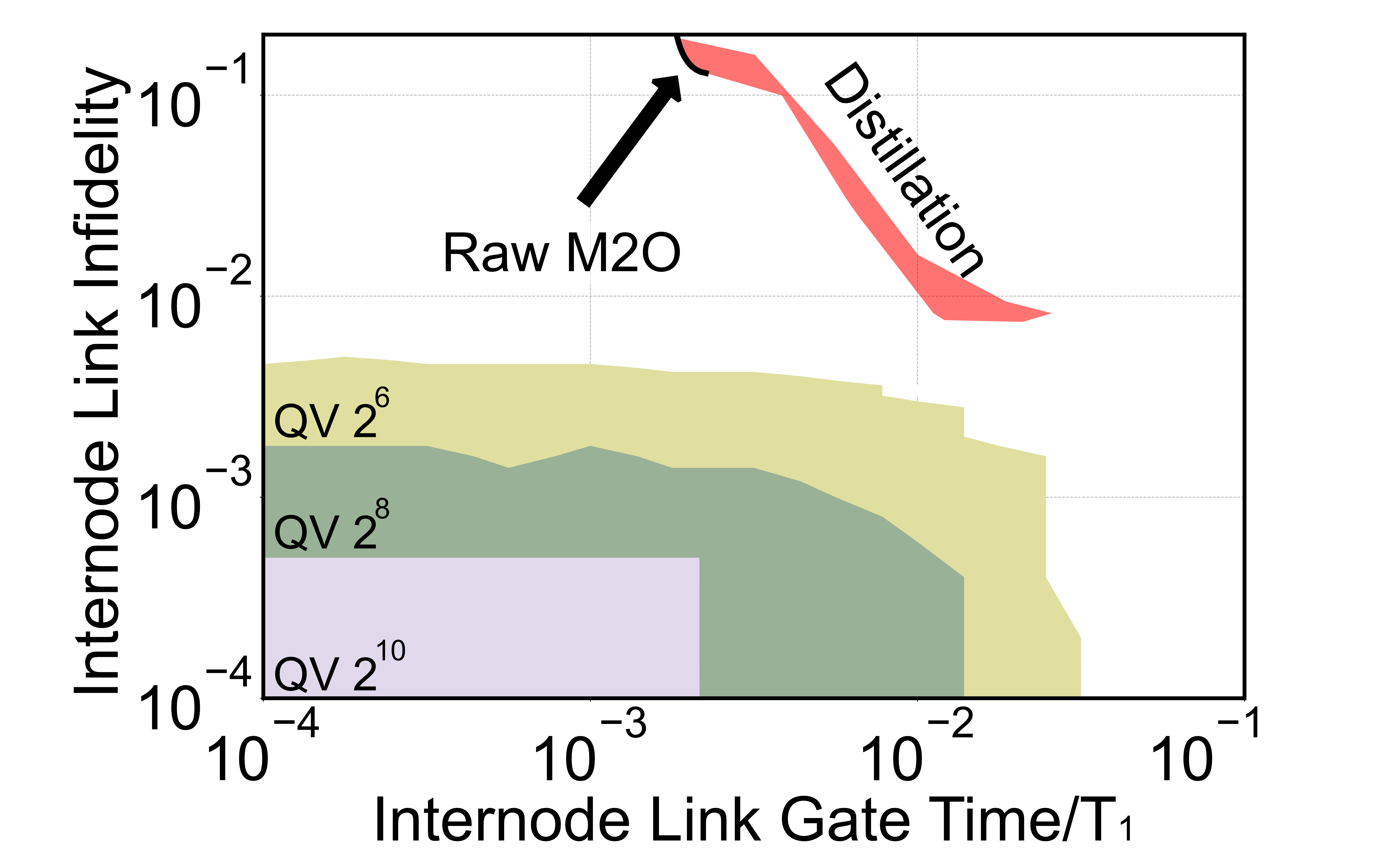}
    \caption{
    Gate-algorithm performance plot of Quantum Volume evaluated on a 10-qubit system comprised of 2 ring QPUs connected with M2O and Entanglement Distillation simulation. Each distillation curve, denoted by the red region, has been truncated at the number of nested rounds at which its performance begins to degrade. The QV with no internode link is $2^5$. With presently available technology, including distillation, the QV of the MNQC does not reach $2^6$.
    }
    \label{fig:QV}
\end{figure}

The results of the QV benchmark are shown on a `Gate-Algorithm Performance' (GAP plot) in Figure \ref{fig:QV}, which allows us to compare the available performance of gates produced by the MNQC network stack with the demands of algorithms we wish to execute. Times and fidelities that lead to successful completion of a QV circuit are denoted by shaded `success' regions. Beginning from the unshaded region, lowering the infidelity and the gate average execution time allows for the successful execution of larger and larger QV circuits. Both parameters are key because while the infidelity of internode links directly causes noise, the long execution times allow errors to accumulate within the nodes. 

On top of the shaded success regions, we overlay the available gate performances in a similar manner as in Figure \ref{fig:Gates}. The black line depicts gates executed using raw M2O generation, while the red lines denote internode gates using entanglement distillation. We can quickly see that the achievable performance is much slower and noisier than needed for QV circuits. Indeed, the rate and infidelity will require significant improvement for the MNQC to be able to achieve a QV that improves on the single node performance at all, and orders of magnitude improvement to achieve the maximum possible QV of $2^{10}$.



\begin{figure}
    \centering
    \includegraphics[width=\columnwidth]{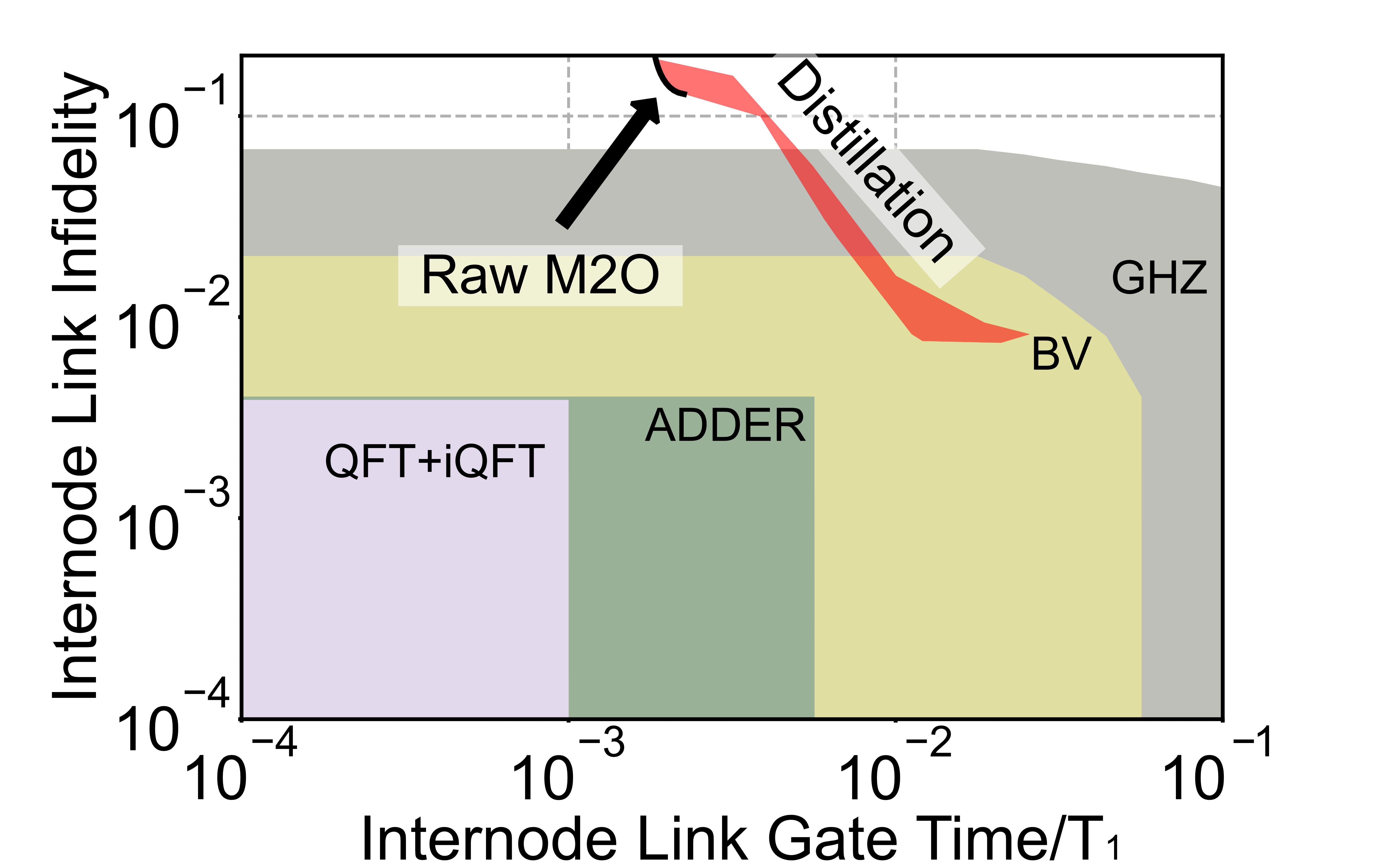}
    \caption{Gate-algorithm performance plot of several benchmarks evaluated on a 10-qubit system comprised of 2 ring QPUs connected with M2O and Entanglement Distillation simulation. Shaded regions indicate performance of $>$90\% for the respective algorithm. Each distillation curve has been truncated at the number of nested rounds at which its performance begins to degrade.}
    \label{fig:TotalSim}
\end{figure}

While QV gives a single number benchmark randomized over circuits built from all possible two-qubit gates, we also create a GAP plot for a benchmark battery \cite{10.1145/3550488, tomesh2022supermarq} to understand performance of the distillation model for specific algorithms. Our battery of tests is composed of: a Quantum Fourier Transform (QFT) benchmark, an ADDER benchmark, the Bernstein-Vazirani (BV) benchmark, and GHZ state distribution, in order of decreasing demands on the internode link. Again we see the need for faster and higher fidelity internode gates. However, using entanglement distillation, the GHZ and BV benchmarks can be achieved with high fidelity. Hence we see that even though entanglement distillation increases the internode gate time, its use is critical for enabling MNQCs to execute algorithms effectively. 

\subsection{Quantum Roofline Model}

Because MNQCs are employed to create large quantum systems, we must be able to understand the scaling behavior of large systems in order to identify and navigate tradeoffs and performance bottlenecks. While the GAP models of the previous section gave us a manner to navigate tradeoffs in the MNQC network stack and determine performance requirements for small systems, they cannot scale to large systems as they require density matrix simulations.

In this section, we introduce a Quantum Roofline (Q-Roofline) model, based on the classical roofline model \cite{williams2009roofline}, which analyzes the scaling behavior of large systems. The Q-Roofline model allows us to determine whether quantum algorithms are bound by internode or local performance. It can then evaluate compiler performance by determining whether the compiler has sufficiently balanced internode and local operations. The Q-Roofline model aims at modeling steady state behavior (e.g., averaging over the entire application) rather than instantaneous manner. However, when an application contains significantly distinct phases, one may draw particular Q-Roofline figures for each individual phase. 

As a first example, we can use the Q-Roofline model to determine whether applications running on an MNQC are bottlenecked by internode or local performance. 
For a compiled circuit, we define the Computation-to-Communication Ratio (CCR) as the ratio of the number of local gates of the algorithm versus remote internode gates over the entire circuit. On the other hand, given a quantum system, we define the machine CCR (MCCR) as the ratio of the rate of execution of local gates to the rate of execution of internode gates. Efficient compilation then seeks to match the balance of internode and local gates in the compiled circuit to that available to the machine, i.e. to match the CCR and MCCR, so as to maximize overall gate throughput while minimize circuit duration for the distributed circuit. We also define the gate density \cite{10.1145/3550488} as the occupancy of gates slots along the time evolution steps of a circuit (i.e., liveness defined in \cite{tomesh2022supermarq}), which provides an upper bound of performance when all remote gates become local.
As an initial study on bound analysis, we assume the execution of computation and communication gates can be fully overlapped through the transpiler or runtime scheduler.

\begin{figure*}[]
    \centering
    \includegraphics[width = .9\textwidth]{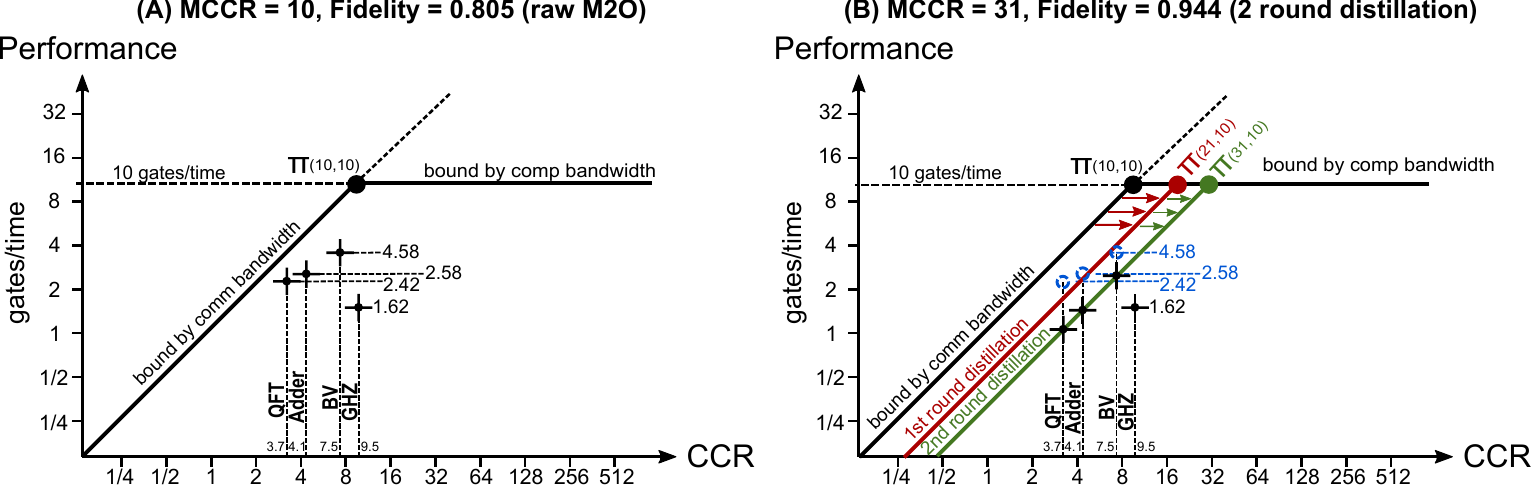}
    \caption{Performance bound analysis for QFT, Adder, BV and GHZ on the 10-qubit MNQR system through Q-Roofline model.}
    \label{fig:roofline}
\end{figure*}

Figure~\ref{fig:roofline} shows the Q-Roofline analysis of the application benchmarks from the previous section on the physical architecture in Figure~\ref{fig:dist_qpu}. The vertical axis shows the rate of single-qubit gate execution. We take the time unit to be the average gate time. Thus, for this 10-qubit system, the computation performance upper-bound is 10 gates/time. We can draw an horizontal line to set the computation performance bound. 

The horizontal axis of Figure~\ref{fig:roofline} denotes the CCR of a circuit. Since there is only one inter-module link (Figure~\ref{fig:dist_qpu}), given the duration of the remote gate is $1.041E^{-6}$s as shown in Figure~\ref{fig:TotalSim}, the internode gate duration is then 10.4 times that of a local gate (i.e., 100ns \cite{wei2022hamiltonian} as used in Section~\ref{sec:GAP}) and so the MCCR is 10.4. Using this (MCCR=10.4 and 10 gates/time) coordinate, we can locate a point $\pi$ in Figure~\ref{fig:roofline}a. From that point, drawing a 45 degree line (following the definition of CCR and MCCR), we can obtain the communication performance bound for the targeted MNQC system.

Using these two bounds, we can understand whether internode or local performance bounds the application. The Roofline shape, showcasing the performance bounds, is purely dictated by the quantum hardware. The ridge point $\pi$ defines the machine's balance point \cite{li2015transit}: if the compiled application's post-transpilation CCR is less than $\pi$, it is communication bound in this machine; otherwise, it is computation bound. To see the exact bound, a vertical line can be drawn from the application's CCR on the horizontal axis; the point it hits on the Roofline shape implies the performance bound. 

 \begin{table}[]
     \centering\footnotesize
     \begin{tabular}{|c|c|c|c|c|c|c|c|}\hline
       Algorithm  & Qubits  & Depth & 1q gate & 2q gate & Comm & CCR & Density \\ \hline
       GHZ & 10 & 13 & 3 & 8 & 1 & 9.5 & 0.162  \\ \hline 
       BV & 10 & 26 & 57 & 24 & 7 & 7.5 & 0.458 \\\hline
       QFT & 10 & 633 & 323 & 439 & 164 & 3.662 & 0.242 \\\hline
       ADDER & 10 & 219 & 101 & 177 & 55 & 4.136 & 0.258 \\\hline
     \end{tabular}
     \caption{Statistics of mapping four 10-qubit algorithm circuits to the small MNQC in Figure~\ref{fig:dist_qpu} using 
     Qiskit (Version 0.33.0) transpiler. Depth refers to circuit depth post-transpilation. 1q gate and 2q gate refer to post-transpilation 1-qubit and 2-qubit gates. Comm refers to the number of internode link gates. Density refers to gate density.}
     \label{tab:profile}
 \end{table}

In particular, let us evaluate the four benchmarks (i.e., BV, GHZ, ADDER and QFT). The GHZ benchmark shows the least demand of communication or the highest CCR, while QFT incorporates frequent entanglement operations through the inter-module link, showing the smallest CCR. The Adder and BV benchmarks display intermediate CCR. This is consistent with the difficulty of each benchmark to reach in Figure~\ref{fig:TotalSim}. In Figure~\ref{fig:roofline}a, all four benchmarks are communication bound given their CCRs in Table~\ref{tab:profile} and the settings of the system. However, none of them can hit the bounds due to their poor gate density. Using QFT as an example, the CCR of QFT is nearly 3.7, but the gate density is merely 0.242, which means the low utilization of the local gates slots (due to application's logic structure, transpiler behavior, and cost of intranode routing, etc.) limits its ability to even fully utilize the inter-module link, i.e., hit the communication bound. With a density of 0.242, in the best case, the computation performance is 2.42 gates/time, below the communication bound. The same conditions apply to the other three circuits. Therefore, in addition to the machine bound, one should also consider the circuit features such as gate density. Figure~\ref{fig:roofline}b shows a different scenario: let's say we want to enhance the inter-node link fidelity from 0.9 to 0.99 through two rounds of distillation (see Figure~\ref{fig:TotalSim}). After the first round, the communication performance halves (MCCR=20.8) and we obtain the red slash by shifting right for a unit. Hence both QFT and ADDER are predicted to be communication bound despite their low gate density. Furthermore, through two rounds of distillation, the machine's communication performance quarters (MCCR=31.2), and we obtain the green slash. Now, except for GHZ, the other three benchmarks QFT, ADDER, BV all become communication bound, with a delivery performance smaller than 2.42, 2.58 and 4.58 gates/time, respectively.

\begin{figure}[]
    \centering
    \includegraphics[width =.85\columnwidth]{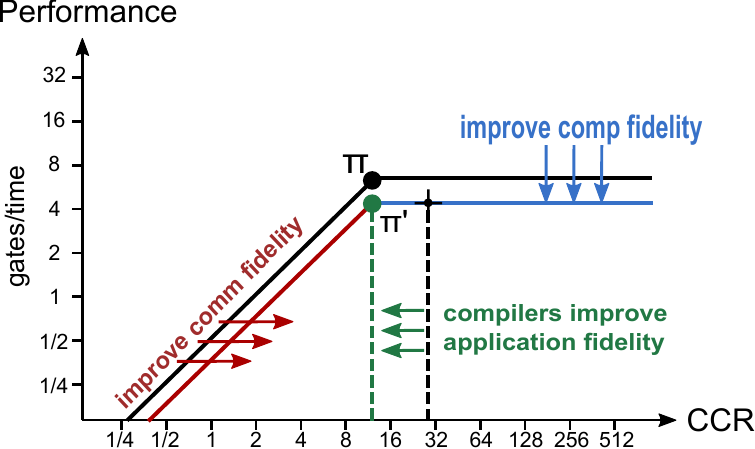}
    \caption{The Q-Roofline model also shows how the tradeoffs in fidelity and internode gate execution time affect performance bottlenecks.}
    \label{fig:roofline_fidelity}
\end{figure}

We can also see how the internode fidelity vs. execution time tradeoff that we have investigated affects the scaling performance of applications. From Figure~\ref{fig:TotalSim}, when the internode link gate time is $1.041 \times 10^{-6}$s, the link fidelity is about 0.805 with raw M2O. This results in an overall circuit execution fidelity of  0.9. With two rounds of distillation, the fidelity increases from 0.805 to 0.842 to 0.944 with the overhead of 3$\times$ communication latency. This shifts the sloped line right by two units, as shown in Figure~\ref{fig:roofline}b. Note that each round of distillation doubles the communication latency, and both axes are in 2-log scale.  

In particular, in the NISQ era, most fidelity enhancement techniques lead to certain performance degradation with overhead, as shown in Figure~\ref{fig:roofline_fidelity}. Nevertheless, the Q-Roofline model shows how the compiler can play a key role in reaching the best scenario for an application circuit by matching the machine's balance point. For example, when the application is communication bound, the compiler can increase the CCR to reach the balance point. On the other hand, when the application is computation bound, it can trade-off performance for fidelity (e.g., through distillation, error-mitigation, etc.) until again reaching the balance point.

\begin{figure}[]
    \centering
    \includegraphics[width =.85\columnwidth]{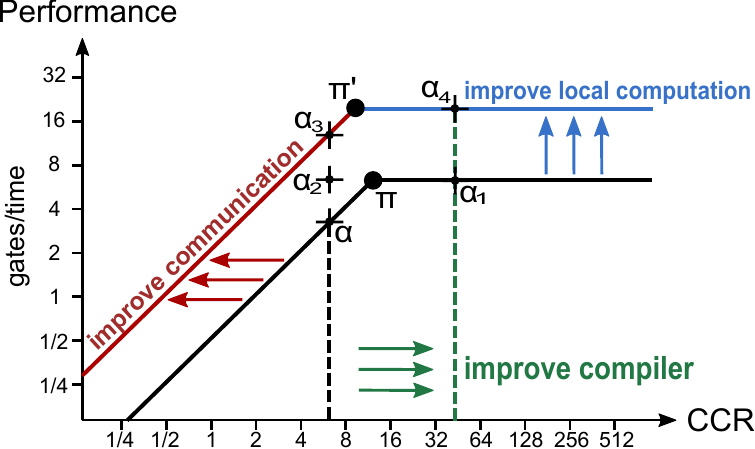}
    \caption{Improvements to the local compute nodes and the communication operations between them increases the area beneath the hardware bounds in the Q-Roofline model. Within the software layer, optimizations performed by the compiler to minimize communication and maximize parallelism will move an application $\alpha$ up and right.}
    \label{fig:roofline_scale}
\end{figure}

Lastly, we may also use the Q-Roofline model to predict the effect of improvements to each layer on the scaling behavior of applications. Figure~\ref{fig:roofline_scale} illustrates how technology advancement of local performance, internode operations (i.e. the MNQC network stack), and compilers would impact an application's performance scaling. As shown in the figure, (i) enhanced internode operations will shift the sloped communication bound of the Q-Roofline to the left, making it less likely that applications will be communication bound; (ii) improved quantum processors will lift the local computation bound up, leading to better system performance; (iii) better quantum compilers which minimize the number of communication operations between processors will contribute to larger CCRs, moving an application to the right along the $x$-axis and decreasing the chances of being communication bound. If an application is computation bound but has not saturated the device's local computation bandwidth, then a compiler which increases the parallelization of the program's instructions will increase the gate throughput and move the application upwards along the $y$-axis. 

For example, through the performance scaling of local quantum devices and quantum interconnects, the machine's balance point $\pi$ moves towards the upper-left to $\pi'$. Meanwhile, if an application is bound by communication at $\alpha$, (i) with only compiler improvement, the larger CCR renders the application from communication bound to computation bound, with a higher performance ($\alpha_1$); (ii) with only communication improvement, the communication bound is lifted and performance improves to $\alpha_2$; (iii) with both computation and communication improvement, the performance further improves to $\alpha_3$; (iv) with all computation, communication and compiler improvement, the performance can arrive at $\alpha_4$. For quantum programs of a sufficiently large size, the compilation problem may become intractable and therefore the reported gate density and computation-to-communication ratio will be lower bounds on the true, optimal values. 

Overall, the Q-Roofline model provides a way to conceptually and quantitatively balance the internode link performance with local performance. It allows us to identify and navigate bottlenecks by trading off internode and local computation intensity (i.e. adjusting the CCR) so that applications are balanced for the machines they are executed on. For hardware designers, this information is useful for deciding whether it is most beneficial to increase compute or communication bandwidth or fidelity. On the software side, the location of an application with respect to both bandwidth bounds will inform the compiler whether it is better to focus on minimizing the number of remote operations to increase the CCR and unblock the application from the communication bound, or focus on maximizing parallelism.

\subsection{Error Mitigation and Circuit Cutting}
We have quantified the performance of MNQCs as a function of the internode gate time and fidelity, shown how to navigate the tradeoff between these two quantities, and examined the role that the Compiler and Application layers have in minimizing the use of the internode link. However, we have also seen the dramatic limitations of near-term MNQCs, whose performance only modestly exceeds that of a single node. Given this, we must determine whether the quantum link is worth building at all, or, more precisely, whether a quantum link can outperform a purely classical link. 

For purely classical links, we could use classical circuit-knitting techniques \cite{tang2021cutqc, tang2022scaleqc, tang2022cutting} which execute circuits separately on the individual nodes many times to replicate a quantum link. On the quantum side, the use of multiple circuit executions allows us to consider error mitigation techniques. Here we compare the number of executions required for error mitigation to those required for circuit knitting in order to quantify the relative performance of quantum links and classical links. The key to achieving this is to combine the MNQC network simulations of internode gate execution time and fidelity from Section \ref{sec:GAP} with models of error mitigation~\cite{temme2017,vanderberg2022} and circuit cutting~\cite{bravyi2016,peng2020simulating,mitarai2020,sutter2022}.

For both error mitigation and circuit knitting, the number of circuits required scales exponentially with the number of circuit uses, i.e. as $O(\gamma^k$), where $k$ is the number of gates across the link and $\gamma$ depends on which method we use and the underlying hardware performance. In the case of error mitigation, more executions are required to mitigate the loss in fidelity from the quantum link. In particular, for probabilistic error cancellation (PEC)~\cite{temme2017,vanderberg2022} the value of $\gamma$ per gate is\footnote{This is $\gamma^2$ from \cite{vanderberg2022}} 
\begin{equation}
    \gamma_{\text{PEC}}(d,F_\text{p}) = \left( \frac{d^2F_\text{p} -1}{d^2-1}\right)^{-4(d^2-1)/d^2},
\end{equation}
where $d$ is the dimension of the gate ($d=4$ for a two-qubit gate) and $F_p$ is the process fidelity. For an internode gate of fidelity $F_{\text{LL}}$ and gate time $T_\text{LL}$ the total error due to the internode gate, including both the error of the operation and the (intranode) noise accumulated during the long internode gate execution time, is $\gamma_{\text{PEC}}(4,F_\text{LL})\gamma^{N_q}_{\text{PEC}}(2,e^{- T_\text{LL}/T_*})$, where $T_{*} = T_{1}T_{2}/(T_1 + T_2)$ is the effective fidelity lifetime of a qubit. 
In the case of circuit cutting or knitting~\cite{bravyi2016,peng2020simulating,mitarai2020,sutter2022}, there is no quantum link, but one can emulate the $2n$-qubit system by running more circuits on the smaller devices and combining the results classically. In Ref.~\cite{sutter2022} it is shown that $\gamma=9$, and that this can be reduced to $\gamma=4$ with local operations and classical communication. Since $\gamma$ for circuit knitting is independent of the link fidelity, there is a crossover regime in which the circuit knitting procedures require less overhead. 

\begin{figure}
    \centering
    \includegraphics[width =\columnwidth]{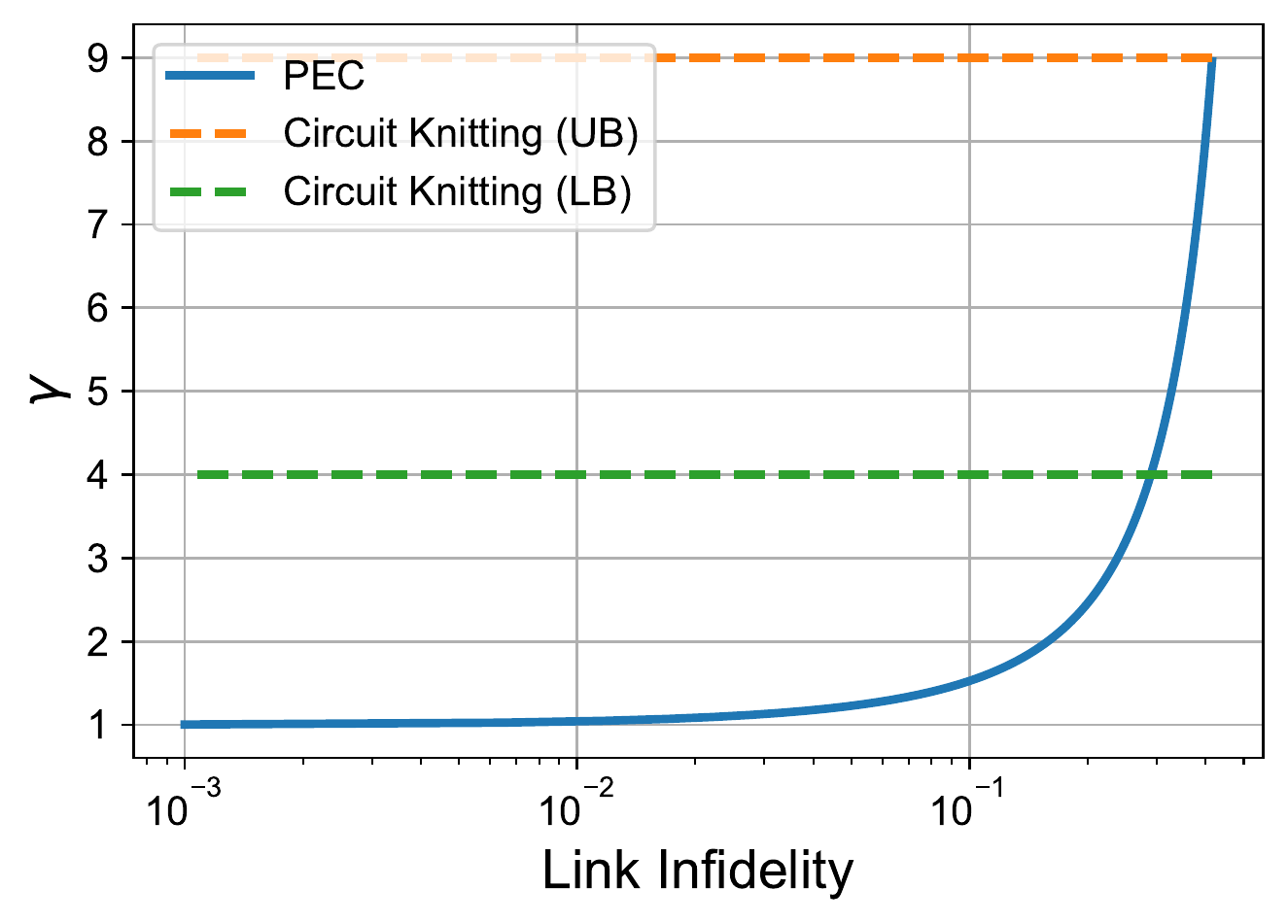}
    \caption{Comparing the scaling of different methods to link circuit subsystems together where the number of circuits requires scales as $O(\gamma^k)$ where $k$ is the number of CX gates between the subsystems. For circuit knitting we give the upper-bound (UB) and lower-bound (LB), see Ref.~\cite{sutter2022}. Circuit knitting methods require no actual connection whereas PEC is a method to mitigate a lossy quantum link between the sections. A quantum link is almost always superior.}
    \label{fig:gamma_fid}
\end{figure}

\begin{figure}
    \centering
    \includegraphics[width = .9\columnwidth]{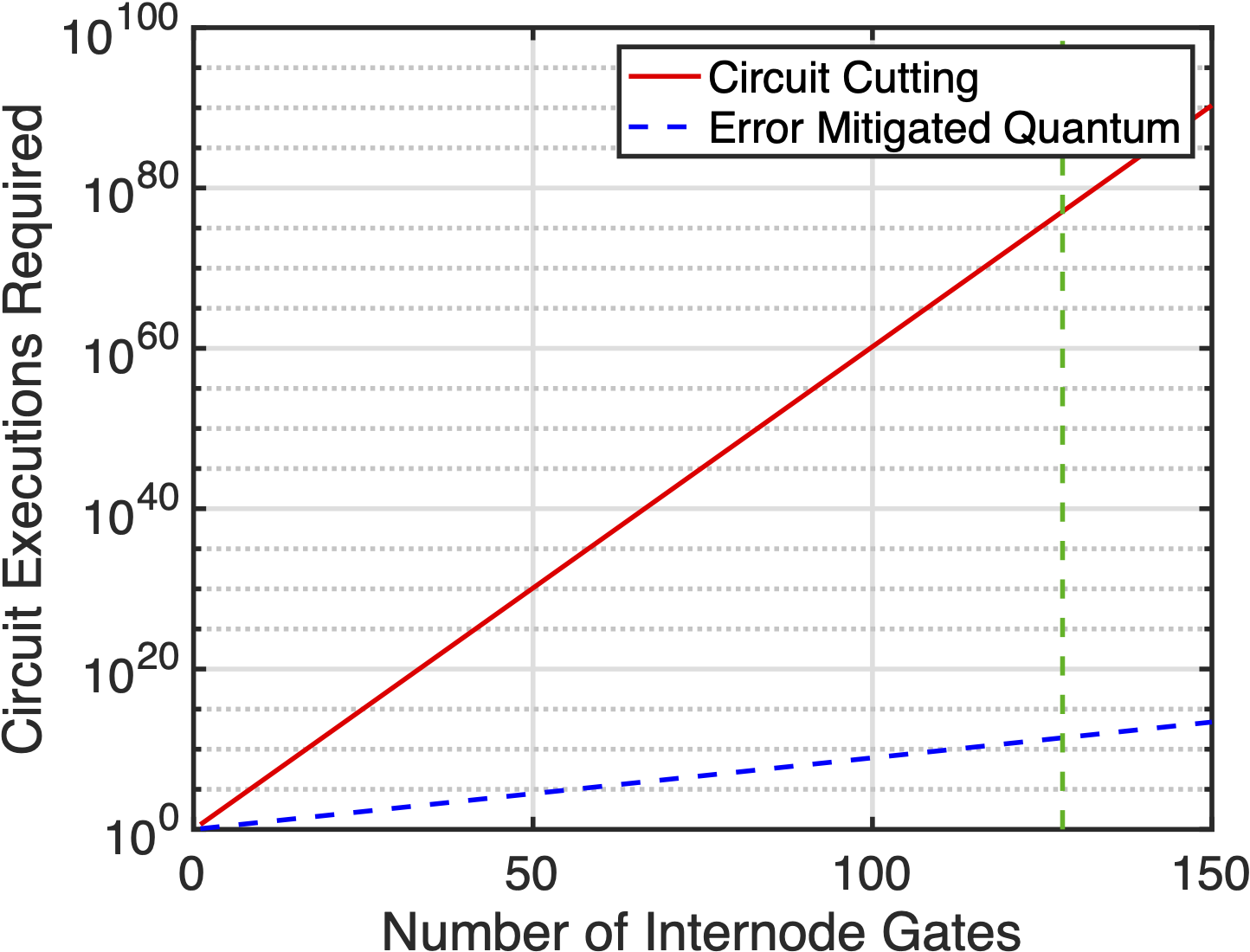}
    \caption{Quantum-Classical Performance Analysis comparing the number of gates required for circuit cutting (red) and PEC (dotted blue) as a function of the number of internode gates. For the QFT benchmark, with 128 internode gates (green), circuit cutting is clearly infeasible.}
    \label{fig:qcpa}
\end{figure}

The contrast between the procedures is summarized in Fig.~\ref{fig:gamma_fid}. Despite the relatively poor performance of the internode link, it still develops a significant advantage over a purely classical link for link infidelity $\lesssim.5$. In the previous subsection, we found that the two-node infidelity is better than this in almost all cases when using M2O and entanglement distillation. Hence the quantum link is advantageous despite the noise and slow gate times. Moreover, this advantage is key when scaling the systems. For example, if we use $k=20$ internode gates during an algorithm, then the circuit cutting requires between $10^{12}$ and $10^{19}$ circuits while a quantum link with an infidelity of 10\% requires only $10^4$ circuits. A classical algorithm that scales as  $2^n$ needs about $10^{18}$ steps. For example, as shown in the QCPA in Figure \ref{fig:qcpa}, the 10-qubit QFT circuit simulated for the benchmarks required $128$ gates across the link, which for PEC at $2.5\%$ infidelity of the link requires about $10^6$ circuits to mitigate while for circuit knitting would require a clearly infeasible $10^{77}$ circuits.

While PEC is advantageous in many cases, it can be at a disadvantage if the internode gate is very long since then the $\gamma_\text{PEC}$ increases due to the infidelity due to decoherence on all the other qubits. This effect is considerable if internode execution time $T_\text{link}$ is on the order of $T_*/N_q$. Managing this effect will require balancing the number of qubits $N_q$ in use, which sets $\gamma$, with the number of uses $k$ of the internode link. Hence compilers that can maintain a high CCR are critical for maintaining the advantage of quantum links.

We have quantified the performance of each layer of the MNQC stack, showed how to navigate tradeoffs in the network layers, developed a model for evaluating Compiler and Application layer performance, and compared classical and circuit knitting approaches. In particular, we found that in order to navigate the tradeoff between the time and fidelity of internode operations at Physical and Distillation layers, the Physical layer should use initial states which depend on the number of rounds of distillation to be performed, which in turn require inputs from algorithm requirements. Furthermore, optimizing MNQC performance requires balancing local computation with internode computation which can be quantified using the Q-Roofline model. Moreover, we also saw the advantage of quantum links over classical circuit knitting approaches, which demonstrated that even with modest fidelity quantum links can reduce the number of circuits needed by many orders of magnitude over classical approaches.

\section{Towards a Distributed Quantum Computer: Research Targets}\label{sec:Roadmap}

In the previous section, we saw that although quantum links outperform their classical counterparts, MNQCs will need considerable improvement to become viable models for scaling quantum computers. Developing MNQCs that can outperform any of their nodes and execute algorithms of practical importance will require improvements in each layer of the MNQC stack. In this section, we propose research directions that can deliver improved performance at each layer and illustrate how these 
improvements combine to improve MNQC performance in terms of the GAP, Q-Roofline, and QCPA models of the previous section.

\subsection{Physical Layer Improvements: M2O Conversion and Multiplexing} \label{sec:MO-conv}

Improving internode gate performance is a key target for enabling performant MNQCs. The analysis of section \ref{sec:FullModels} shows that MNQC performance is significantly bottlenecked by the low fidelity and generation rate of EPs, which lead to gate times and infidelities 10-1000x worse than what we expect from local gates. However, achieving improvements of this magnitude will require significant progress in current technology, or entirely new paradigms all together. Here we briefly describe three potential improvements: iterating on current M2O approaches, developing robust multiplexing, and, in the long term, developing a high-fidelity coupling between superconducting qubits and trapped ions. The speculative effects of these improvements are summarized in Table \ref{tab:physical_improvements}. 

\begin{table*}[]
    \centering
    \begin{tabular}{|p{\mywidthL}|p{\mywidthR}|p{\mywidthR}|p{\mywidthR}|}\hline
        & \begin{center}
            \Large GAP
        \end{center} & \begin{center}
            \Large Q-Roofline
        \end{center} & \begin{center}\Large QCPA\end{center} \\ \hline
        \raggedright \textbf{M2O Improvements:}
Iterated improvements 
to M2O devices and 
protocols can yield 10x 
higher rate and 10x 
lower infidelity.
5x or more & \centering\raisebox{-\totalheight}{\includegraphics[width = \mywidthR]{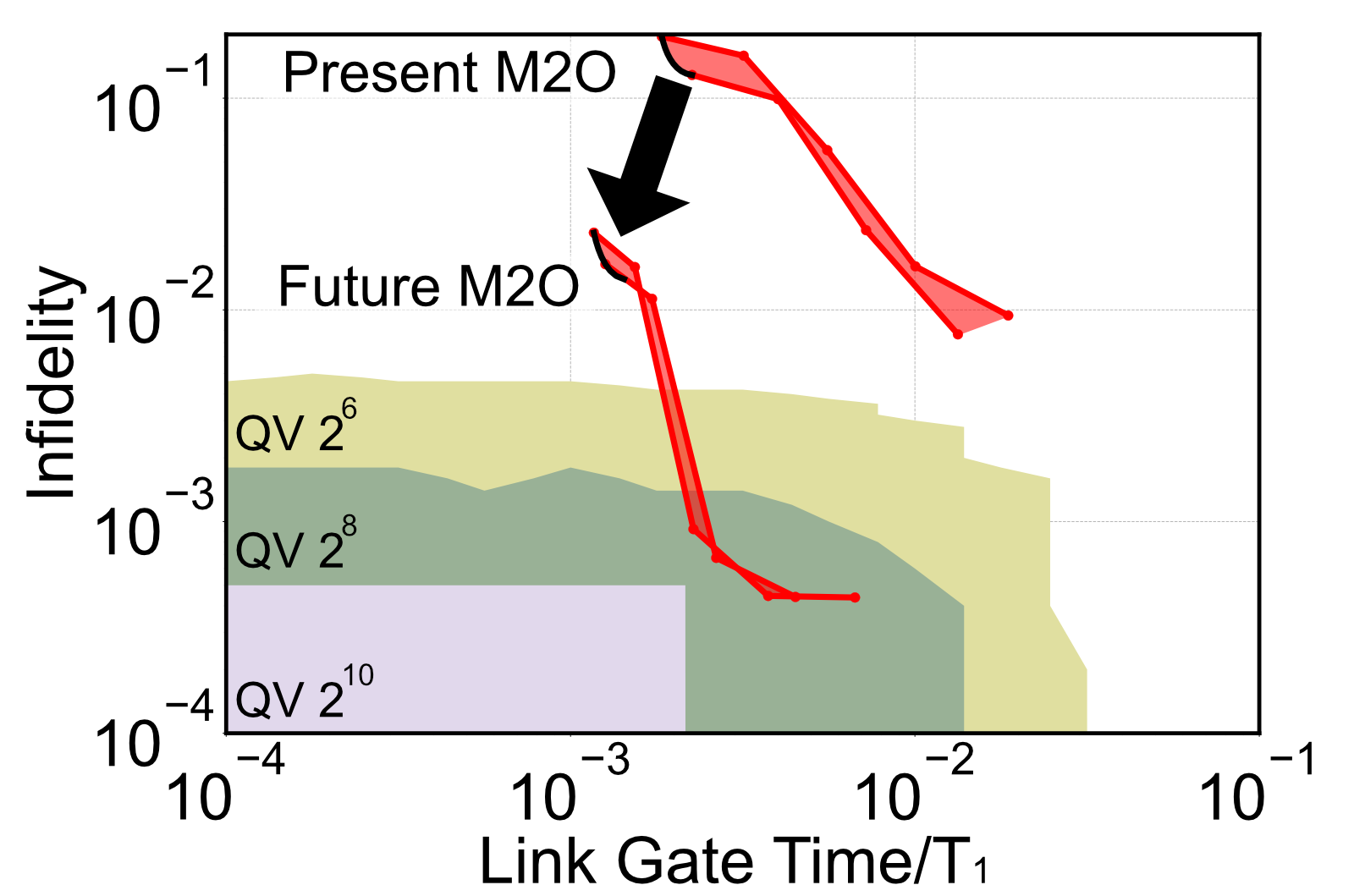}} & \centering\raisebox{-\totalheight}{\includegraphics[width = \mywidthR]{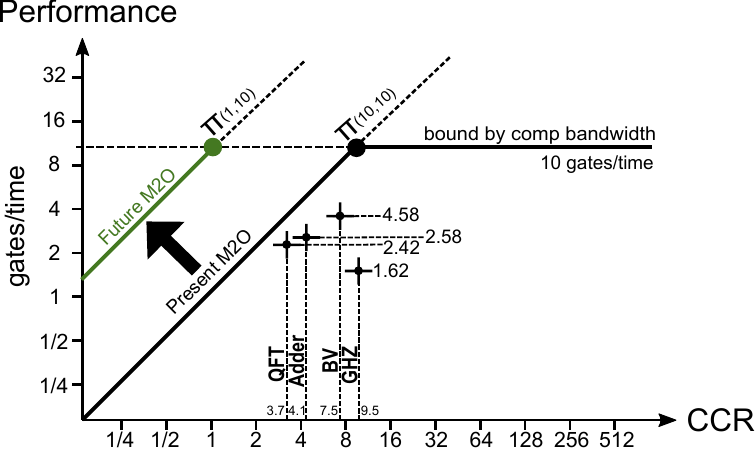}} &  \raisebox{-\totalheight}{\includegraphics[width = \mywidthR]{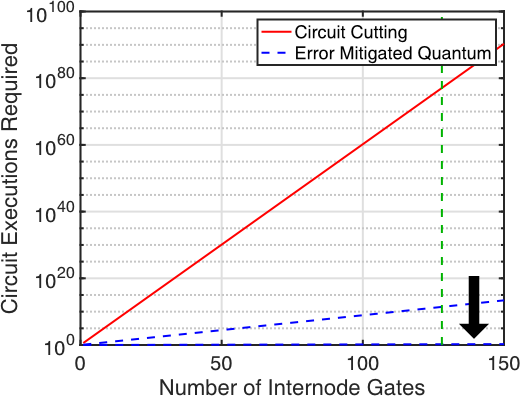}} \\ \hline 
\raggedright \textbf{M2O Multiplexing:}
Frequency and spatial
multiplexing can 
increase effective EP 
generation rate 100x. &\centering\raisebox{-\totalheight}{\includegraphics[width = \mywidthR]{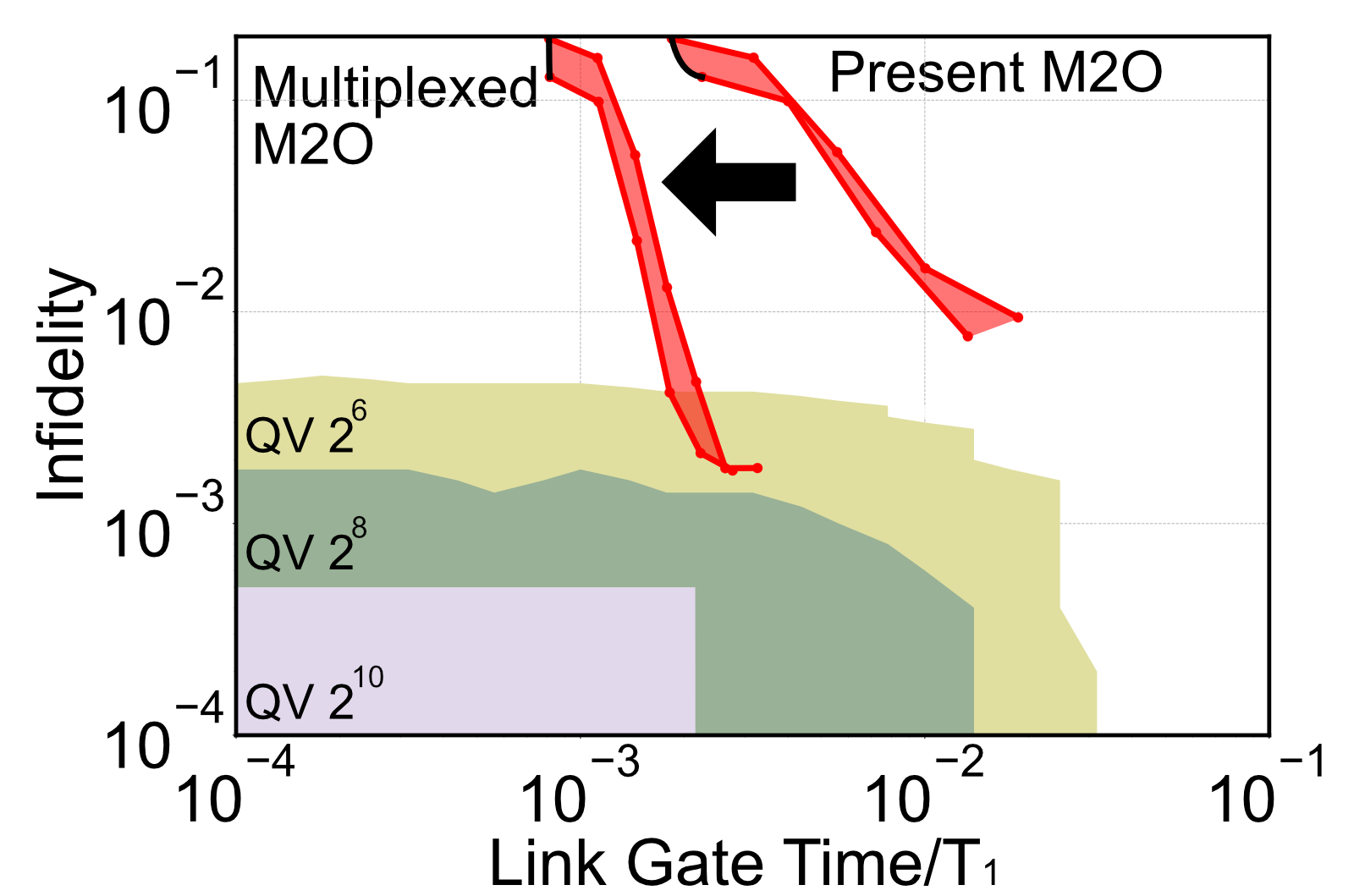}}&\centering\raisebox{-\totalheight}{\includegraphics[width = \mywidthR]{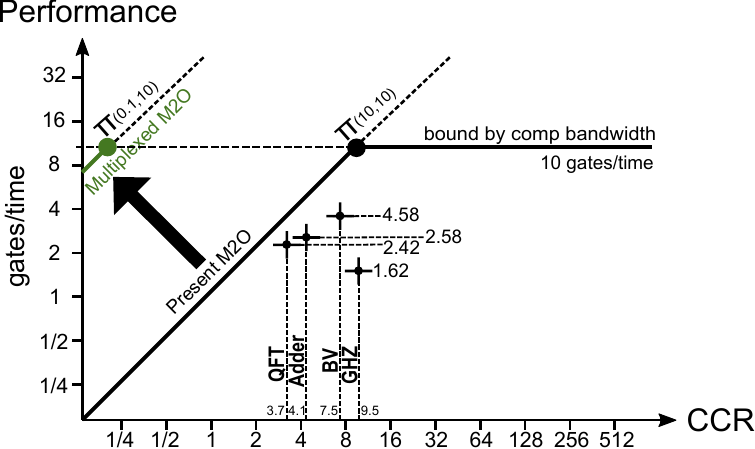}} & \raisebox{-\totalheight}{\includegraphics[width = \mywidthR]{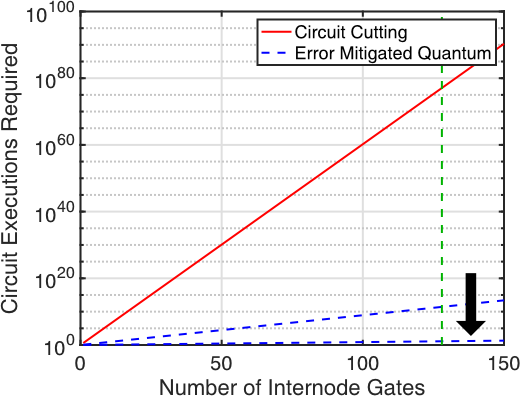}} \\ \hline 
\raggedright \textbf{Hi-fidelity Ion M2O:}
Ion-superconducting
qubit coupling could
increase rate 1000x using buffering
and reduce infidelity
1000x using ion-ion links.   & \centering\raisebox{-\totalheight}{\includegraphics[width = \mywidthR]{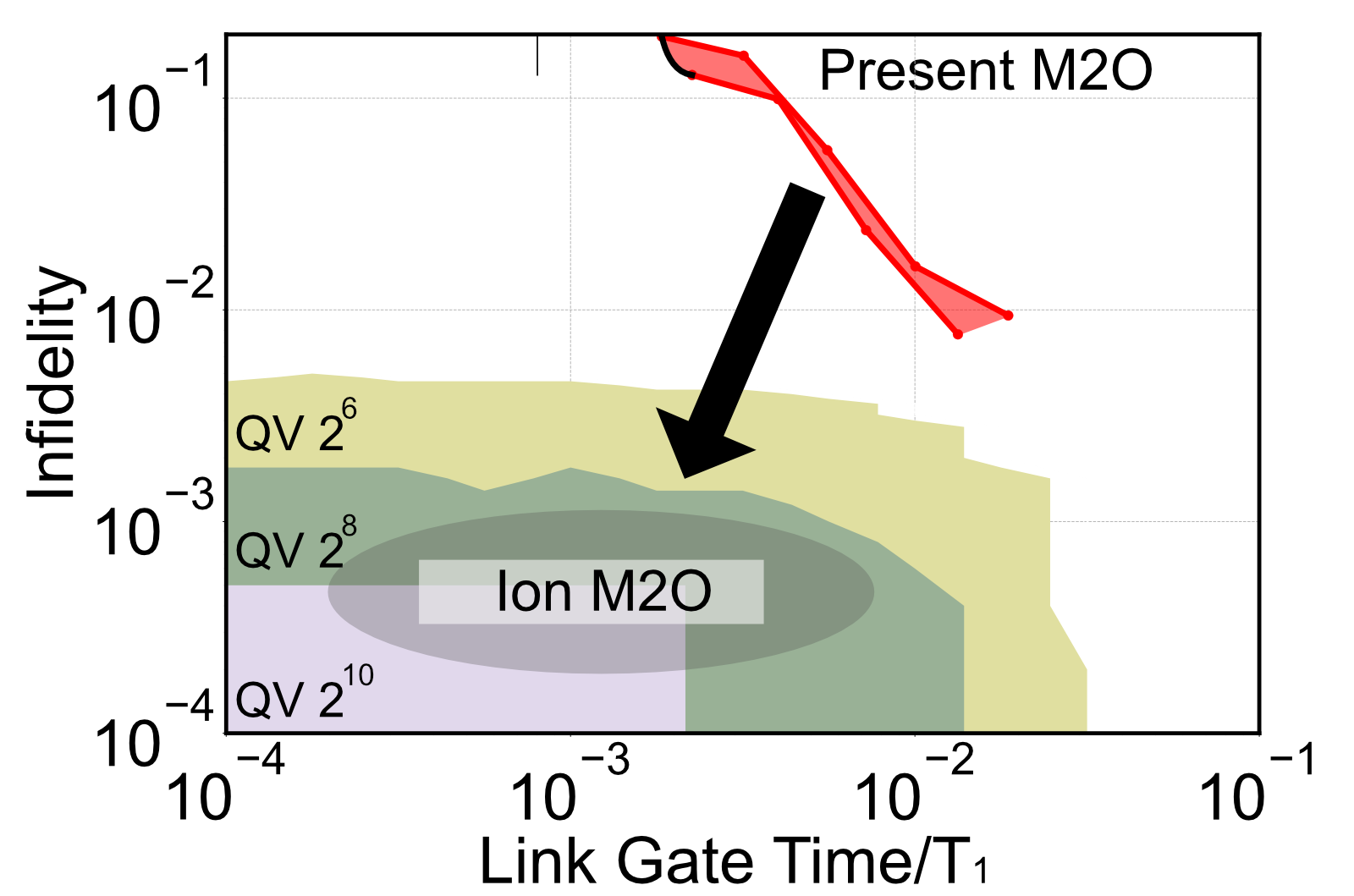}} & \centering\raisebox{-\totalheight}{\includegraphics[width = \mywidthR]{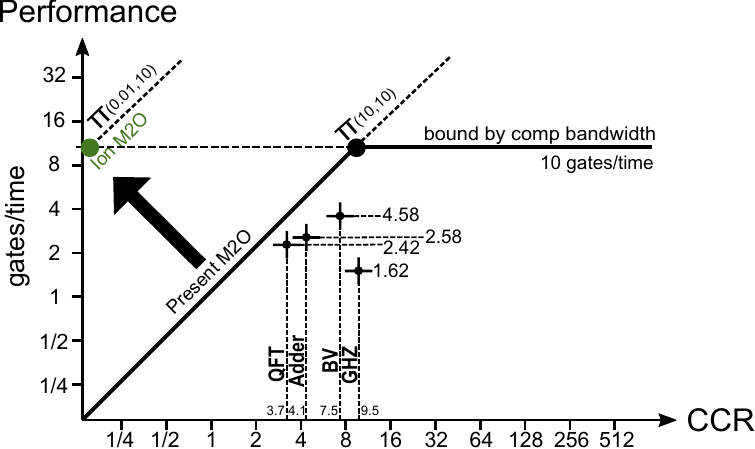}} & \raisebox{-\totalheight}{\includegraphics[width = \mywidthR]{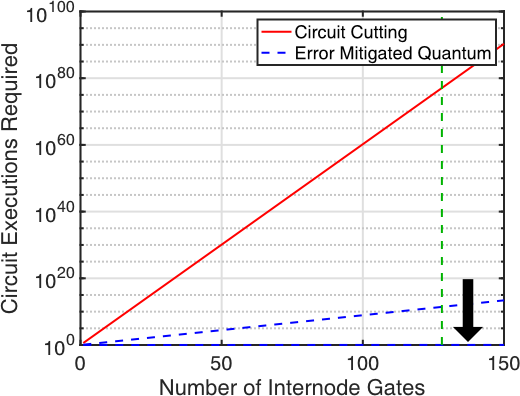}} \\ \hline 
    \end{tabular}
    \caption{The effects of three classes of improvements to the physical layer, as demonstrated in the GAP, Q-Roofline, and QCPA analyses.}
    \label{tab:physical_improvements}
\end{table*}


\begin{figure}[t]
    \centering
    \includegraphics[width=\linewidth]{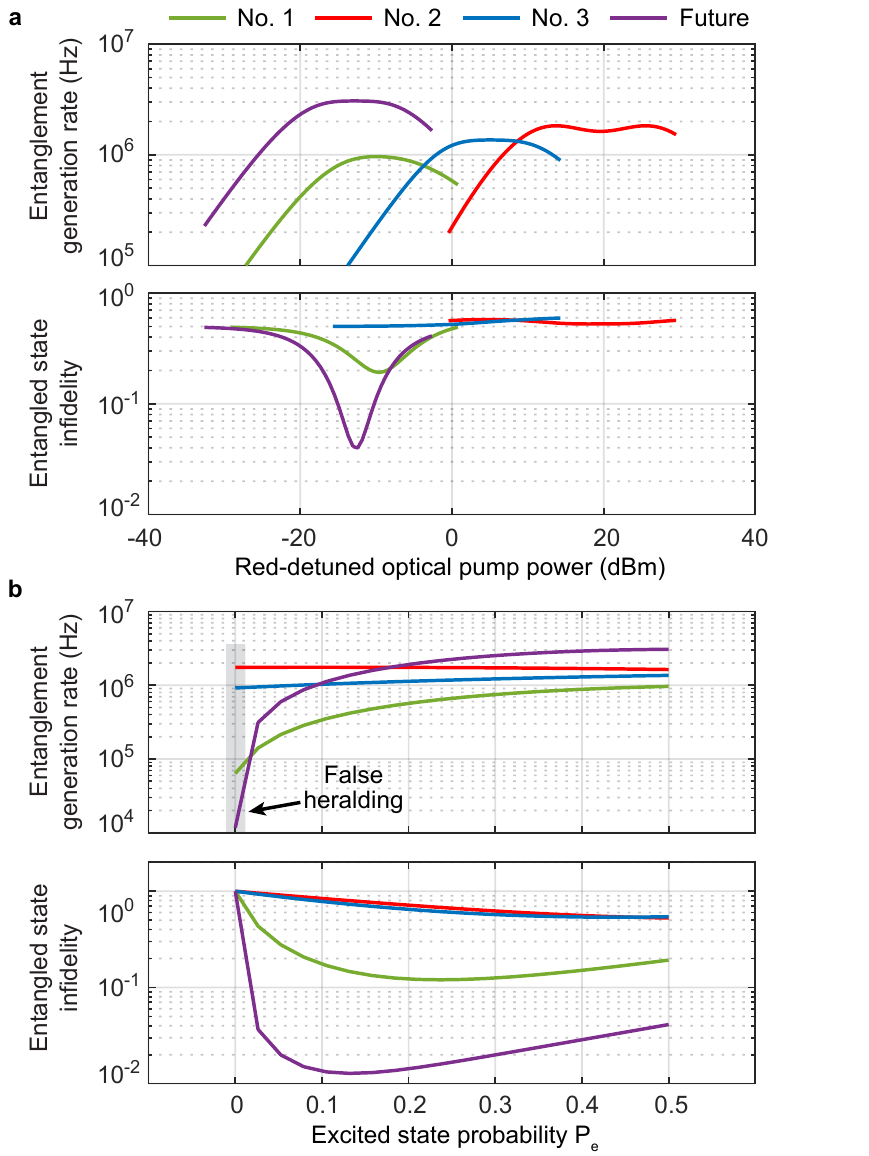}
    \caption{The performance of a hypothetical M2O converter is shown as the purple curve. The data from Fig.~\ref{fig:M2O_state} are also shown as a comparison.}
    \label{fig:estimate}
\end{figure}

Considerable progress may be made in the continued development of M2O devices. Metal reflectors \cite{kang2020high, krasnokutska2019high} and spot size converters \cite{nauriyal2019fiber, bakir2010low} have been experimentally demonstrated to minimize the insertion photon loss of grating couplers and edge couplers respectively, which can be applied to on-chip M2O converters to reduce fiber-to-chip coupling loss. Enhancement of the single-photon interaction rate is also critical, which requires further material and device optimization such as the minimization of mode volume \cite{li2020lithium, honl2022microwave} and the improvement of optical and microwave resonator quality factors \cite{sahu2022quantum}. In addition, thermal added noise induced by optical pump heating needs to be well suppressed to reduce the conversion infidelity. Possible heat dissipation methods to be investigated include radiative cooling \cite{xu2020radiative, wang2021quantum}, the use of superfluid helium for cooling \cite{lebrun2015cooling}, and the use of epitaxially grown superconducting materials \cite{yan2018gan, cheng2020epitaxial}. The bandwidth of the converters can be increased by operating the resonators in the overcoupled regime. Waveguide-based converters rather than resonator-based converters also present a potential route to broadband conversion. As a target for development, we present the performance of a hypothetical M2O converter (see Appendix~\ref{app:M2O} Table~\ref{tab:M2O_sets} in Appendix~\ref{app:M2O}) as the purple curve in Fig.~\ref{fig:estimate}. Such a M2O converter can be used to achieve $>$1~MHz production rate with an infidelity as low as $\sim$0.05, which might be available in the future if the bandwidth, photon loss, and the single-photon nonlinear coupling rate of existing converters can be improved by one to two orders of magnitude. In the first row of Table \ref{tab:physical_improvements}, we simulate these improvements using the method of the previous section; The lower time and infidelity of internode communication allow the ADDER benchmark to be executed on small architectures, while the balance between local and internode gates shifts towards allowing more internode gates and the gap between circuit cutting and quantum gates widens.

Besides experimental efforts, the development of protocols is another way to enhance the performance of current experiments. The fidelity of the direct conversion heralded scheme is primarily limited by the photon loss and thermal noise. The SPDC heralded scheme, however, is additionally limited by the possibility of multi-photon excitations in the resonator during the SPDC process \cite{guha2015rate}. Multi-photon excitations could potentially be suppressed through the use of an anharmonic resonator \cite{krastanov2021optically}. In both schemes, the small probability that a photon is emitted simultaneously at both nodes, combined with the optical loss, will lead to a false heralding signal. One potential solution based on double-heralded detection has been proposed by Barrett and Kok~\cite{Barrett2005} and experimentally realized with defects in crystals and trapped ions \cite{Bernien2013, Pfaff2014, Hensen2015, Casabone2013} and superconducting circuits \cite{2016PhRvX...6c1036N}. While boosting the fidelity, this design requires two successful photon detections, and thus the success probability--as well as the entanglement generation rate--scales with the square of the photon detection probability. Alternative emerging protocols designed for M2O interfaces have also been proposed, such as the adaptive control protocol for reducing thermal noise \cite{zhang2018quantum}, the active quantum feedback for deterministic entanglement generation \cite{Martin2015}, the continuous-variable quantum teleportation \cite{wu2021deterministic, rueda2019electro} for high-fidelity state transfer, and time-bin \cite{zhong2020proposal} and frequency-bin \cite{zhong2020entanglement} encoding for improved entanglement generation rate. One advantage of M2O links is that it may be possible to use the quantum control techniques available in circuit QED to use error correctable bosonic codes, several of which have recently exceeded the break-even point as quantum memories~\cite{Ofek2016, 2022arXiv221109116S, 2019NatPh..15..503H}, for the communications.

Another key direction for improvement at the Physical layer is the use of multiplexing. As we saw in Section \ref{sec:FullModels}, increasing the rate of pair generation is a key goal of the Physical layer. By operating multiple entanglement generation devices in parallel, we can increase the effective rate of entangled pair generation. Because decoherence accumulated while waiting for further EPs is a major source of internode noise, increasing the effective rate of EP generation reduces both the time and infidelity of internode communication, resulting in the dramatic effects shown in the second row of Table \ref{tab:physical_improvements}.

Multiplexing M2O EP generation requires routing entangled photons generated in parallel channels into a superconducting node's distillation module in real time. 
There are several methods for multiplexing flying qubits into a superconducting node. One multiplexing method that is promising for long-distance entanglement uses the ``pitch-and-catch'' framework, where the flying qubit is caught by a linear bus and swapped into the qubit coupled to that bus \cite{2016PhRvX...6c1036N, Campagne2018,Leung2019,Burkhart2021}. Frequency-multiplexing the flying qubits would allow multiple flying qubits to be caught in parallel by the corresponding modes in the send/receive bus, and distributed into various coupled qubits. One advantageous choice of bus-qubit coupler is the SNAIL (Superconducting Nonlinear Asymmetric Inductive eLement) \cite{Frattini2017} rather than currently used transmon couplers; the three-wave mixing interaction has reduced susceptibility to unwanted transitions/parametric processes compared to the four-wave mixing in transmon-based couplers. The SNAIL has been used to demonstrate successful all-to-all routing among 4 quantum modules \cite{Zhou:2021cri}. The SNAIL can also be used as an alternative method for multiplexing flying qubits which is relevant for physically compact quantum computing within a single fridge. In this modality, a nonlinear SNAIL bus passively couples together all the qubits extending from it \cite{Zhou:2021cri,McKinney2022}.


Finally, hybrid technologies promise the greatest potential improvements, but also pose the most severe technical challenges \cite{2009PhST..137a4001W}. In particular, a hybrid system using ions coupled to superconducting qubits \cite{doi:10.1146/annurev-conmatphys-030212-184253, 2016QuIP...15.5385D, 2012PhRvL.108m0504K} could allow for optical ion-ion links \cite{2020Natur.586..538N, 2020Natur.586..533M, 2009PhRvL.102y0502M},  between chips in separate dilution refrigerators. Significant technical challenges accompany hybrid ion-superconducting qubits \cite{PhysRevLett.103.043603}. However, techniques using molecular ions coupled to superconductors 
\cite{PhysRevA.83.012311, 2006quant.ph..5201A, 2010ApPhL..97x4102W, PhysRevLett.97.033003, PhysRevA.76.042308}, while the use of ion chains for mode matching \cite{PhysRevLett.107.030501} can improve this coupling. As shown schematically in the third row of Table \ref{tab:physical_improvements}, a superconducting-ion coupling would enable the rapid production of high-fidelity internode entangled pairs, likely limited chiefly by the rate and fidelity of the superconducting-ion coupling \cite{2012PhRvL.108m0504K}, and the use of ions as an extremely long-lived memory would also have significant implications for entanglement distillation and is discussed in the following section. 

\subsection{Distillation Layer Improvements}\label{sec:EPAlgs}

In Section \ref{sec:FullModels} we saw the key role entanglement distillation plays in enabling MNQC performance by improving the fidelity of EPs produced during the M2O process. However, entanglement purification performance is currently limited by the low yield of the purification protocols as well as by qubit decoherence during the purification. Potential improvements to this performance include careful co-design of protocols to adapt for the noise profile of M2O generation, the use of memory to prevent decoherence during the distillation process, and the use of long-term memories to allow buffering and effectively remove the fidelity bound. We tabulate these approaches and their speculative effects in Table \ref{tab:distillation_improvements}. 

\begin{table*}[]
    \centering
    \begin{tabular}{|p{\mywidthL}|p{\mywidthR}|p{\mywidthR}|p{\mywidthR}|}\hline
        & \begin{center}
            \Large GAP
        \end{center} & \begin{center}
            \Large Q-Roofline
        \end{center} & \begin{center}\Large QCPA\end{center} \\ \hline
        \raggedright \textbf{Distillation Protocol
Co-Design:}
Careful
tailoring of distillation 
protocols to M2O noise
may reduce infidelity
2x or more & \centering\raisebox{-\totalheight}{\includegraphics[width = \mywidthR]{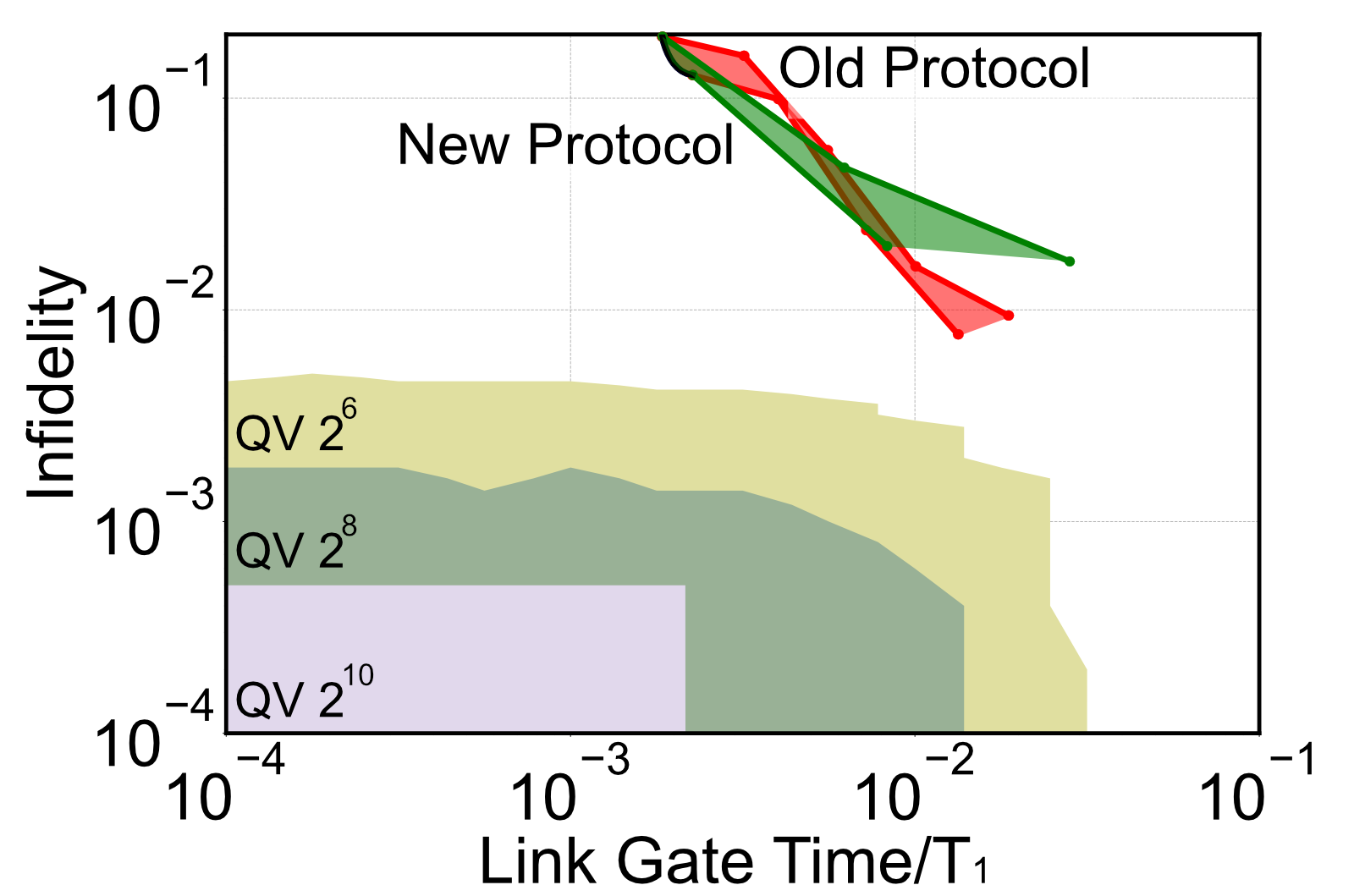}} & \centering\raisebox{-\totalheight}{\includegraphics[width = \mywidthR]{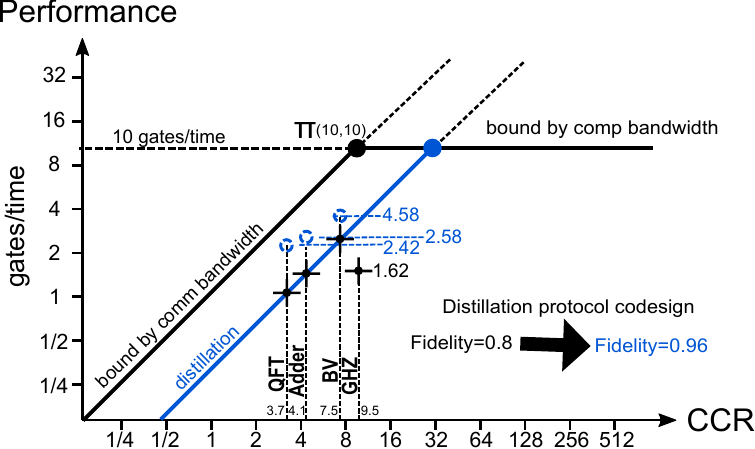}} & \raisebox{-\totalheight}{\includegraphics[width = \mywidthR]{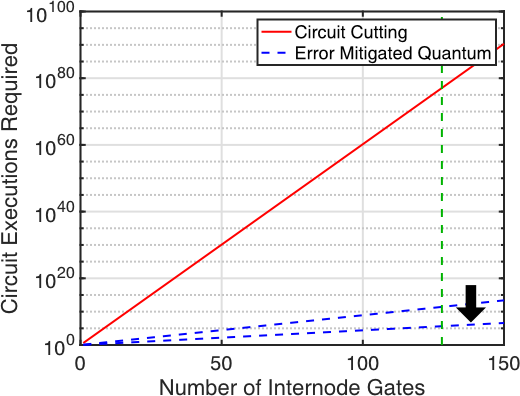}} \\ \hline 
\raggedright \textbf{1ms Memory:}
Protection from decoherence greatly improves 
distillation performance
allowing new regimes
of algorithms.  &\centering\raisebox{-\totalheight}{\includegraphics[width = \mywidthR]{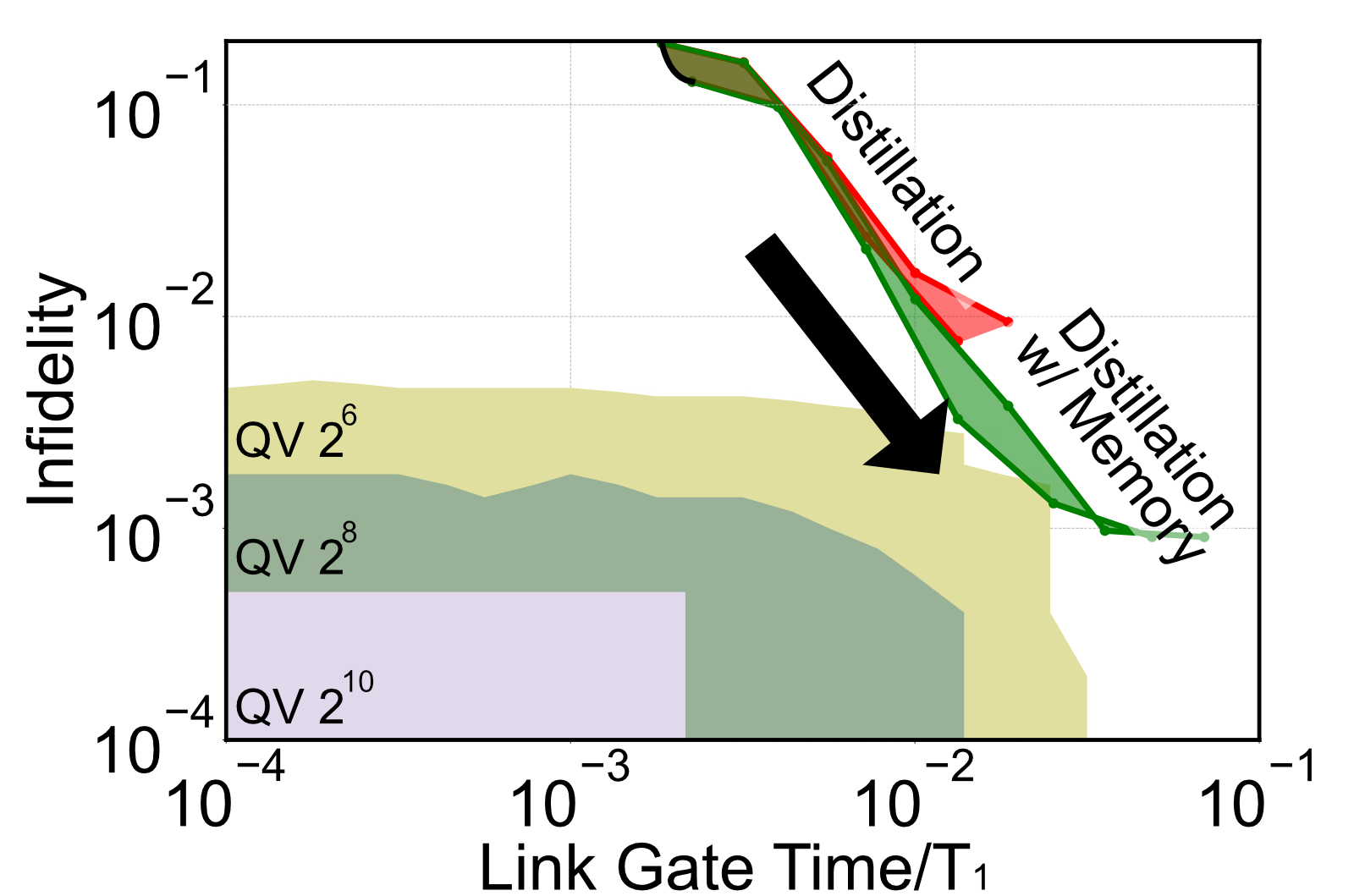}}& \centering\raisebox{-\totalheight}{\includegraphics[width = \mywidthR]{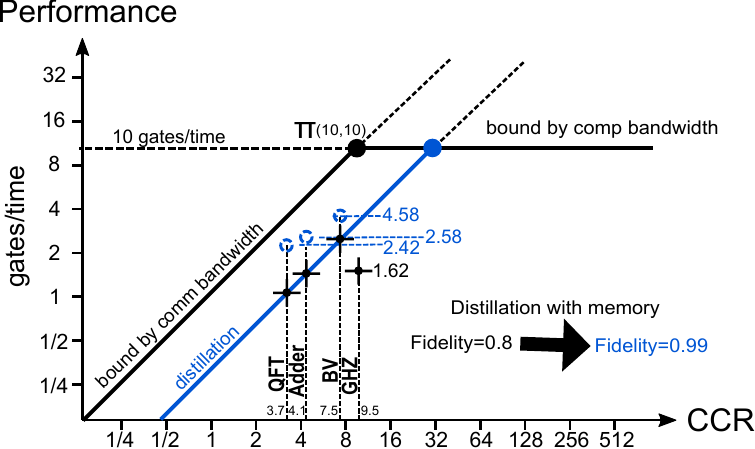}} & \raisebox{-\totalheight}{\includegraphics[width = \mywidthR]{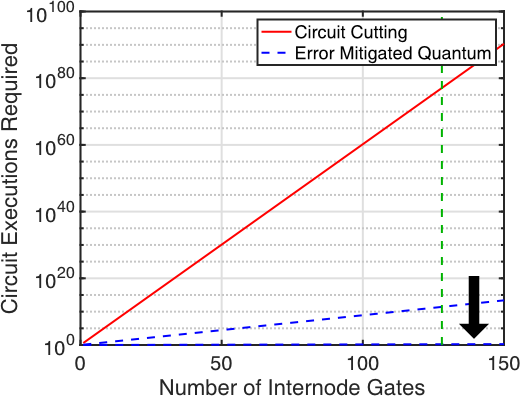}} \\ \hline 
\raggedright \textbf{10ms Memory:}
Memory protects from decoherence and
allows buffering of
many EPs, allowing for
high-fidelity, cheap 
internode gates.  & \centering\raisebox{-\totalheight}{\includegraphics[width = \mywidthR]{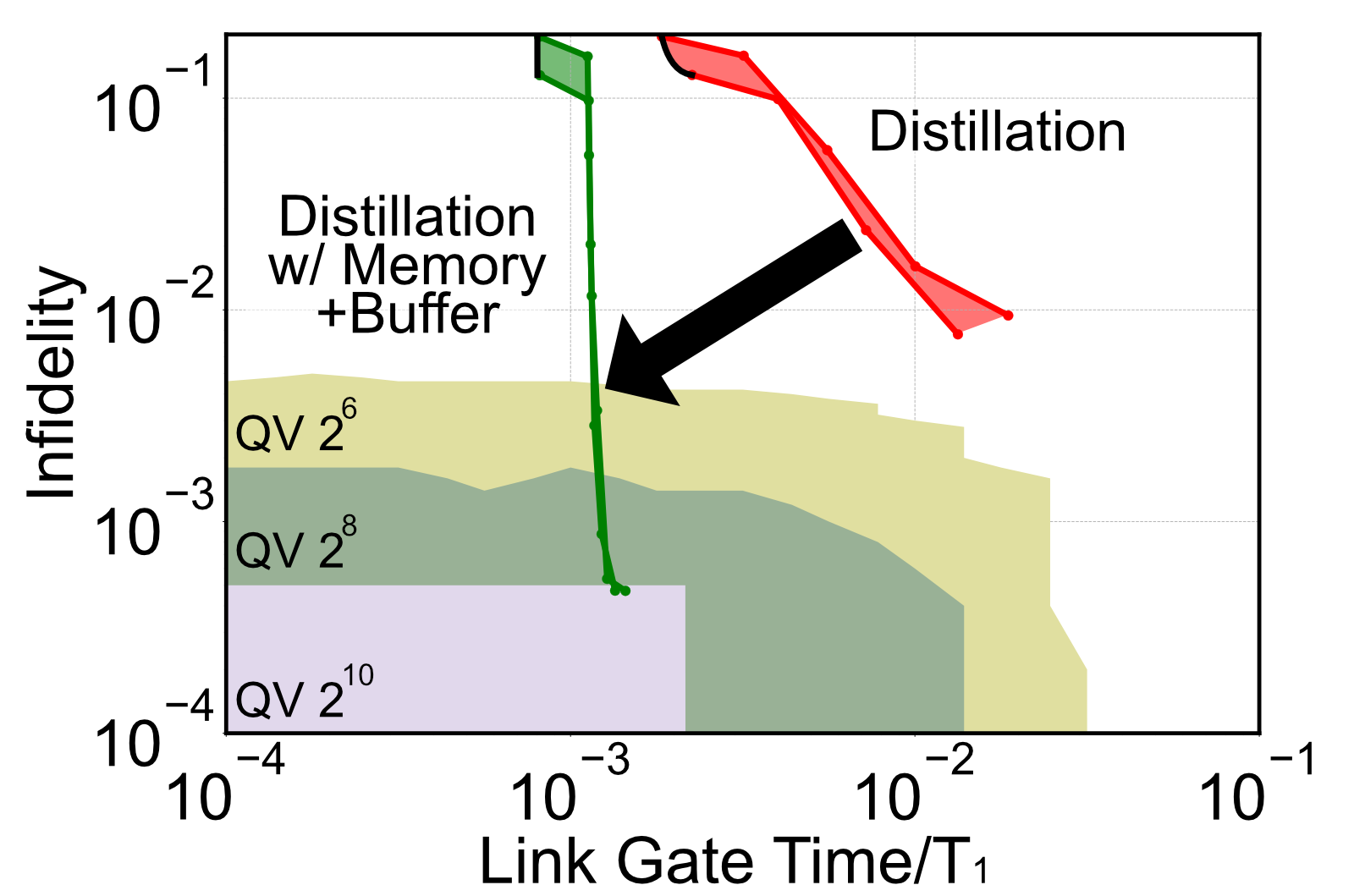}} & \centering\raisebox{-\totalheight}{\includegraphics[width = \mywidthR]{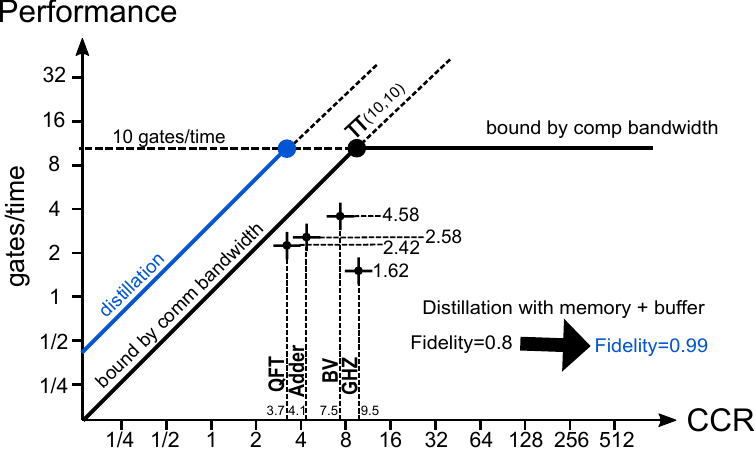}} & \raisebox{-\totalheight}{\includegraphics[width = \mywidthR]{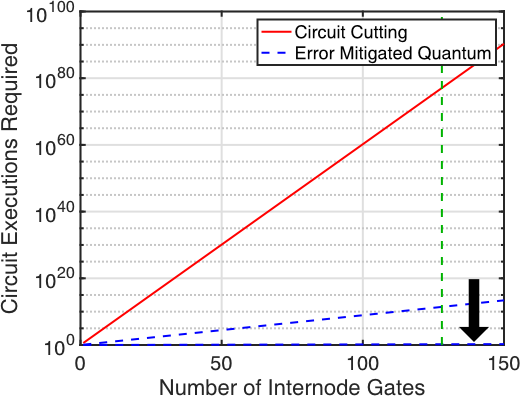}} \\ \hline 
    \end{tabular}
    \caption{The effects of three classes of improvements to the Distillation layer, as demonstrated in the GAP, Q-Roofline, and QCPA analyses.}
    \label{tab:distillation_improvements}
\end{table*}

\begin{figure}[ht]
    \centering
    \includegraphics[width = 0.8 \columnwidth]{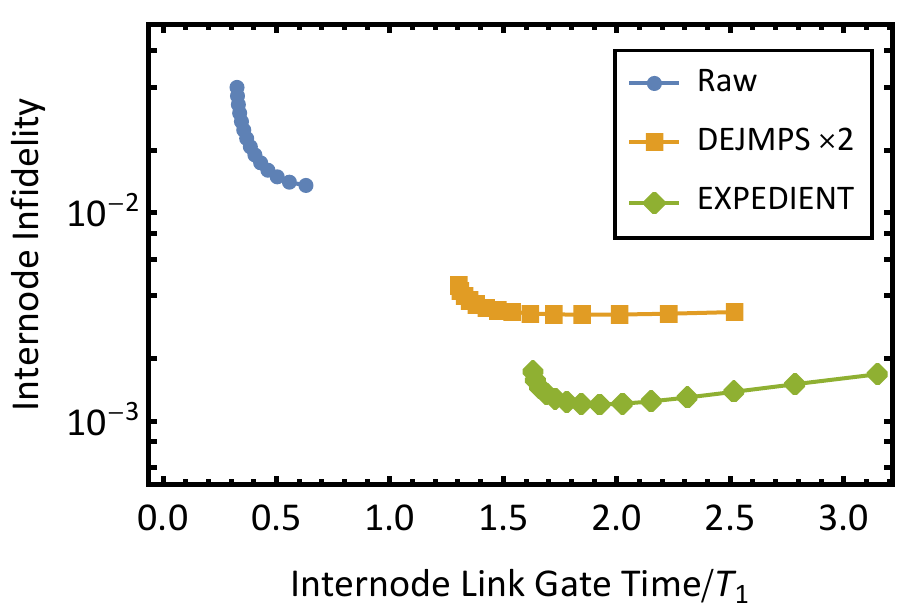}
\caption{Comparison of the EXPEDIENT purification protocol \cite{Nickerson2013} with the two-round nested DEJMPS protocol~\cite{Deutsch1996}. }
    \label{fig:expedient}
\end{figure}

\begin{figure*}[!htbp]
\centering
\subfloat[]{\includegraphics[width = 0.45 \textwidth]{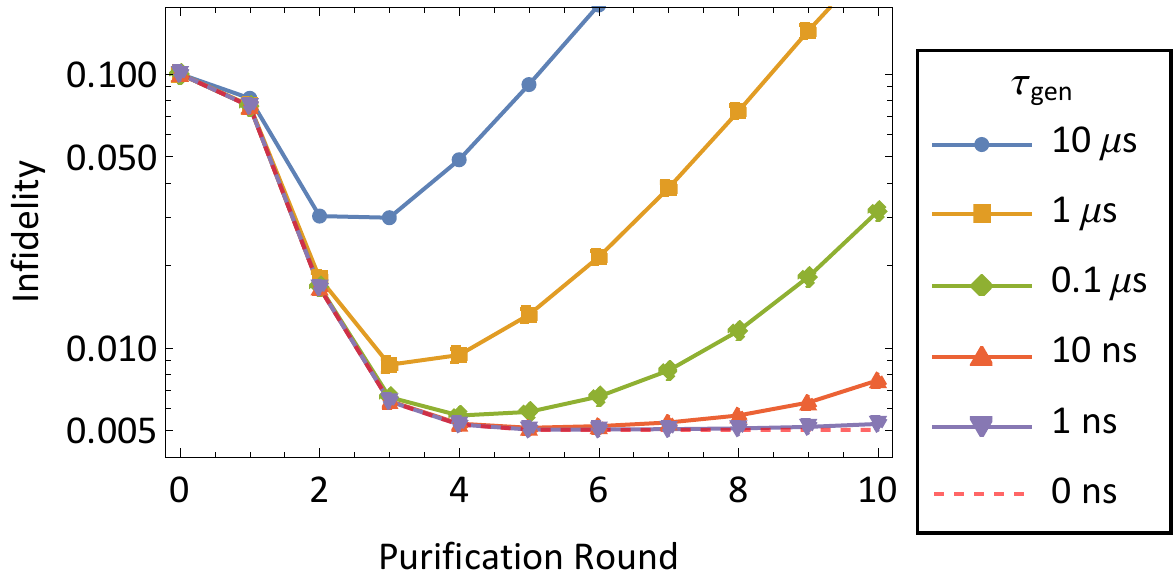}} \;
\subfloat[]{\includegraphics[width = 0.35 \textwidth]{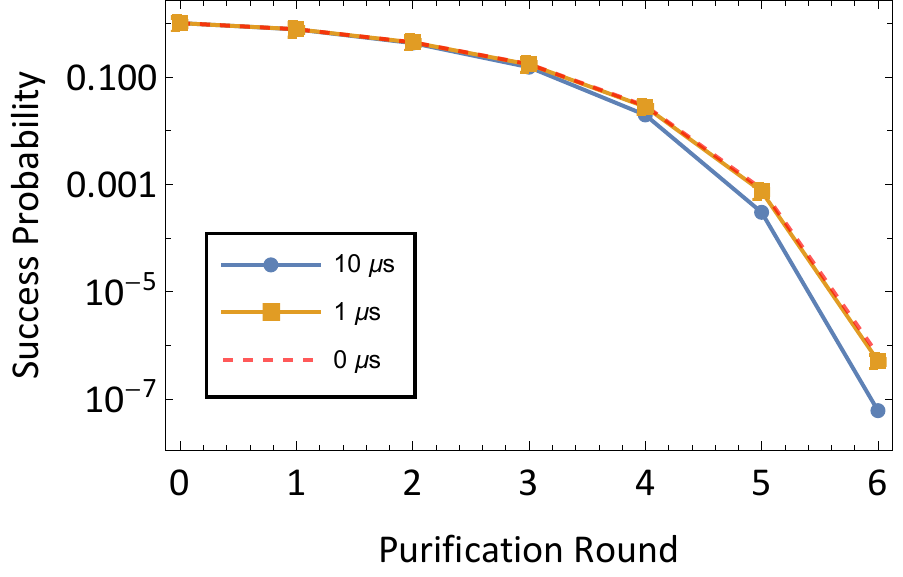}}
\caption{The performance of the DEJMPS purification protocol~\cite{Deutsch1996} with sequential raw EP generation. In (a), we plot the infidelity of the output state from $n$ rounds of purification using DEJMPS protocol. The generated raw EPs have fidelity $0.9$. The superconducting qubits have $T_1 = T_2 = 1$~ms. In (b), we plot the one-shot success probability as a function of nested purification rounds. The one-shot success probability decreases exponentially.}
\label{fig:practicle_purification}
\end{figure*}

To improve the fidelity of these EPs, several entanglement purification protocols have been invented. In Appendix~\ref{app:Distillation} we briefly introduce two purification protocols: the BBPSSW protocol~\cite{Bennett1996} and the DEJMPS protocol~\cite{Deutsch1996}. In Section \ref{sec:LayerModels}'s full stack simulation the DEJMPS protocol is used, as it provides more efficient purification compared to the BBPSSW protocol~\cite{Deutsch1996} and uses a small number of EPs to perform purification. We also notice that more advanced purification protocols, e.g., double selection purification protocol~\cite{Fujii2009}, EXPEDIENT and STRINGENT purification protocols~\cite{Nickerson2013}, may give better output EP fidelities after purification. In Fig.~\ref{fig:expedient}, we compare the performance of the two-round nested DEJMPS protocol with the EXPEDIENT protocol. The input EP state is from the M2O calculation. We notice the EXPEDIENT protocol uses 5 EPs in total to generate one EP with higher fidelity. Compared to the 2-round nested DEJMPS protocol, the EXPEDIENT protocol can give $\sim 2$ times improvement. However, as it requires more EPs for each purification operation, the time for remote gate operation will be longer, and it will suffer more from the decoherence error if the EP generation is slow.  However, even the DEJMPS protocol has a low purification yield. This is because for each purification, one of the two input EPs is destroyed. One direction for future work is to design new protocols for more efficient entanglement purification. Both the BBPSSW and DEJMPS protocols accept any raw EPs whose fidelity to the target state is greater than $0.5$, without utilizing any other information about those states. One way to improve purification efficiency is to construct a precise error model for the raw EPs generated from the physical layer, and use that error information to design a more efficient purification protocol. This new protocol can either use hashing protocols with high finite yield~\cite{Dur2007,Nickerson2013, Krastanov2019optimized} or require fewer rounds of nested purification to achieve high-fidelity EPs, so it could be used to implement more complex distributed algorithms. The improved performance of the EXPEDIENT protocol is shown in the first row of Table \ref{tab:distillation_improvements}.

In the full stack simulation (see Fig.~\ref{fig:QV} and~\ref{fig:TotalSim}), with the finite raw entangled pair generation rate, it is only practical to perform a few rounds of nested purification. In Fig.~\ref{fig:practicle_purification}a, we calculate the fidelity of the output entangled pair after $n$ rounds of purification. 
In this calculation, we especially show the effect of the finite rate of raw EP generation on the purification protocol.
We observe a steady increase in output state infidelity due to the qubits relaxing and dephasing while waiting for more raw EPs to be generated. As discussed in Section \ref{sec:MO-conv}, the fastest raw EP generation rates are currently on the order of 1 MHz, so to make purification robust, effort must be made to reduce the raw EP generation time and increase qubit coherence times. In Fig.~\ref{fig:practicle_purification}b we plot the single-shot success probability of $n$ rounds of purification [see Eq.~\eqref{eq:single_shot_p} in Appendix~\ref{app:Distillation}]. Due to the DEJMPS protocol's low yield, even though the success probability of each single purification of two raw EPs can be close to unity, the overall single-shot success probability decreases exponentially as the number of nested purification rounds increases. This can also be seen from the fact that the number of terms in Eq.~\eqref{eq:single_shot_p} increases exponentially as the round increases. So the purification protocol's low yield limits the practical benefit of doing many purification rounds in the entanglement distillation layer.

\begin{figure}[b]
    \centering
    \includegraphics[width = \columnwidth]{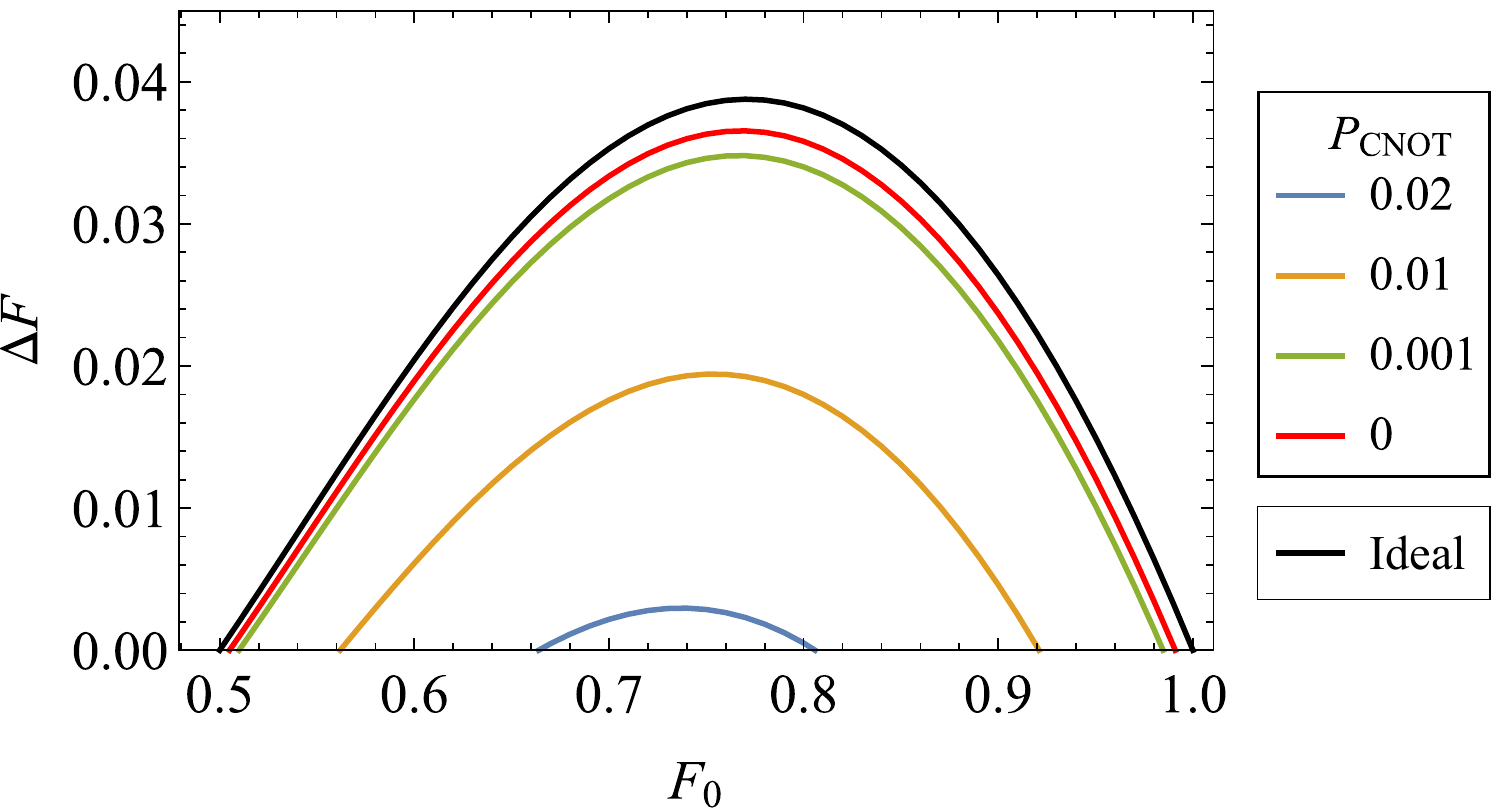}
    \caption{The performance of the DEJMPS purification protocol with imperfections. We consider a single round of entanglement purification in the presence of imperfect  gates between superconducting qubits and finite qubit coherence times. The CNOT gate error is modeled by depolarizing error channels with error probability $P_{\text{CNOT}} = .0001$. The fidelity of the imperfect EPs before purification is $F_0$, and after the purification we calculate the fidelity gain $\Delta F = F_{\text{new}} - F_{0}$. In addition, the superconducting qubits have finite relaxation and dephasing times for all but the case labeled ``Ideal''. The qubit relaxation and dephasing times are $T_1 = T_2 = 1~$ms. Each round of purification takes $1~\mu$s. The ``Ideal'' line is for the performance of the purification protocol with neither CNOT gate error nor decoherence error.}
    \label{fig:fid_gain}
\end{figure}

Furthermore, in our full-stack simulation, we assume that the local gates have depolarization error with probability $0.1\%$. However, in reality, the local gates between the compute qubits used to purify EPs may have larger errors. In Fig.~\ref{fig:fid_gain}, we consider the fidelity gain ($\Delta F$) by performing a single round of entanglement purification to explore the effect of local imperfections. We consider the effect of CNOT gate error as well as qubit relaxation and dephasing during the purification protocol. 
We notice that even with CNOT gate error $P_{\text{CNOT}}=0.01$~\cite{Wei2022}, the efficiency of entanglement purification is noticeably affected compared to $P_{\text{CNOT}} = 0.001$ case. 
To improve the performance of the purification layer and fully leverage the power of entanglement purification, local gate error needs to be kept low. 

One likely way to suppress relaxation and dephasing during purification is to use dedicated quantum memory elements. When the compute qubits are waiting for the next raw EP to arrive, their states can be swapped into quantum memory elements that have longer coherence times. This is particularly helpful for later rounds of distillation when the idle time on one of the two EPs from the previous round is substantial.  In order to achieve this goal, the quantum memory elements need to have fast and high-fidelity SWAP gates with the compute qubits and they need to be stabilized against relaxation and dephasing, either by having naturally long coherence times or via active or autonomous quantum error correction ~\cite{Gertler2021}.  The effect of a memory with a $1$ms coherence time is shown in the second row of Table \ref{tab:distillation_improvements}, where we see that it dramatically reduces the achievable internode infidelities. 

Transmon and fluxonium superconducting qubits have demonstrated high-fidelity two-qubit entangling gates ~\cite{Sung2021,Wei2022, Ficheux2021,Bao2022,Dogan2022}, which make them good computing elements. 
Recent improvements in material processing and shielding/filtering have also boosted their coherence times towards 1 ms (as assumed in our simulations). However, 2D qubit coherence is often limited by dielectric loss from the substrate \cite{Read2022} and the interfaces \cite{Wang2015}, while 3D microwave modes can serve as even better memory elements \cite{Reagor2016} with potential lifetimes up to seconds \cite{Romanenko2020}. 
Furthermore, 3D multimode cavities are a promising form of quantum memory element because a memory buffer with many storage modes can be created out of a single physical cavity, and high-fidelity SWAP gates in and out of the buffer can be performed by a single transmon \cite{chakram2021seamless, Chakram2022}. 

For these memory cavities to be effective in distillation protocols, an important area of improvement is the fidelity of cavity-qubit \cite{chakram2021seamless} or cavity-cavity SWAP \cite{Gao2018} gates.  These previous demonstrations rely on the shared nonlinearity of transmons to activate relatively slow four wave mixing processes.  More recent experiments have shown that by using purpose-built parametric couplers one can perform much faster SWAP operations (100 ns or less) regardless of the nonlinearity of the swapping modes \cite{deGraaf2022, Guinn2022}, analogous to parametric two-qubit gates \cite{Reagor2018}.  Implementation of these gates may allow storage of retrieval of EPs to and from quantum memory elements with infidelity at $10^{-4}$ level.

An effective quantum memory exceeding $1$ms, such as the hybrid superconducting-ion system discussed in the previous section, can have paradigm-shifting effects on both the fidelity and rate of internode communication because it can allow for the buffering \cite{wu2022collcomm} of entangled pairs. The effects of a $10$ms quantum memory are schematically shown in the third row of Table \ref{tab:distillation_improvements}, where the buffering of memory reduces the time to execute internode gates during an algorithm to be comparable to that of local computation and thus results in a dramatic improvement in MNQC performance. In order to further increase the coherence time of the memory qubit, one could consider encoding the quantum information into a bosonic error correction code and implementing error correction \cite{Terhal2020, Cai2021, Joshi2021}. Recently, active and autonomous stabilization of bosonic codes has been demonstrated close to or beyond the break-even point including the cat code \cite{Ofek2016, Gertler2021}, the binomial code \cite{Hu2019}, and the GKP code \cite{Campagne2020, 2022arXiv221109116S}. However, in these experiments the coherence of the error-corrected quantum memory is limited by that of the nonlinear ancilla element used for stabilization. Eliminating this limitation and realizing a fault-tolerant bosonic memory well beyond the break-even point is an active topic of research \cite{Rosenblum2018,Puri2019,Grimsmo2021}. 
\subsection{Compiler and Application Improvements}\label{sec:CompilerAndApp}

While the lower levels of the MNQC stack determine the properties of the internode gates, it is the application and compiler layers that determine the use of internode gates and thus the performance of applications on a future MNQC. Much as in classical computing, developing compilers that can efficiently optimize around weaker internode links and applications that are adapted to multinode architectures will be critical for the success of MNQCs, and we must understand how improvements to these layer intersect with those of the rest of the stack. Determining the potential improvements for these layers involves considerable uncertainty as we do not have bounds for the performance achievable by compilers that have not been built (in the language of Section \ref{sec:FullModels}, we do not have bounds on achieveable CCRs), nor can we estimate the potential of algorithms yet to be discovered. Hence for these layers we take a schematic approach that still allows us to lay out a research agenda towards effective MNQCs.

\begin{table*}[]
    \centering
    \begin{tabular}{|p{\mywidthL}|p{\mywidthR}|p{\mywidthR}|p{\mywidthR}|}\hline
        & \begin{center}
            \Large GAP
        \end{center} & \begin{center}
            \Large Q-Roofline
        \end{center} & \begin{center}\Large QCPA\end{center} \\ \hline
        \raggedright \textbf{Efficient Compiling:}
Compiler improvements
reduce internode 
gate use by a factor of 
2x or more and scale
to large sizes & \centering\raisebox{-\totalheight}{\includegraphics[width = \mywidthR]{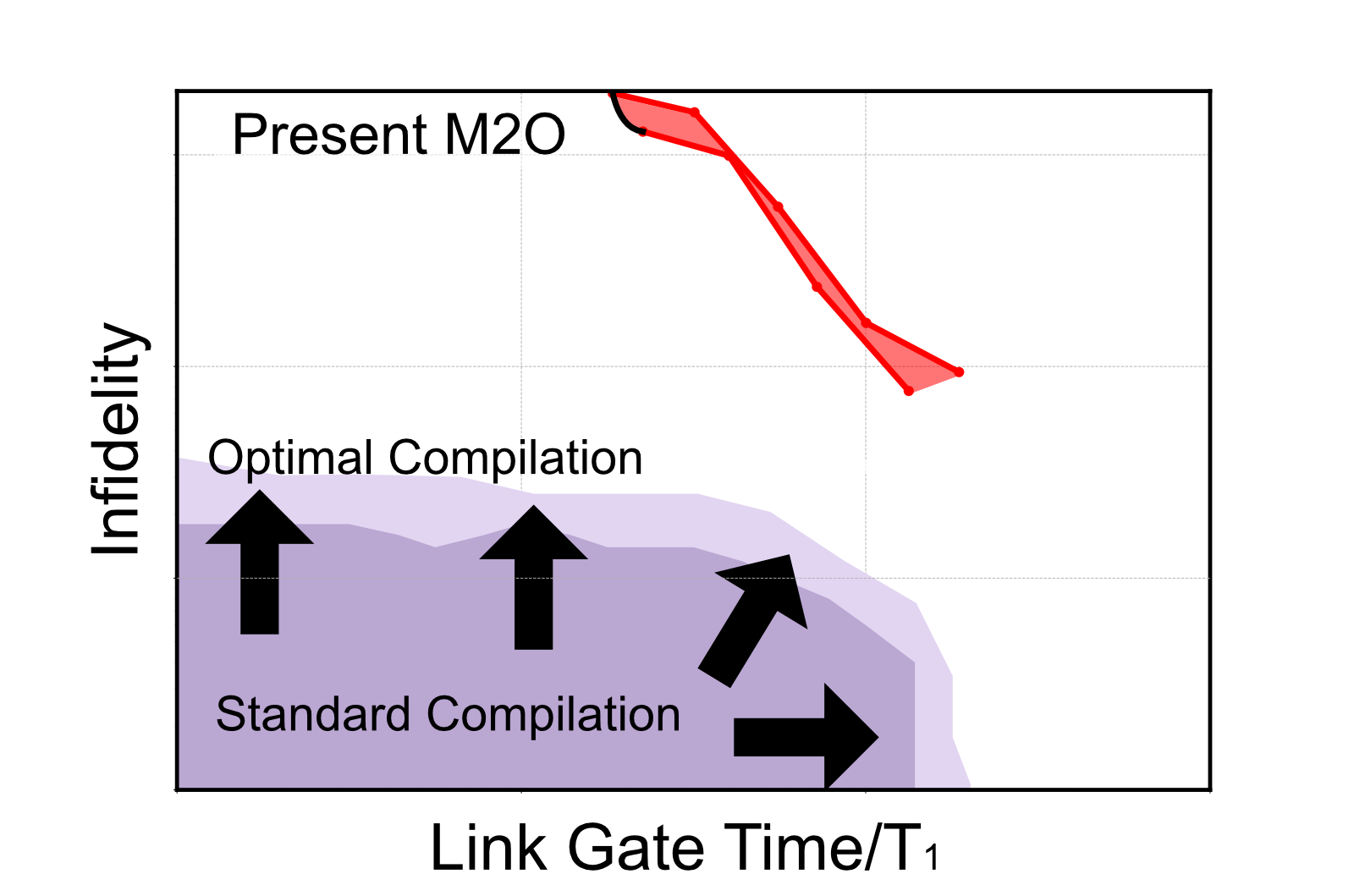}} & \centering\raisebox{-\totalheight}{\includegraphics[width = \mywidthR]{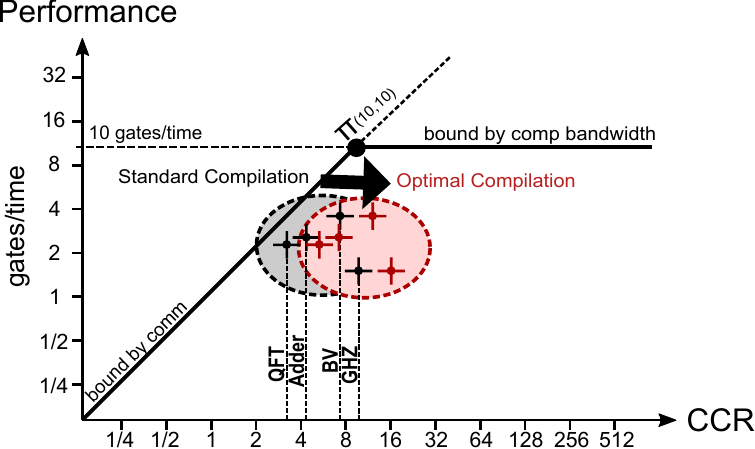}} & \raisebox{-\totalheight}{\includegraphics[width = \mywidthR]{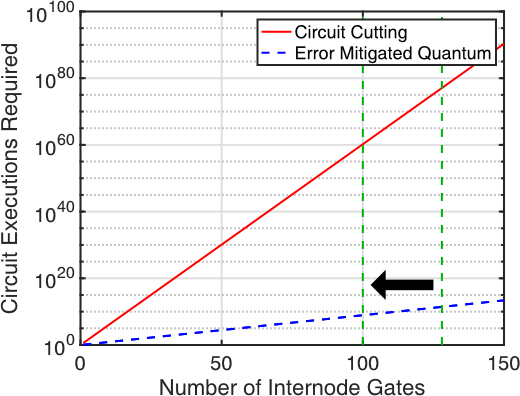}} \\ \hline 
\raggedright \textbf{New Distributed 
Algorithms:}
Large gains can be obtained with algorithms  designed for distributed systems, e.g. distributed QPE, reducing internode 
gates by 100x or more. &\centering\raisebox{-\totalheight}{\includegraphics[width = \mywidthR]{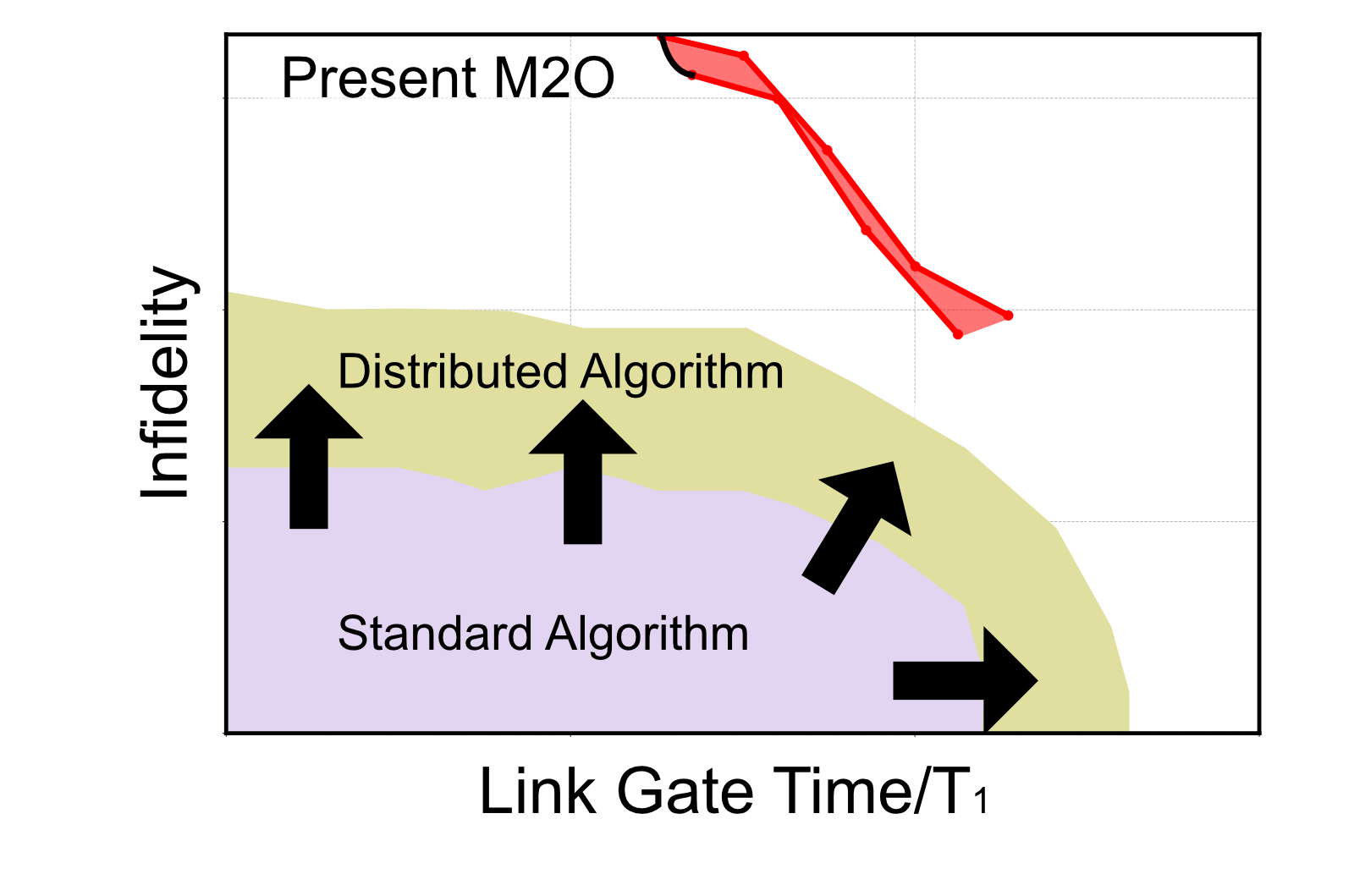}}& \centering\raisebox{-\totalheight}{\includegraphics[width = \mywidthR]{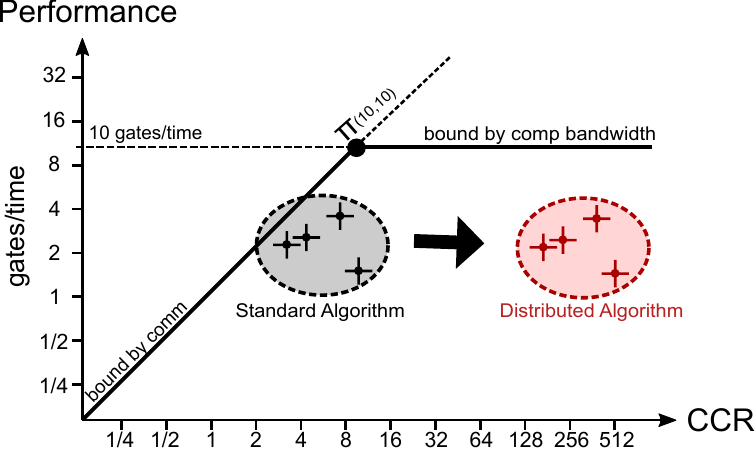}} & \raisebox{-\totalheight}{\includegraphics[width = \mywidthR]{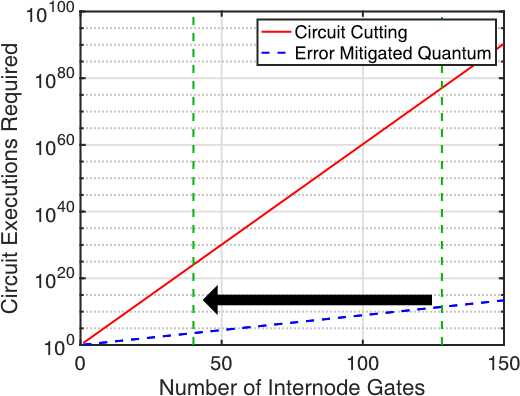}} \\ \hline 
    \end{tabular}
    \caption{Schematic depiction of the effects of improvements to the compiler and application layers, as demonstrated in the GAP, Q-Roofline, and QCPA analyses.}
    \label{tab:upper_improvements}
\end{table*}

At the compiler level, the perennial issues of qubit placement and routing must be overcome in addition to the complexities introduced by modular architectures containing heterogeneous qubit implementations and gate operations. 
Between the compiler and application layers, questions surrounding the software infrastructure responsible for workload management and resource sharing must be addressed.
To overcome these issues we point to the similarities between distributed QC and classical HPC and discuss ways in which the strategies developed in the classical domain might be adapted to the quantum case.
Finally, at the top of the stack we emphasize the need to profile and better understand distributed applications such that the information learned at the application level might help inform the co-design of the lower layers of the MNQC stack.

\subsubsection{Compiler Improvements}

Multinode systems pose a significant challenge for compilers due to both their scale and the complexities of balancing internode gates, local gates, and circuit cutting gates. As we have seen, internode operations are likely to remain more expensive and error-prone than local quantum gates, and therefore minimizing the communication overhead incurred during compilation will remain a primary concern.

Scale poses a challenge because assigning logical qubits to physical qubits, scheduling complex multi-qubit interactions, and routing physical qubits while respecting connectivity constraints become intractable as the number of qubits and gates in the program increase \cite{siraichi2018qubit, cowtan2019qubit, bonnet2018complexity}. Current compilers are capable of translating large programs (containing more than $10^6$ logical qubits and gates) into hardware-\textit{agnostic} assembly programs \cite{javadi2017towards}, but are currently limited in their ability to map this to a hardware-compatible executable \cite{tan2020optimality}. This problem is similar to the situation within classical high-performance computing where empirical studies have shown that the communication overhead quite often accounts for a larger portion of the program runtime than compute~\cite{kumar2008optimization, fujiwara2017visual, orenes2022tiny}. Quantum compilers may look to the field of classical HPC where load balancing has been extensively studied and efficient heuristic methods have been developed \cite{pellegrini2012scotch, karypis1998fast}. Compiler-oriented partitioning, where circuit partitioning algorithms \cite{andres2019automated, davarzani2020dynamic, dadkhah2022reordering} are applied during compilation, can also be applied to optimize for minimal communications, maximum fidelity, and balanced workloads \cite{ferrari2020compiler}. Once a program has been partitioned, distribution binds circuit partitions to module nodes and schedules inter-node communications. This is similar to qubit mapping, but at a coarser grain. The goal is to shorten the critical path (e.g., hide communication latency with local computation) and maximize program success rates while respecting communication dependencies 
constraints \cite{ferrari2020compiler}.
The architecture and metrics introduced in Figure \ref{fig:sQOSI_Schematic} help to quantify the entanglement distillation process such that this information may be incorporated into a compiler to optimize distillation scheduling.
Additional optimizations include buffer management \cite{wu2022collcomm}, aggregation \cite{wu2022autocomm}, and collective communication \cite{haner2021distributed, wu2022collcomm}. 

Furthermore, good system performance may be achieved through efficient load balancing by boosting local occupancy and minimizing communication overheads. As discussed in Section \ref{sec:FullModels}, for large quantum programs we use the CCR of the compiled program as the performance metric for comparing among compilers, algorithms, and runtimes. Theoretically, the CCR is bounded by the number of local computations when no communication is ever needed. However, as shown in the Q-Roofline models of Figure \ref{fig:roofline}, the connectivity constraints of the hardware may lead to an application becoming communication bound, and Figure \ref{fig:roofline_scale} demonstrates how compiler optimizations may be used to mitigate this overhead. Developing techniques to incorporate gate fidelities into the Q-Roofline model to estimate program success rates of large scale quantum programs is a promising area of future research.

As a feature of user access, designing clusters of distributed quantum computers presents new and interesting challenges with regards to their software infrastructure. First, the appropriate level of abstraction for distributed quantum systems is an open question. Recent work has shown that quantum program success rates can be greatly improved by breaking layers of abstraction~\cite{shi2020resource} and thus it will be necessary to balance quauntum program success rates with user efficiency when designing distributed quantum systems. Secondly, while Section~\ref{sec:FullModels} is concerned with optimizing the compute throughput for a single application, multiple users submitting multiple dependent or independent job requests to a QC cluster presents set of challenges for efficient workload scheduling.
In the classical paradigm, workload managers such as \texttt{SLURM}~\cite{yoo2003slurm} are responsible for scheduling the available hardware resources to best meet the needs of the users.
In a distributed quantum cluster, it appears that shared entanglement is likely to be the most precious resource but the exact optimization objective itself and the specific management method still remain open questions.

Finally, MNQC systems lead to an interesting problem of software-hardware co-design because they may naturally support diverse heterogeneous architectures.
Heterogeneity may manifest within the computation or communication within a distributed architecture. Individual nodes may consist of memory and compute regions implemented via different qubit modalities, and diverse technologies, implementing both quantum sensors and computers, may be used within a single quantum network. In this work we focused our analysis on combined quantum-classical communication channels which enable entanglement distribution \cite{Bennett1996} and teleportation \cite{meter2008arithmetic}. However, other protocols may be used requiring only classical channels as in quantum circuit cutting \cite{peng2020simulating, tang2021cutqc, tang2022scaleqc} and entanglement forging \cite{eddins2022doubling}, or solely quantum channels such as shuttling \cite{rowej2002shuttling}, direct state transfer \cite{Axline2018} and cross-chip two-qubit gates \cite{2021npjQI...7..142G}. 
Each protocol presents unique tradeoffs between fidelity, speed, and ease of implementation that any future compiler for a distributed system must consider.

\subsubsection{Application and Algorithm Improvements}

The design of a distributed quantum architecture will be heavily influenced by the workloads it is expected to encounter in practice. Profiling quantum programs to better understand the similarities and differences in their resource requirements is a critical area of future work.
Prior work evaluating the performance of potential quantum architectures demonstrated that the match between hardware and application is important because quantum programs display different levels of computation versus communication~\cite{thaker2006quantum, tomesh2022supermarq}. Our work in Section \ref{sec:FullModels} and Figure \ref{fig:TotalSim} supports this view by demonstrating quantum applications' sensitivity to the parameters which characterize the quantum communication channels.
Taking an example from classical computing, most applications can be assigned to one of a small number of application classes such as dense linear algebra, sparse linear algebra, $N$-body methods, and so on \cite{asanovic2006landscape}.
An important open question is understanding whether most quantum algorithms can similarly be grouped into a small number of general computational motifs.

In addition to profiling existing quantum applications, algorithm development -- especially algorithms developed specifically for distributed systems -- will play a critical role in the evolution of the field.
Early investigations into distributed quantum applications include quantum telecomputation \cite{grover1997quantum}, distributed Shor's algorithm and arithmetic \cite{yimsiriwattana2004distributed, meter2006architecture, meter2008arithmetic}, distributed VQE (via classical networks \cite{stein2022eqc} or quantum interconnects \cite{diadamo2021distributed}), and distributed phase estimation \cite{reiher2017elucidating}. 

In Section \ref{sec:FullModels}, we noted that while many complex algorithms were unachievable using present technology, GHZ creation could be performed with high fidelity. In turn, this actually implies that Quantum Phase Estimation (QPE) is a good candidate for execution on early MNQCs. In fact, despite the fact that QPE is viewed as a high circuit depth algorithm, the multinode architecture can be used to increase the phase kickback coming from the controlled unitary operation and thus forms a good candidate for evaluation on an MNQC. Two strategies exist for such parallelism: the fully coherent approach of~\cite{knill2007optimal} which gives a reduction in the depth of phase estimation that is linear in the number of nodes and the approach that uses classical communication (found in the supplementary material of~\cite{reiher2017elucidating}). Both of these are reviewed in detail in Appendix \ref{app:QPE}. In the case of an MNQC with quantum links, then we can use $O(\frac{1}{\epsilon})$ nodes to perform phase estimation to accuracy $\epsilon$ in $O(1)$ time; in the case of purely classical links, then $O(\frac{1}{\epsilon^2})$ nodes suffices to achieve the same bound.

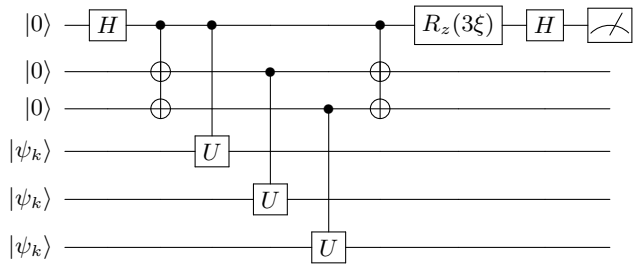
\begin{figure}
    \centering
    \[
\Qcircuit @C=1em @R=.7em {
\lstick{\ket{0}} & \gate{H}& \ctrl{2}&\ctrl{3} &\qw&\qw&\ctrl{2}&\gate{R_z(3\xi)} &\gate{H}&\meter\\
\lstick{\ket{0}} & \qw & \targ&\qw&\ctrl{3}&\qw&\targ&\qw&\qw&\qw\\
\lstick{\ket{0}} & \qw & \targ&\qw&\qw&\ctrl{3}&\targ&\qw&\qw&\qw\\
\lstick{\ket{\psi_k}} & \qw &\qw& \gate{U}&\qw&\qw&\qw&\qw&\qw&\qw\\
\lstick{\ket{\psi_k}} & \qw &\qw& \qw&\gate{U}&\qw&\qw&\qw&\qw&\qw\\
\lstick{\ket{\psi_k}} & \qw &\qw& \qw&\qw&\gate{U}&\qw&\qw&\qw&\qw\\
}
\]
    \caption{Distributed Quantum Phase Estimation circuit on $p=3$ nodes. Although QPE is a high-depth algorithm, the highly communication-efficient structure of the distributed QPE circuit renders it a natural candidate for early MNQCs.}
    \label{fig:QPE}
\end{figure}

In brief, the fully coherent version of distributed quantum phase estimation takes the form in Figure \ref{fig:QPE}~\cite{knill2007optimal}. It then follows from noting that the circuit returns the phase $e^{i3\theta_k}$ from the phase kickback effect that in general this idea can be repeated $p$ times to obtain $p$ times the phase that would be seen with a single step of an iterative phase estimation procedure. However, the error in the internode link must be $O(\frac{\epsilon^2}{\log(1/\epsilon)})$, placing a significant demand on the performance of the MNQC stack. These properties thus make the QPE an intriguing early primitive for future early multinode machines using both classical and quantum links. 

For quantum simulation, there are physical systems and model Hamiltonians that exhibit hybrid quantum-classical characters that can be naturally parallelized. One example of these model Hamiltonians is the quantum embedding descriptions of complex materials with multiple inequivalent impurity-bath subsystems~\cite{RevModPhys.78.865, PhysRevX.5.011008, PhysRevX.6.031045, PhysRevResearch.3.013184,welborn2016bootstrap} and quantum minimal entanglement typical thermal state sampling for finite-temperature simulations~\cite{2020NatPh..16..205M}.  Additionally, in complex chemical systems such as metal-organic framework \cite{zhou2012introduction} and protein-ligand binding \cite{fu2022accurate,warshel2014multiscale}, reaction centers typically contain transition metal species that exhibit strong quantum effects while the rest of molecular backbone are largely classical and thus enables natural parallelization in simulation by utilizing locality of chemical processes.

Besides these rather straightforward quantum parallelizations, quantum algorithms and physical systems can also be tailored for calculations on the multinode quantum architecture with weak linkages. For example, the impurity-bath model in dynamical mean-field theory calculations can be optimized to minimize the direct interactions between the impurity and bath subsystems~\cite{DMFTNotes}, and quantum transport systems of leads through nanocontacts naturally minimize the number of inter-node nonlocal gates~\cite{1998Natur.391..156G, 2009SurSR..64..191G}. In quantum embedding calculations, the size of the fragment or cluster can be reduced and the level of theory for treating the bath may be performed at a lower mean field level which can minimize the number of entangled degrees of freedom with the fragment. Furthermore, simulations of the full electronic structure of periodic materials may lend themselves well to MNQC architectures. Electronic structure calculations at different in reciprocal momentum space can be parallelized with limited inter-node communication required \cite{giannozzi2009quantum}.  Alternatively, real-space Wannier function representations to achieve compact encodings of electronic orbitals \cite{clinton2022towards}  may allow for parallelization of neighboring periodic cell images over separate nodes. Such approaches could facilitate electronic property calculations in the thermodynamic limit with smaller simulation cells, and thus a reduction in the qubit requirements per node.

These parallel schemes on electronic structure calculation can be directly applied to semi-classical \textit{ab initio} molecular dynamics simulation such as the Born-Oppenheimer molecular dynamics \cite{marx2009ab}. A key step of the molecular dynamics simulation is to evaluate the electronic structure at different nuclei coordinates repeatedly, which naturally benefits from the distributed QPE protocol in Appendix \ref{app:QPE}.
Beyond semi-classical dynamics, it is possible to rephrase quantum dynamics as finding the ground state of a composite Hamiltonian \cite{mcclean2013feynman}, where parallelization protocols for embedding schemes as discussed above are promising to accelerate quantum dynamics. VQE approaches can also be designed to have structure that can take advantage of systems in which disjoint degrees of freedom are connected by small terms in a Hamiltonian (weak linkages) with only weak correlations between the subregions. Recent advances in classical simulations have been able to exploit this type of correlation structure with cluster algorithms~\cite{vibin2020}.  A recent quantum algorithm has demonstrated that a VQE approach can be designed with the same advantages as a cluster algorithm~\cite{yuzhang2021}.  Further research based on clustering algorithms may allow for new VQE approaches that are well suited for MNQC  architectures.

\section{Outlook}\label{sec:Outlook}

We have quantified the potential performance of internode gates by building a layered architecture for internode link execution in MNQCs and developing a detailed quantitative model of each layer. By uniting these models, we were able to compare the available internode gate performance with the demands of algorithms in the GAP analysis, then reveal the relative costs of internode gates relative to local gates with the Q-Roofline model and relative to circuit cutting with the QCPA analysis. Our results paint a picture of the improvement in internode link performance needed to realize MNQCs links capable of competing with monolithic systems, and we laid out a research roadmap towards MNQCs, displaying potential improvements for each of the Physical, Distillation, and Application and Compiler layers in terms of the GAP, Q-Roofline, and QCPA models. 

Going forward, these models provide benchmarks to quantify the impact of actual research developments as they are achieved. For future improvements in M2O technology improves, we can now directly predict the algorithms unlocked by improved fidelity and rate of M2O conversion. Similarly, as quantum memories are developed, we can determine how entanglement distillation will be improved, or how buffering of entangled pairs will allow new, more demanding computations to be completed successfully. For distributed compilers and algorithms, the Q-Roofline model provides a tangible metric for compiler performance, while the GAP and QCPA analyses demonstrate the impacts of improved efficiency and reduced reliance on internode communication. Uniting these analyses, we can now measure progress towards MNQCs that outperform monolithic systems. 

More broadly, one of the most exciting future directions is extending this analysis to other platforms. While we have focused on superconducting systems with M2O interlinks as an MNQC, there is a wide array of platforms which have been envisioned as potential realizations of MNQCs. Borrowing from classical co-design~\cite{co-design1, foster2011high}, our models are designed in a way which allows for different interconnection platforms to be analyzed in future work by changing only the Physical layer, while new distillation protocols can be used in the Distillation layer simulations and new local architecture can be changed in the Application and Compiler layer simulations. By interchanging these models, our approach can quantify the available performance across a range of systems from large scale quantum networks~\cite{2020arXiv200211808C, 2018Sci...362.9288W} for distributed computing to smaller networks using cryogenic microwave links~\cite{2020PhRvL.125z0502M, bravyi2022,Zhou:2021cri, Yan2022,2021npjQI...7..142G}, which show considerable promise in the nearest term. Similarly, our approach can also characterize modular trapped ion systems~\cite{2019ApPRv...6b1314B, 2021Natur.592..209P, 2020AVSQS...2a4101K, 2021Natur.592..209P} and neutral atom platforms~\cite{Young:2022cyz, Bluvstein:2021jsq}. Hybrid systems consisting of several of these technologies~\cite{doi:10.1146/annurev-conmatphys-030212-184253, 2016QuIP...15.5385D, 2012PhRvL.108m0504K, 2016NatCo...710352S, 2021npjQI...7..121N} are a promising route for future networked systems and can also be treated within this framework, allowing this approach to treat the full range of inter-operable distributed quantum systems. Just as our analysis revealed key tradeoffs and pointed the way for future technology development in superconducting M2O systems, a similar analysis of each of these other candidate platforms can quantify the performance currently available technology can offer as an MNQC, reveal key tradeoffs and interactions between components that may be make-or-break for multinode systems, and help develop research roadmaps that point the way towards the successful realization of MNQCs.

\section{Key Author Contributions}
M. A. DeMarco led the project, developed the MNQC architecture, and guided simulation pipeline development. 
S. Stein built the simulation pipeline and performed simulations of entanglement distillation, remote gates, and benchmark execution. 
Y. Zhou performed simulations of the remote entanglement generation based on M2O converters and led writing on M2O simulation. 
C. Liu built models for the distillation layer and led the proposals for distillation improvements. 
T. Tomesh provided input on the MNQC architecture and led the proposals for compiler and application improvements.
S. Sussman modeled coupling flying qubits into transmons and developed the proposals for the use of quantum memory for distillation.
Y. Chen drafted the sections on protocols for entanglement generation.
W. Tang drafted the sections on quantum compilers and multicomputing. 
P. Hilaire developed language on entanglement generation protocols and distributed architecture. 
D. McKay performed the error mitigation and circuit cutting analysis and edited the final draft. 
A. Li proposed the quantum roofline model, performed calculations for it, and guided the project.
N. Wiebe wrote the analysis for distributed QPE.
I. Chuang supervised the project and provided guidance in the architecture and writing.

\section{Acknowledgements}
This material is based upon work supported by the U.S. Department of Energy, Office of Science, National Quantum Information Science Research Centers, Co-design Center for Quantum Advantage (C2QA) under contract number DE-SC0012704.

\bibliography{ref}

\clearpage

\appendix

\section{Distributed QPE}\label{app:QPE}
Quantum phase estimation (QPE) is perhaps surprisingly one of the easiest subroutines in quantum computing to distribute in MNQC systems.  This is in spite of the fact that QPE is often viewed as a high circuit depth algorithm.  In this section we will discuss two approaches for distributing quantum phase estimation.  Two strategies exist for such parallelism: the fully coherent approach of~\cite{knill2007optimal} which gives a reduction in the depth of phase estimation that is linear in the number of nodes and the approach that uses classical communication (found in the supplementary material of~\cite{reiher2017elucidating}).  Our aim in this section is to review these approaches and place bounds on the channel fidelity needed to see an advantage from the former approach.

The task of phase estimation is to provide an estimate of an eigenphase of a unitary operation $U$.  Specifically, assume that for unitary $U \in \mathbb{C}^{2^n \times 2^n}$ that $\ket{\psi_n}$ are eigenvectors such that $U\ket{\psi_n} = e^{i\theta_n} \ket{\psi_n}$ for real valued $\theta_n$.  The aim of the phase estimation problem is to find, for any $\epsilon>0$ and probability of success at least $2/3$, an estimate $\hat{\theta}_k$ such that there exists $\theta_k$ that obeys $|\hat{\theta}_k - \theta_k| \le \epsilon$.  In practice, the phase estimation problem is usually more specific and a particular eigenphase is desired.  In which, case the user must provide a quantum state that has high-overlap with the target eigenstate for this protocol to succeed with high probability.

The central challenge of phase estimation is that the optimal scaling  is given by the Heisenberg limit $|\hat{\theta}_k - \theta_k| \le \frac{\pi}{T}$ for any quantum algorithm that estimates $\theta_k$ using $T$ applications of the unitary $U$.  For applications in chemistry, these errors need to be on the order of $10^{-4}$ or smaller~\cite{reiher2017elucidating}, necessitating a large number of applications of the underlying unitary.  Our aim is to distribute these executions of the unitary over the network in such a way so that the phase estimation can be performed in low depth.  Specifically, if we will see that if we have an MNQC then in the most extreme case we can use $T$ nodes to perform phase estimation in $O(1)$ time and in the case where classical interconnects are used then $T^2$ nodes suffices to achieve the same bound.

We will begin with the simplest case wherein each node can only communicate classically with each of the other nodes.  Let us assume that in each case a quantum state $\ket{\phi_k}$ can be prepared such that $|\langle\phi_k|\psi_k\rangle|^2=1-\delta$ for target state $\ket{\psi_k}$.  Further, let us assume that for all $j\ne k$, $|\theta_j - \theta_k| \ge \epsilon_\theta$.  We begin our protocol by applying phase estimation to prepare each of the states on the $T^2$ nodes.  This requires $O(\log(1/\gamma)/\epsilon_{\theta})$ number of applications of the underlying $U$ on each node to ensure the correct eigenvalue with probability of failure at most $1-\gamma$ using conventional phase estimation.  The number of trials needed per node before a successful state preparation is observed is geometrically distributed.  If $\gamma \ge 2/3$ then the probability distribution function shows that the number of trials needed before the probability of observing no successful preparations is $O(1/T^2)$ is $O(\log(T^2)/\delta)$.  From the union bound, the probability of any of the runs requiring more than this is $O(1)$ and thus the total depth (as quantified by the number of unitary circuits applied to prepare the $O(T^2)$ independent eigenstates is in
\begin{equation}
    O\left(\frac{\log(T)}{\delta \epsilon_{\theta}} \right)
\end{equation}

Next we need to invoke the parallelized phase estimation procedure of~\cite{reiher2017elucidating}.  Each experiment in this algorithm involves communicating to each of the nodes and requesting it to perform $U^p$ for some value of $p$ and measuring the result in the iterative phase estimation circuit, which takes the following form
\begin{equation}
\Qcircuit @C=1em @R=.7em {
\lstick{\ket{0}} & \gate{H} &\ctrl{1} &\gate{R_z(\xi p)} &\gate{H} &\meter \\
\lstick{\ket{\psi_k}}&\qw &\gate{U^p}& \qw&\qw&\qw
}\label{eq:itpe}
\end{equation}
The likelihood of measuring zero for this circuit is $\cos^2(p(\theta-\xi)/2)$.  The circuit is repeated $T^2$ times, one for each node, and the results are communicated back to the classical head node.  From the central limit theorem, the distribution on the number of zeros observed out of the $T^2$ measurements is approximately a Gaussian with mean $T^2 \cos^2(p(\theta - \xi)/2)$.  As the likelihood is approximately Gaussian, the method of conjugate priors can be used to efficiently update an initially Gaussian prior distribution on $\theta$ to find a posterior distribution in polynomial time (alternatively if one does not want to use the central limit theorem the Monte-Carlo methods of~\cite{reiher2017elucidating} can be employed).  By choosing $p$ adaptively using the heuristic in~\cite{reiher2017elucidating} they show that $O(\log(T))$ suffices to achieve error in the estimate $O(1/T)$.  Further each such experiment requires evolution time at most $O(T/\sqrt{C})$ where $C$ is the number of nodes devoted to the phase estimation.  In cases where $C=O(T^2),$ such as our setting, this implies that the depth of each phase estimation job (as quantified by the number of sequential operations of $U$) is $O(1)$.  

The above tasks are repeated $O(\log(T))$ times by each node and therefore the depth of the phase estimation algorithm once the state $\ket{\psi_k}$ is prepared on each node.  The depth of the classical communication version of the QPE algorithm appropriate for a MNQC setting with $T^2=O(1/\epsilon^2)$ is dominated by the state preparation step, which can be performed using $T^2$ workers in depth
\begin{equation}
    {\rm Depth}_{U,{\rm Cl}} = O\left(\frac{\log(1/\epsilon)}{\delta \epsilon_\theta} \right).
\end{equation}
Note that through the use of fixed point amplitude amplification rather than statistical sampling, the depth can further be reduced by to $O\left(\frac{\log(1/\epsilon){\rm polylog}(1/\epsilon_{\theta})}{\sqrt{\delta} \epsilon_\theta} \right)$; however, the use of this technique will require additional ancillae and comparison logic to implement the required reflections about the estimated energy returned by a coherent (as opposed to iterative) phase estimation procedure and thus we focus our attention on the non-amplified case.

As no quantum communication is needed for this algorithm there are no further errors if we assume that we are working in a model wherein all intra-node operations are error free but inter-node operations have intrinsic error associated with them.  This also makes this application a good baseline comparison to judge the quantum version of phase estimation.

The fully coherent version of distributed quantum phase estimation takes the form in Figure \ref{fig:QPE}~\cite{knill2007optimal}. It then follows from noting that the circuit returns the phase $e^{i3\theta_k}$ from the phase kickback effect that in general this idea can be repeated $p$ times to obtain $p$ times the phase that would be seen with a single step of an iterative phase estimation procedure.  Specifically, in both cases the probability of measuring zero is $\cos^2(p(\theta_k -\xi)/2)$ per experiment.

The protocol for implementing this circuit on a quantum MNQC works as follows.  
\begin{enumerate}
    \item In parallel prepare a state $\ket{\phi_k}$ on each of $T$ nodes on the quantum computer.
    \item Use the above phase estimation procedure and prior knowledge of $\theta_k$ to ensure that each state is $\ket{\psi_k}$ with probability $1-O(1/T)$.
    \item For each invocation of the circuit of~\eqref{eq:itpe} in the implementation of an iterative phase estimation algorithm (such as Robust Phase Estimation) replace the circuit with the following procedure:
    \begin{enumerate}
        \item Prepare a $T$ qubit GHZ state on the head node.
        \item Send one qubit of the GHZ state to each of the $T$ worker nodes.
        \item For each worker, apply controlled $U$ using the share of the GHZ state to their state $\ket{\psi_k}$.
        \item Return all qubits to head node.
        \item Apply single qubit rotation (if required by iterative phase estimation protocol), invert GHZ preparation and measure qubit $0$ and return result as outcome of measurement for the step of ITPE.
    \end{enumerate}
\end{enumerate}

The above protocol works because the likelihood as argued above is precisely the same in the distributed algorithm as it would be in the ordinary algorithm for phase estimation.  As the core element of an iterative phase estimation procedure is the inference of the most likely eigenphase given a set of experimental data, the inference procedure will take precisely the same form since the likelihood function is the same.  Thus the protocol allows us to trivially parallelize any iterative phase estimation procedure over the $T$ workers.

Iterative phase estimation procedures such as Robust Phase Estimation require $O(\log(T))$ rounds if we desire an error of $O(1/T)$.  Each such round can be executed in constant depth (as measured by the number of layers of controlled $U$ gates executed).  It further requires $2T$ applications of a communication channel from the head node to the workers.  For simplicity, let us assume that the interaction graph is star graph wherein the root is the head node so that all workers can directly communicate with the head node.  In settings where a more restricted topology is present, the communication will need to be chained between the workers involved to distribute the GHZ state.  Regardless, the total number of bits that need to be sent by the protocol is in $O(T\log(T))$ and the overall depth as mentioned is logarithmic.  Thus, assuming that the cost of any entanglement distillation is negligible, the overall depth of the algorithm is also
\begin{equation}
    {\rm Depth}_{U,{\rm Qm}} = O\left(\frac{\log(1/\epsilon)}{\delta \epsilon_\theta} \right);
\end{equation}
however, the number of workers needed to achieve this limit is quadratically smaller than the case where only classical communication is permitted.

Next let us assume that the channel that describes communication between the head node and the workers is within diamond distance $\Delta$ from the ideal channel.  That is to say if $\Lambda$ is the ideal channel that swaps a qubit between the two nodes and $\tilde{\Lambda}$ is the actual quantum channel then $\|\Lambda - \tilde{\Lambda}\|_\diamond \le \Delta$.  Here the diamond norm is the supremum of the induced trace norm between the inputs and the outputs of the channel when all possible input states (including states that are entangled with qubits that are not put through the channel) are considered.  An important property of the diamond norm is that it is sub-additive meaning that for any positive integer $m$ the composition of $m$ channels obeys
\begin{equation}
    \|\Lambda^{\circ m} - \tilde{\Lambda}^{\circ m}\|_\diamond \le m \Delta.
\end{equation}
Thus by the von Neumann trace inequality, for any observable $Q$ and input state $\rho$
\begin{equation}
    \|{\rm Tr}(\Lambda^{\circ m} (\rho) Q) - {\rm Tr}(\tilde{\Lambda}^{\circ m} (\rho) Q)\| \le m\|Q\|\Delta.
\end{equation}
Thus as the observable for phase estimation has norm at most $\pi$ it follows that the maximum error that is observable from the invocation of the channel in this fashion is $m\pi \Delta$.  This implies that if we wish the error in the estimated phase to be at most $\epsilon$ from communication between the head node and the workers then it suffices to take 
\begin{equation}
    \Delta = \frac{\epsilon}{m\pi} = O\left(\frac{\epsilon}{T\log(T)} \right)
\end{equation}
Setting $T=O(1/\epsilon)$ as well suffices to remove the $O(1/\epsilon)$ overhead from phase estimation from the circuit depth while guaranteeing that we hit a fixed accuracy target
\begin{equation}
    \Delta = \frac{\epsilon}{m\pi} = O\left(\frac{\epsilon^2}{\log(1/\epsilon)} \right)
\end{equation}
This suggests that the error in the quantum communication channel must be exceptionally small in order guarantee (without further assumptions) that the overall error in the phase estimation protocol is small. Further,  such applications are likely to be impractical without entanglement distillation or possibly virtual distillation.

Given that $\epsilon$ is sufficiently low, entanglement distillation can be used to implement this channel.  In order to distill states with this level of error we need $O(T{\rm polylog}(T\log(T)/\epsilon))= {\widetilde O}(T{\rm polylog}(1/\epsilon))$ noisy uses of a channel connecting the head node with the workers in order to distill high enough fidelity states to teleport within the desired accuracy~\cite{fang2019non}. Thus if we assume that the depth of required to communicate between the nodes is $\gamma \ge 0$ times the depth required to implement $U$ (where $\gamma$ will often but not always be less than $1$) the total depth of the algorithm using $T$ workers is
\begin{equation}
    {\rm Depth}_{U, Dist} = { O}\left(\frac{\log(1/\epsilon)}{\delta \epsilon} + \frac{1}{\epsilon T} +\gamma T{\rm polylog}(T/\epsilon)\right)
\end{equation}

This shows that as the number of workers increases, a favorable tradeoff in the depth of the circuit can be achieved.  Specifically, such an optimal tradeoff is obtained when $T\approx \Theta(\sqrt{1/(\epsilon \gamma)})$.  Given this choice, the optimized depth reads

\begin{equation}
    {\rm Depth}_{U, Dist}^{\rm opt} = { \tilde O}\left(\frac{\log(1/\epsilon)}{\delta \epsilon} + \sqrt{\frac{\gamma}{\epsilon}}\right).
\end{equation}
Thus if $\gamma$ is viewed as a constant, then this approach can attain quadratically better depth scaling than with the error tolerance than the na\"ive phase estimation algorithm permits.  However, this is not necessarily better than the case where no quantum interconnects are used if  $\gamma$ is not sufficiently small.

In order to understand the gulf between this let us assume that the phase estimation step used to validate the eigenstate has circuit depth $\alpha \log(1/\epsilon)$, where in the case of Hamiltonian simulation $\alpha$ would be the sums of the absolute values of the coefficients and corresponds to a simulation method such as qubitization being used.  Next, let us assume that each of the workers uses a low-order method such as Qdrift~\cite{campbell} to perform the simulation.  In this case, we would take the Qdrift approximation to $e^{-i H t}$ for a sufficiently short value of $t$ and perform phase estimation on the result to precision $\epsilon t$.  

The work of~\cite{lee2021even} shows that $O(\alpha^4/\epsilon^4)$ exponentials need to be simulated to perform phase estimation to within error $\epsilon$ using Qdrift.  If we assume that we can parallelize $T$ of them over our workers, the combined cost of phase estimation becomes

\begin{equation}
    {\rm Depth}_{U, Dist} = { O}\left(\frac{\alpha\log(1/\epsilon)}{\delta E_{\rm gap}} + \frac{\alpha^4}{\epsilon^4 T} +\gamma T{\rm polylog}(T/\epsilon)\right)
\end{equation}
In the limit of negligible $\gamma$, this protocol can achieve depth $\alpha\log(1/\epsilon)/\delta E_{\rm gap}$ by choosing $T= \alpha^3(\delta E_{\rm gap})/\epsilon^4$.  This shows that in a regime where parallelism is cheap that a simulation experiment can be carried out whose depth scales only with that required to verify that each worker possesses a copy of the groundstate.  Note that by replacing QDrift with another simulation algorithm, such as Trotter formulas or Qubitization, we cannot get the same depth optimal result because we will be limited by the circuit depth needed to implement those protocols.  A single segment of QDrift can be executed in constant depth and thus is the only known algorithm that can meet the above scaling.

\section{M2O Conversion Simulation}\label{app:M2O}

\begin{figure}[b]
    \centering
    \includegraphics[width=\linewidth]{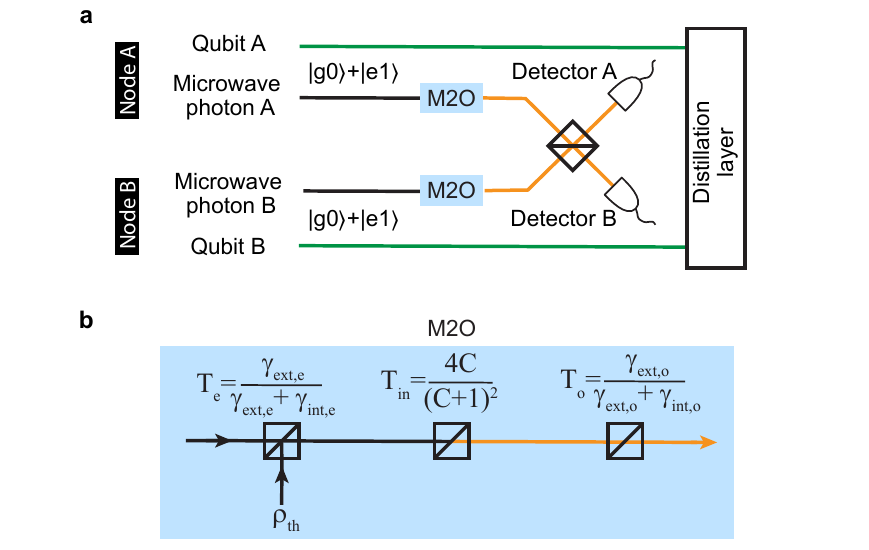}
    \caption{(a), the diagram of the scheme used for remote entangled qubit generation. (b), the M2O converter is modeled by a series of beamsplitters. The first beamsplitter represents the microwave resonator extraction efficiency, the second beamsplitter represents the intracavity M2O conversion efficiency, and the third beamsplitter represents the optical resonator extraction efficiency. Thermal added noise is modeled by a thermal state $\rho_{\text{th}}$ at the first beamsplitter.}
    \label{fig:M2O_circuit}
\end{figure}

The simulation model used for entangled generation is shown in Fig.~\ref{fig:M2O_circuit}(a) \cite{2016PhRvX...6c1036N}. At both nodes, the qubit and a microwave photon are prepared in an entangled state $\ket{\phi_0} = \sqrt{1-P_e} \ket{g0} + \sqrt{P_e}\ket{e1}$, where $0 \le P_e \le 0.5$ is the probability of excited qubit-microwave photon state and is experimentally tunable \cite{2016PhRvX...6c1036N}. The microwave photons are up-converted to optical photons and then interfere in a beamsplitter. The M2O converters are phenomenologically modeled as a series of beamsplitters as shown in Fig.~\ref{fig:M2O_circuit}(b). The first beamsplitter has a power transmission of $T_e= \gamma_{\text{ext,e}} / \gamma_{\text{tot,e}} $, where $T_e$ is the extraction efficiency of the microwave resonator, $\gamma_{\text{ext,e}}$ is the external coupling rate of the microwave resonator, $\gamma_{\text{int,e}}$ is the intrinsic decay rate of the microwave resonator, and $\gamma_{\text{tot,e}} = \gamma_{\text{ext,e}} + \gamma_{\text{int,e}}$ is the total decay rate of the microwave resonator. Due to the pump-induced heating, the microwave resonator suffers from thermal added noise, and we model it as a thermal state $\rho_{\text{th}}(n_{\text{add}})$ at another input port of the beamsplitter, and its mean photon number is $n_{\text{add}}$. In our simulation, we assume $n_{\text{add}}$ depends linearly on the optical pump power $P$ as $n_{\text{add}} = k_{\text{add}} P$, and we assume $n_{\text{add}} = 1$~photon when $P=1$~mW (i.e., $k_{\text{add}}=$1~photon\slash1~mW) \cite{fu2021cavity}. The second beamsplitter has a power transmission of $T_{\text{in}}= 4C / (C+1)^2$, where $T_{\text{in}}$ represents the intracavity M2O conversion efficiency for electro-optics converters, $C=4 g^2 / (\gamma_{\text{tot,e}} \gamma_{\text{tot,o}})$ is the cooperativity \cite{fan2018superconducting}, $\gamma_{\text{tot,o}} = \gamma_{\text{ext,o}} + \gamma_{\text{int,o}}$ is the total decay rate of the optical resonator, $\gamma_{\text{ext,o}}$ is the external coupling rate of the optical resonator, $\gamma_{\text{int,o}}$ is the internal decay rate of the optical resonator, $g=g_0 \sqrt{n_p}$ is the nonlinear coupling rate, $g_0$ is the single-photon nonlinear coupling rate, $n_p= 4\gamma_{\text{ext,o}} P /[\hbar \omega (\gamma_{\text{ext,o}} + \gamma_{\text{int,o}})^2]$ is the intracavity pump photon number, and $\omega$ is the pump photon frequency. The last beamsplitter has a power transmission of $T_o= \gamma_{\text{ext,o}} / (\gamma_{\text{ext,o}} + \gamma_{\text{int,o}})$ which represents the optical resonator extraction efficiency. We assume an optical detector dark count rate of 50~Hz \cite{natarajan2012superconducting}.

\begin{table*}[]
\resizebox{\textwidth}{!}{
\begin{tabular}{|c|c|c|c|c|c|c|c|}
\hline
No.                                                                           & 1                                                   & 2                   & 3              & 4             & 5               & 6          \\ \hline
Platform                                                                      & Electro-optomechanics                                & Electro-optics & Electro-optics & Optomagnonics & Rare-earth ions & Cold atoms \\ \hline
\begin{tabular}[c]{@{}c@{}}Single-photon\\ coupling rate (Hz)\end{tabular} & \begin{tabular}[c]{@{}c@{}} $\frac{g_{\text{om}}}{2\pi} = 60$ \\$\frac{g_{\text{em}}}{2\pi} = 1.6$\end{tabular}            &      $\frac{g_{\text{eo}}}{2\pi} = 37$   &  $\frac{g_{\text{eo}}}{2\pi} = 750$        &    $\frac{g_{\text{mago}}}{2\pi} = 17.2$    &   -     &      -    \\ \hline
\begin{tabular}[c]{@{}c@{}}Cavity\\ decay rate (Hz)\end{tabular} & \begin{tabular}[c]{@{}c@{}} $\frac{\gamma_{\text{ext,o}}}{2\pi}= 2.1\times 10^6 $ \\ $\frac{\gamma_{\text{int,o}}}{2\pi}= 5.6\times 10^5 $ \\ $\frac{\gamma_{\text{ext,e}}}{2\pi}= 1.4\times 10^6 $ \\ $\frac{\gamma_{\text{int,e}}}{2\pi}= 1.3\times 10^6 $\end{tabular}  & \begin{tabular}[c]{@{}c@{}} $\frac{\gamma_{\text{ext,o}}}{2\pi}= 1.5\times 10^7 $ \\ $\frac{\gamma_{\text{int,o}}}{2\pi}= 1.1\times 10^7 $ \\ $\frac{\gamma_{\text{ext,e}}}{2\pi}= 5.6\times 10^6 $ \\ $\frac{\gamma_{\text{int,e}}}{2\pi}= 8.1\times 10^6 $ \end{tabular} & \begin{tabular}[c]{@{}c@{}} $\frac{\gamma_{\text{ext,o}}}{2\pi}= 3.3\times 10^7 $ \\ $\frac{\gamma_{\text{int,o}}}{2\pi}= 1.4\times 10^8 $ \\ $\frac{\gamma_{\text{ext,e}}}{2\pi}= 3.2\times 10^6 $ \\ $\frac{\gamma_{\text{int,e}}}{2\pi}= 5.8\times 10^6 $ \end{tabular}  & \begin{tabular}[c]{@{}c@{}} $\frac{\gamma_{\text{ext,o}}}{2\pi}=4.8\times 10^7 $ \\ $\frac{\gamma_{\text{int,o}}}{2\pi}= 1.5\times 10^9 $ \\ $\frac{\gamma_{\text{ext,e}}}{2\pi}=1.7\times 10^8 $ \\ $\frac{\gamma_{\text{int,e}}}{2\pi}= 8.5\times 10^7 $\end{tabular} & - & - \\
\hline
Cooperativity                                              & \begin{tabular}[c]{@{}c@{}} $C_{\text{om}}=4.5\times 10^4$ \\$ C_{\text{em}}=1.0 \times 10^4 $\end{tabular} &   $C_{\text{eo}}=0.92$ &  $C_{\text{eo}}=0.04$     &    \begin{tabular}[c]{@{}c@{}} $C_{\text{mago}}=4.1 \times 10^{-7}$ \\$ C_{\text{mage}}=0.8$ \end{tabular}  &       -         &    -     \\ \hline
Efficiency                &  \begin{tabular}[c]{@{}c@{}} $\eta_{\text{tot}}=0.19$ \\$\eta_{\text{in}}=0.59$\end{tabular} &     \begin{tabular}[c]{@{}c@{}} $\eta_{\text{tot}}=0.14$ \\$\eta_{\text{in}}=0.99$ \end{tabular}           &  \begin{tabular}[c]{@{}c@{}} $\eta_{\text{tot}}=0.01$ \\$\eta_{\text{in}}=0.15$ \end{tabular}  & \begin{tabular}[c]{@{}c@{}} $\eta_{\text{tot}}=1.1 \times 10^{-8}$ \\$\eta_{\text{in}}=5.2 \times 10^{-7}$ \end{tabular} &     $\eta_{\text{tot}} = 1.26 \times 10^{-5}$    &    $\eta_{\text{tot}}=0.82$   \\ \hline
Bandwidth (Hz)                                                                &   $6.1\times 10^{3}$                                       &     -     & -  &    $1.6\times 10^7$      &       -      &     $1\times 10^6$  \\ \hline
\begin{tabular}[c]{@{}c@{}}Added noise\\ $n_{\text{add}}$\end{tabular}     &     $1.4$                           &    $0.41$    & -   &     -     &     -     &     $0.8$     \\ \hline
 \begin{tabular}[c]{@{}c@{}} Environment \\ temperature (K) \end{tabular}                                                          &     $0.04$                                     &       $0.01$     &  $1.9$ & $300$      &    $4.6$     &     $300$    \\  
\hline
Reference                                                                     &   \cite{delaney2022superconducting}       &   \cite{sahu2022quantum} &   \cite{xu2021bidirectional}      &    \cite{zhu2020waveguide}       &   \cite{bartholomew2020chip}    &    \cite{tu2022high}     \\ \hline
\end{tabular}}
\caption{Summary of M2O converter performances on different experimental platforms. The definitions of parameters are discussed in Appendix~\ref{app:M2O}.}
\label{tab:M2O_efficiency}
\end{table*}

\begin{table}[]
\begin{tabular}{|c|c|c|c|c|c|c|c|}
\hline
No.                                       & 1                 & 2                 & 3                 & Future          \\ \hline
$\frac{g_0}{2\pi}$ (Hz)                   & 60                & 37                & 750               & $10^3$          \\ \hline
$\frac{\gamma_{\text{ext,o}}}{2\pi}$ (Hz) & $2.1\times 10^6$  & $1.5\times 10^7$  & $3.3\times 10^7$  & $10^7$          \\ \hline
$\frac{\gamma_{\text{int,o}}}{2\pi}$ (Hz) & $1.1\times 10^5$  & $2.2\times 10^6$  & $2.8\times 10^7$  & $2\times 10^5$  \\ \hline
$\frac{\gamma_{\text{ext,e}}}{2\pi}$ (Hz) & $1.4 \times 10^6$ & $5.6 \times 10^6$ & $3.2 \times 10^6$ & $ 10^7$         \\ \hline
$\frac{\gamma_{\text{int,e}}}{2\pi}$ (Hz) & $2.6 \times 10^5$ & $ 1.6\times 10^6$ & $1.2 \times 10^6$ & $2 \times 10^5$ \\ \hline
\end{tabular}
\caption{Parameter sets used for M2O entanglement generation simulation. The No.~1, No.~2 and No.~3 parameter sets come from Table~\ref{tab:M2O_efficiency}. The optical and microwave resonator intrinsic decay rate are made 5 times lower than the original values. The last column presents a hypothetical parameter set which we wish to be available in the future.}
\label{tab:M2O_sets}
\end{table}

In our simulation, we begin with an initial state $\ket{\phi_0}_A \ket{\phi_0}_B$ and numerically evolve the state with the Python QuTiP package \cite{johansson2012qutip} to obtain the density matrix $\rho_{\text{f}}$ after the interfering beamsplitter. We assume that it takes $t_1 = 50$~ns to prepare the initial states by local gate operations. We also assume the microwave photon and optical photon transmission loss is zero. We note that high photon transmission loss can decrease the entanglement generation rate and the fidelity and thus fails the next Distillation layer. Although zero transmission loss is experimentally unavailable yet, we still make this assumption for the purpose of illustrating the workflow of our stack model, and the consequent rate and fidelity can be understood as on-chip metrics. The nonzero transmission loss can be easily included into the model by incorporating the transmission loss to the optical/microwave cavity extraction efficiency. The converter bandwidth can be approximated as $B \approx\gamma_{\text{tot,e}}$ \cite{fan2018superconducting}, and we thus assume a pump pulse duration of $t_2 = 1/B$ and a resonator reset time $t_3 = 1/B$. Hence, the total time duration for one period is $t_{\text{tot}} = t_1 + t_2 + t_3$. The event that detector A measures 1 photon while detector B measures 0 photon is considered a successful heralding, and the probability of a successful heralding can be calculated as $P_{\text{herald}} = \text{Tr} \bra{1,0} \rho_{\text{f}} \ket{1,0}$. Thus, the entanglement generation rate can be computed as $R = P_{\text{herald}} / t_{\text{tot}}$. In the case of a successful heralding, the corresponding qubit state is $ \rho_{q}= \bra{1,0} \rho_{\text{f}} \ket{1,0} / \text{Tr} \bra{1,0} \rho_{\text{f}} \ket{1,0}$, and the entanglement fidelity is $F = \bra{\Psi^+} \rho_{q} \ket{\Psi^+}$, where $\ket{\Psi^+} = (\ket{ge}+\ket{eg})/\sqrt{2}$ is the target qubit Bell state. 

The parameter sets used for simulation are shown in Table~\ref{tab:M2O_sets}. The first three parameter sets come from Table~\ref{tab:M2O_efficiency}, but both the microwave and optical intrinsic decay rate are 5 times smaller than the original values to allow a higher generation rate and lower infidelity and thus enable the next Distillation layer. We assume these parameter sets are experimentally available relatively soon given the recent progress in low-loss nonlinear optical material fabrication \cite{gao2022lithium, zhuang2022high, jin2021hertz} and hence we still refer to them as `current M2O' in the manuscript, despite several optimistic assumptions made above. In addition, although the No.~1 converter in Table~\ref{tab:M2O_efficiency} is based on electro-optomechanical effects which require different formulas to calculate its conversion efficiency and bandwidth \cite{arnold2020converting}, we treat it as an electro-optic converter for simplicity, because this work aims at presenting a simulation model rather than a comprehensive analysis on various types of converters. We also present a hypothetical parameter set that we wish to be available in the future. The entangled pair generation rate, entanglement fidelity, and the density matrices are used as the input of the next distillation layer.

The result of simulation is shown in Fig~\ref{fig:M2O_state}. We first set $P_e=0.5$ and sweep the pump power as shown in Fig~\ref{fig:M2O_state}(a). For the No.~1 parameter set, one can see that the highest generation rate and the lowest infidelity are obtained at the pump power corresponding to $C=1$, where the conversion efficiency is maximized. The entangled qubit state generation rate can reach 1 MHz with an infidelity near 0.2. However, for No.~2 and No.~3 parameter sets, the infidelity remains above 0.5, because a high pump power is needed for $C=1$, and the generation rate is dominated by the false heralding triggered by the thermal added noise. The false heralding rate can be observed in Fig~\ref{fig:M2O_state}(b), where we fix the pump power such that the cooperativity $C=1$ while sweeping $0 \le P_e \le 0.5$. The entanglement generation rate at $P_e=0$ is thus the false heralding rate, which dominates for No.~2 and No.~3 parameter sets. The tuning of $P_e$ reveals a rate-infidelity tradeoff regime, which is highlighted as the green shaded area, where the rate increases but the infidelity also increases with an increasing $P_e$. In this regime, a larger $P_e$ allows more optical photons to be generated, but it also increases the error of having two nodes in the excited states simultaneously. The simulation results including the `future' parameter set are shown in Fig.~\ref{fig:estimate}. It can be observed that a large bandwidth, low loss, and low thermal noise are key to a high generation rate and low infidelity to enable the MNQC. 

\section{Entanglement Distillation Simulation}\label{app:Distillation}

In the entanglement distillation layer, we use raw EPs generated from the physical layer and perform entanglement distillation on them to generate higher fidelity EPs, at the cost of a slower generation time. Specifically, we take the heralding raw entangled state generation rate and the density matrix as inputs, perform the entanglement distillation, and report the distillation results to the Data layer. The output information to the Data layer includes the success distilled state density matrix, the distillation time and the success probability to the specified number of rounds of nested distillation.

To improve the quality of the remote entanglement is one of the key problems in the community of quantum communication. Historically, Bennett {\it et al.} proposed a protocol, to purify the imperfect Bell state and improve the fidelity of the Bell state to unity~\cite{Bennett1996}. In this protocol, each round of purification protocol will consume a pair of imperfect Bell states to generate an imperfect Bell state with higher fidelity and entanglement with less than unit fidelity. Suppose the remote superconducting qubits are in a Bell state (spin singlet) with imperfection and the state fidelity is $F$, after one round of entanglement purification, the fidelity is improved to
\begin{equation}
F_{\text{new}} = \frac{F^2 + (1-F)^2/9}{F^2 + 2F(1-F)/3 + 5(1-F)^2/9}. 
\label{eq:BBPSSW}
\end{equation}
Following this work, Deutsch {\it et al} proposed a similar method (DEJMPS), which improves the efficiency of the purification protocol~\cite{Deutsch1996}. This protocol avoids random bilateral single-qubit rotations to depolarize the imperfect state but uses the local operation to change into the Bell-diagonal basis. The outcome fidelity depends on the overlap to the other three Bell basis states~\cite{Deutsch1996}.

With the above two purification protocols, one way to generate Bell states of remote superconducting qubits close to unit fidelity is to use the recurrence purification scheme~\cite{Dur2007}. In this scheme, in order to perform $n$ rounds of entanglement purification, we need to prepare $2^n$ EPs of imperfect Bell states of superconducting qubits. In each round, the states from the last step undergo the pairwise entanglement purification to get states with higher entanglement. 

In our purification layer, the entanglement purification is performed based on DEJMPS protocol in Ref.~\cite{Deutsch1996}, while the effects of experimental imperfections are also considered. The experimental imperfection can come from two sources, (1) the error on the local two-qubit gates between superconducting qubits, and (2) the decay and decoherence error on the qubits while the qubits are idling. Specifically, in (2), we consider the idling from either waiting for the qubits are being measured during the purification process or waiting for the generation of required raw EPs. 

In the purification simulation, we take the superconducting qubits to have lifetime $T_1$ and coherence time $T_2$. From the simulation of the physical layer, we extract the density matrix ($\rho_0$) of the raw Bell pair with the generation rate ($r$). We assume that even with multiplexing, the raw Bell pair generation can still be considered sequential. Therefore, the average generation time of each pair is $\tau = 1/r$. For the error source (1), we assume unit fidelity local operations, as throughout our simulation stack we assume that local operations are perfect. We consider an instantaneous purification operation described by the quantum channel,
\begin{equation}
    \rho_{\text{new}} = \mathcal{P}[\rho_{\text{old},1} \otimes \rho_{\text{old},2}],
\end{equation}
where two EPs of entangled states with density matrices $\rho_{\text{old},1}$ and $\rho_{\text{old},2}$ are purified and get a single pair of qubits in the state $\rho_{\text{new}}$. Again, the actual implementation of the entangled purification is based on Ref.~\cite{Deutsch1996}.

For the error source (2), we need to estimate the idling time for superconducting qubits. Providing the two EPs of Bell states are ready for purification, we assume the local gate operations between the superconducting qubits and the measurements take $t_p$ time. This is modeled by a decay and decoherence error channel,  noted as $\mathcal{E}(t_p)[\rho]$, applied to the Bell state after the purification process. To estimate the total time for $n$ nested rounds of entanglement purification, we assume the time for $(n-1)$ rounds of purification takes $t_{n-1}$ time. In the $n$-th round, the first pair of Bell states is generated from the $(n-1)$-th round, which takes $t_{n-1}$ time. The second bell state used in the $n$-th round starts from $t_{\text{idle},n-1} = 2^{n-1} \tau$, while it also takes another $t_{n-1}$ to generate the second Bell state for the $n$-th round. Therefore, the first Bell pair needs to wait for another $t_{\text{idle},n-1}$ time. This is also modeled by a decay and decoherence error channel applied to the first Bell states used for the $n$-th round of purification. Further, we can construct the following recurrence relation for the $n$ rounds of purification
\begin{equation}
t_{n} = 2^{n-1} \tau + t_{n-1} + t_p.
\end{equation}
The state of the EPs after $n$ rounds of success purification is
\begin{equation}
    \rho_{(n)} = \mathcal{E}(t_p)\left[ \mathcal{P} \left[ \mathcal{E}(t_{\text{idle},n-1}) \left[\rho_{(n-1)}\right] \otimes \rho_{(n-1)} \right]\right]
\end{equation}

The probabilistic nature of entanglement purification is from the measurement on one pair of Bell states. As pointed out in Ref.~\cite{Deutsch1996}, if the measurement outcomes on two qubits in a single pair of input states coincide, the purification is considered as a success. The probability of having coincident measurement outcomes is the success probability for each round of purification, noted as $p_j$ for the $j$-th round. The overall success probability of $n$ rounds of purification can be calculated as 
\begin{equation}
P_{n} = \prod_{j=1}^{n} p_j^{2^{j-1}}.
\label{eq:single_shot_p}
\end{equation}
After the distillation calculation finishes, the required time $t_{n}$, the success state density matrix $\rho_{(n)}$, and the success probability $P_{n}$ of $n$ rounds of purification is passed to the Data layer for Internode gate simulation.

\section{Internode Gate Simulation}\label{app:INGate}

Having generated a distilled EP between modules, the next step in performing multinode quantum computing is inter-node operations. As a CX gate is computationally complete communication between nodes, here we focus on the case of only internode CX gates. Gate teleportation of the CX gate can be accomplished via the consumption of one EP, two measurements, and two local CX gates.  Simulation of the inter-node gate requires the use of a gate teleportation protocol, combined with the EP generated via M2O simulation under Section \ref{sec:LayerModels}, and optionally distilled by the protocol underlined in Section \ref{sec:LayerModels}. Simulation of the internode CX gate comprises beginning with an EP, represented by the density matrix output from entanglement distillation. We model local operations involved in the execution of the remone CX gate as having a local gate time of $100$ns, suffering depolarizing errors with a probability of $.0001$, and taking $T_1 = T_2 = 1\mu$~s.  Having performed the protocol, the density  matrix is captured at the protocol output, and reduced to represent the two data qubits. This density matrix is compared with the ideal simulation of a CX gate between two qubits, and used to report an overall teleported gate fidelity, and over all inter-module CX gate time.  

\section{Quantum Phase Estimation Benchmark}

\begin{figure}[]
    \centering
    \includegraphics[width =\columnwidth]{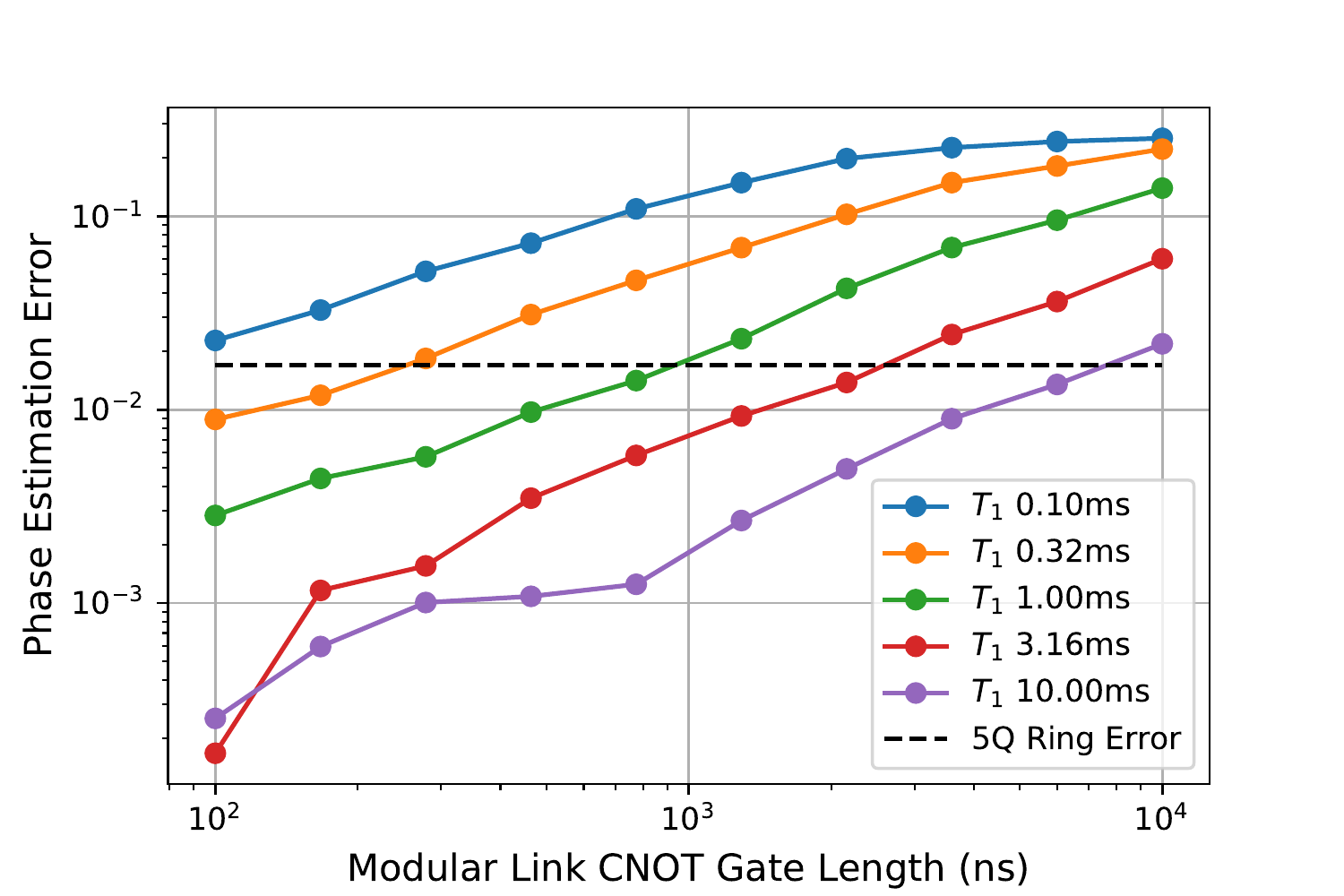}
    \caption{QPE Error for a 10-qubit algorithm as a function of the modular gate length for different values of $T_1$.}
    \label{fig:qpe_error}
\end{figure}

Another approach to the question of when to add a quantum link is by looking at algorithms where the answer improves in precision as the size of the system increases. For example, in quantum phase estimation (QPE) the number of ancilla $n_A$ sets the precision of the phase estimate as $1/2^{n_A}$. For this problem, our phase unitary is a Z rotation with phase $\phi=0.658203$. We setup the problem on the grid defined by (a) of Fig.~\ref{fig:dist_qpu} where the phase is applied to Q4. If there is no modular link then we solve on the $n=5$ qubit ring with $n_A=4$. Assuming the gates on this ring are zero length and perfect, the minimum relative error is 1.7\%. If we add the inter-module link to the problem to add 5 more qubits ($n_A=9$) then we can improve the relative error to zero. However, the error will increase if the fidelity of the link is less than one. We assume the link has a finite time to operate during which the link qubits, and all other qubits, will incur an error. The results are shown in Fig.~\ref{fig:qpe_error}. Again, this gives a estimate on the order of gate errors where adding a link will lead to improvement in the problem space. Although the QPE problem has more optimal solutions on small systems, such as iterative phase estimation, it is an example that is easily extended into other problem spaces. For example, when using VQE to estimate molecular energies using more qubits allows for more molecular orbitals, which may lead to improved accuracy. 

\section{Success Region Shapes}
We can understand the roughly rectangular shape of the success regions in the GAPPs as follows. The axes of these plots are in terms of the logarithmic internode infidelity, $\xi_I = \log I_{\text{local}}$, and the logarithmic average execution time, $ \xi_T = \log (T_{\text{link}})$. The infidelity due to local errors during the internode gate is $ \log I_{\text{local}}\sim N_q (\xi_T - \log T_*$ where $N_q$ is the number of local qubits and $T_{*} = T_{1}T_{2}/(T_1 + T_2)$ is the effective fidelity lifetime of a local qubit. Requiring a given overall logarithmic fidelity of $\log F$ then requires:
\begin{equation}
\log F = \text{const.} \sim log(e^{\xi_I} + e^{\xi_T}) 
\end{equation}
which can be seen to lead to a roughly rectangular shape, as $\log(e^x + e^y) = \text{const.}$ leads to a roughly rectangular curve.  

\section{MNQC Networks and Layout}

\begin{figure*}
    \centering
    \includegraphics[width = 2\columnwidth]{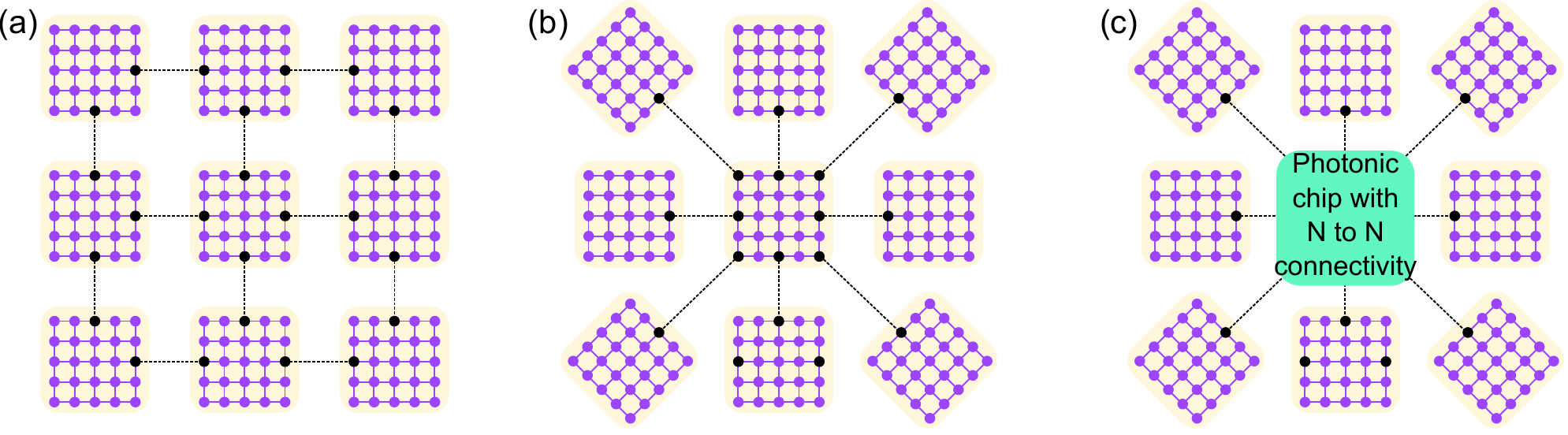}
    \caption{(a) Array-like, (b) star-like architectures of a distributed quantum computer. (c) Star-like architecture with an N-to-N photonic chip enabling EP generation between any communication qubits from other superconducting chips.}
    \label{fig:fid_archi}
\end{figure*}

In the long run, having low-loss telecom communications between multiple QPUs could add even more flexibility on the overall architecture of the MNQC. Finding the QPU connectivity layout that best facilitates the implementation of quantum algorithms is an important question to be addressed. Should the QPUs be arranged in an array-like structure (see Fig.~\ref{fig:fid_archi}(a))? This solution might be more appealing for implementation since the number of communication qubits per QPU remains constant. However, the distance between different QPUs might be an issue for compilation since a CX gate between two QPUs situated far appart might require a large number of inter-fridge CX gates. Should a better option be a  star-like structure  (see Fig.~\ref{fig:fid_archi}(b)), using a central node which is specialized for communication between the other nodes? This solution might reduce the average distance between QPUs and we discuss its potential in the following.

In the previous discussions, we have envisioned multiple QPUs communicating thanks to communication qubits, M2O converters, and fixed fiber links between different QPUs. In that setting however, we may not exploit to its full benefits the flexibility and adaptivity that allow photonic communications.
Over the years, the integrated photonics community has developed reconfigurable silicon photonic hardware, allowing to manipulate photons more efficiently~\cite{Peruzzo2014, Chen2018, Bogaerts2020}. Using a central photonic node has been for example envisioned for trapped ions~\cite{Monroe2014}. A universal photonic chip is an $N \times N$ linear interferometer based on phase-controled Mach-Zehnder interferometer and phase-shifters which allows to realize arbitrary unitary transformation on the input ports. By connecting the communication qubits to the input ports of such a chip and the output ports to single-photon detectors, we can centralize the heralded entanglement generation protocols between communication qubits through that interferometer. Moreover, contrary to the fixed structure where communication qubits are paired, such photonic chips should enable $N$-to-$N$ connectivity: each communication qubit can be connected to any other one. Therefore, using a central photonic node could allow to distribute easily EPs between any QPUs and thus drastically increase the overall modular quantum computer architecture (see Fig.~\ref{fig:fid_archi}(c)).

\end{document}